\documentclass[twocolumn,nofootinbib,aps,prd, 10pt]{revtex4-1}
\usepackage{amsmath,graphicx,multirow,hyperref,url,color,rotating,amssymb,bm}

\def\hmpcinv{h\,{\rm Mpc}^{-1}}
\def\hinvmpc{\,h^{-1}{\rm Mpc}}

\newcommand{\mnras}{MNRAS}

\newcommand{\ds}{\displaystyle}

\begin{document}

\title{Banana Split: Testing  the  Dark Energy Consistency with Geometry and Growth}
\author{Eduardo J. Ruiz}
\email{ejruiz@umich.edu}

\author{Dragan Huterer}
\email{huterer@umich.edu}
\affiliation{Department of Physics, University of Michigan, 450 Church St, Ann Arbor, MI 48109-1040}

\begin{abstract}
We perform parametric tests of the consistency of the standard $w$CDM model in
the framework of general relativity by carefully separating information
between the geometry and growth of structure. We replace each late-universe
parameter that describes the behavior of dark energy with two parameters: one
describing geometrical information in cosmological probes, and the other
controlling the growth of structure.  We use data from all principal
cosmological probes: of these, Type Ia supernovae, baryon acoustic
oscillations, and the peak locations in the cosmic microwave background
angular power spectrum constrain the geometry, while the redshift space
distortions, weak gravitational lensing and the abundance of galaxy clusters
constrain both geometry and growth.  Both geometry and growth separately favor
the $\Lambda$CDM cosmology with the matter density relative to critical
$\Omega_M\simeq 0.3$.  When the equation of state is allowed to vary
separately for probes of growth and geometry, we find again a good agreement
with the $\Lambda$CDM value ($w\simeq -1$), with the major exception of
redshift-space distortions which favor less growth than in $\Lambda$CDM at
3-$\sigma$ confidence, favoring the equation of state $w^{\rm grow}\simeq
-0.8$. The anomalous growth favored by redshift space distortions has been
noted earlier, and is common to all redshift space distortion data sets, but may well be caused by
systematics, or be explained by the sum of the neutrino masses higher than that
expected from the simplest mass hierarchies, $m_\nu \simeq 0.45$ eV.  On the
whole, the constraints are tight even in the new, larger parameter space due
to impressive complementarity of different cosmological probes.
\end{abstract}

\maketitle

\section{Introduction}\label{sec:intro}
The discovery of the acceleration of the universe's expansion
\cite{Riess_1998, Perlmutter_1999} has brought about one of the most
interesting and important questions in modern physics: what is the nature of
dark energy responsible for the acceleration? Arguably the simplest and
certainly the most popular candidate is vacuum energy, responsible for the
cosmological constant term in Einstein's equations. The cosmological
constant-dominated universe ($\Lambda$CDM), where the energy density today is
dominated by $\sim 75\%$ dark energy and $\sim 25\%$ matter, is well fit by
essentially all current data. Nevertheless, many alternatives to vacuum energy
have been discussed over the past 15 years or so. Some of these alternatives
involve scalar fields or other light degrees of freedom which obey the
standard equations of general relativity but lead to a richer dynamics and a
different expansion rate and growth of structure than $\Lambda$CDM and,
therefore, can in principle be distinguished from the latter. Nevertheless, in
all such explanations the growth of linear structures (matter density contrast
$\delta\equiv\delta\rho_M/\rho_M\ll 1$) evolves independently of the spatial
scale $k$ and can be obtained, well within the Hubble radius, by solving the
equation
\begin{equation}
\ddot\delta + 2H\dot\delta-4\pi G\rho_M \delta = 0,
\label{eq:growth}
\end{equation}
where $H$ is the Hubble parameter and dots are derivatives with respect to
time. For a review of dark energy observations and theory, see e.g.\ \citet{Frieman:2008sn}.  

A very different class of explanations fall in the category of modified
gravity (for an excellent review, see \cite{Joyce:2014kja}). Here the 
acceleration of the universe is caused by the corrections to general
relativity at large scales. These corrections obviously have to be suppressed
at Solar-System-size and perhaps galactic-size scales, and there are several known
mechanisms that do just that. Because the gravity theory is truly modified,
the growth is generally {\it not} given by Eq.~(\ref{eq:growth}), and moreover
the growth is not necessarily scale independent any more. Therefore, for a
fixed expansion rate $H(t)$ --- or, for that matter, the comoving distance as a
function of redshift $r(z)$ or any other geometric quantity --- the growth of
linear structures is different in standard and modified gravity. Moreover,
the time dependence of $\delta$ is in general $k$-dependent in modified gravity.

Comparing the geometrical quantities to the growth of structure is, therefore,
an excellent way to test the consistency of the fiducial standard-gravity
cosmological model; this was pointed out soon after the discovery of the
accelerating universe
\cite{Ishak:2005zs,Zhan:2008jh,Mortonson:2008qy,Mortonson:2009hk,Acquaviva:2010vr,Vanderveld:2012ec}. The
idea is to separately measure the redshift evolution of the geometrical
quantities such as distances on the one hand, and growth of structure on the
other, and test whether or not they are related by Eq.~(\ref{eq:growth}). This
approach is the same in spirit to a much more extensive body of work on
parameterizing the nonrelativistic and relativistic gravitational potentials,
$\Phi$ and $\Psi$ (which govern the motion of matter and of light,
respectively), and testing in whether they are the same or not
\cite{Zhao:2010dz,Bean:2010zq,Zhao:2011te,Hojjati:2011xd,Dossett:2011zp,Dossett:2011tn,Silvestri:2013ne}. In
practice and implementation, however, the two approaches are very
complementary.

Our goal is to make a major step forward in developing the first one of the
aforementioned consistency tests --- testing the consistency of $w$CDM (the
generalization of $\Lambda$CDM where the dark energy equation of state $w$ is
allowed to take constant values other than the $\Lambda$CDM value of -1) by
separately constraining the geometry and growth in major cosmological probes
of dark energy. This program has been started very successfully by
\citet{Wang_split} (see also \citep{Zhang:2003ii,Chu:2004qx,Abate:2008au}
which contained very similar ideas), who used data available at the time; the
constraints however were weak. Our overall philosophy and approach are similar
as those in Refs.~\cite{Wang_split,Zhang:2003ii,Chu:2004qx,Abate:2008au}, but
we benefit enormously from the new data and increased sophistication in
understanding and modeling them, as well as the availability of a few
additional cosmological probes not available in 2007.

The paper is divided as follows: we present the reasoning behind our approach 
in Sec.~\ref{sec:philosophy}. In Sec.~\ref{sec:probes} we review the
cosmological probes used in the analysis. A review of the
analysis method is provided in Sec.~\ref{sec:analysis}, and we
present our constraints on parameters in Sec.~\ref{sec:results}. 
We discuss these results in Sec.~\ref{sec:discussion}, and give
final remarks in Sec.~\ref{sec:conclusion}.

\section{Philosophy of our Approach}\label{sec:philosophy}

We would like to perform stringent but general consistency tests of the
currently favored $\Lambda$CDM cosmological model with $\sim$25\% dark
plus baryonic matter and $\sim$75\% dark energy, as well as the more
general $w$CDM model.  The $\Lambda$CDM model, favored since even before the
direct discovery of the accelerating universe (e.g.\ \cite{Krauss_Turner_95}),
is in excellent agreement with essentially all cosmological data, despite
occasional mild warnings to the contrary
(\cite{Scolnic:2013efb,Cheng:2013csa,Xia:2013dea,Shafer_Huterer}).  There has
been a huge amount of effort devoted to tests alternative to $w$CDM -- most
notably, modified gravity models where modifications to Einstein's General
Theory of Relativity, imposed to become important at late times in the
evolution of the universe and at large spatial scales, make it appear as if
the universe is accelerating if interpreted assuming standard general relativity.

Here we take a complementary approach, and study the internal consistency of the
$w$CDM model itself, without assuming any alternative model. We split the
cosmological information describing the late universe into two classes:
\begin{itemize}
\item Geometry: expansion rate $H(z)$ and the comoving distance
  $r(z)$, and associated derived quantities.
\item Growth: growth rate of density fluctuations in linear
  ($D(z)\equiv\delta(z)/\delta(0)$) and nonlinear regime.
\end{itemize}

Regardless of the parametric description of the geometry and growth sectors,
one thing is clear: in the standard model that assumes general relativity with
its usual relations between the growth and distances, the split parameters
$X_{i}^{\rm geom}$ and $X_{i}^{\rm grow}$ have to agree -- that is, be
consistent with each other at some statistically appropriate confidence
level. Any disagreement between the parameters in the two sectors, barring
unforeseen remaining systematic errors, can be interpreted as the violation of
the standard cosmological model assumption.

The split parameter constraints provide very general, yet powerful, tests of
the dominant paradigm.  They can be compared to more specific parameterizations
of departures from general relativity --- for example, the $\gamma$ parametrization
\cite{Linder_gamma}, or the various schemes of the aforementioned comparison
of the Newtonian potentials. Our approach is complementary to these more
specific parameterizations: while perhaps not as powerful in specific
instances, it is equipped with more freedom to capture departures from the
standard model.

\begin{table}[t]
\begin{center}
\begin{tabular}{c c c}
\hline \hline
Cosmological Probe & Geometry                                       & Growth \\ \hline 
SN Ia              & $H_0 D_L(z)$                                     &  ----- \\[0.2cm]
BAO                & $\ds\left(\frac{D_A^2(z)}{H(z)}\right )^{1/3}/r_s(z_d)$ &  ----- \\[0.35cm]
CMB peak loc.\     & $R\propto \sqrt{\Omega_m H_0^2}\,D_A(z_*)$         &  ----- \\[0.2cm]                
Cluster counts     & $\ds\frac{dV}{dz}$                             & $\ds\frac{dn}{dM}$ \\[0.25cm]
Weak lens 2pt      & $\ds\frac{r^2(z)}{H(z)} W_i(z) W_j(z)$         & \,\, $P\left (k=\ds\frac{\ell}{r(z)}\right)$\,\, \\[0.35cm]
RSD                & $F(z)\propto D_A(z)H(z)$                        &  $f(z)\sigma_8(z)$\\
\hline
\end{tabular}
\end{center}
\caption{Summary of cosmological probes that we used and aspects of geometry
  and growth that they are sensitive to. The assignments in the second and
  third column are necessarily approximate given the short space in the table;
  more detail is given in respective sections covering our use of these
  cosmological probes. Here $r_s(z_d)$ refers to the sound horizon evaluated
  at the baryon drag epoch $z_d$.}
\label{tab:summary}
\end{table}

Most of the cosmological measurements involve large amounts of raw data, and
their information is often compressed into a very small number of
meta-parameters. For example, weak lensing shows the two-point
correlation function, cluster number counts are given in mass bins, while
baryon acoustic oscillations, cosmic microwave background, and redshift space
distortion information is often captured in a small number of meta-parameters
which are defined and presented below. [Type Ia supernovae are somewhat of an
  exception, since we use individual magnitude measurements from each SN from
  the beginning.] Given that in {\it some} cases one assumes the cosmological
model (often $\Lambda$CDM) to derive these intermediate parameters, the
question is whether we should worry about using the meta-parameters to
constrain the wider class of cosmological models where growth history is
decoupled from geometry.  Fortunately, in this particular case our constraints
are robust: certainly for surveys that specialize in either geometry and
growth alone, the meta-parameters are {\it de facto} correct by construction,
and capture nearly all cosmological information of interest. For probes that
are sensitive to both growth and geometry, e.g. weak lensing and cluster
counts, the quantities used for the analysis --- correlation functions and
number counts, respectively --- provide a general enough representation of the
raw data that one can relax the assumption that growth and geometry are
consistent without the loss of robustness and accuracy.

\section{Observational Probes}\label{sec:probes}

We now discuss, in turn, the various cosmological probes used in this work: Type
Ia supernovae, the cosmic microwave background fluctuation power
spectrum, baryon acoustic oscillations, cluster counts, weak
gravitational lensing, and redshift space distortions.

In Table \ref{tab:summary} we summarize quantities or aspects of each
cosmological probe that are sensitive to geometry, and those that depend on
growth. In the following subsections, we describe in more detail the
cosmological probes, the quantities that they measure, and the data sets that
we use.

\subsection{Type Ia Supernovae}\label{sec:SNIa}

Type Ia supernovae (SNIa) are the principal probes of geometry of the
universe, as they directly measure the luminosity distance. Thus SNIa are
specialized in probing the geometrical parameters.

Each SNIa provides an independent measurement of the magnitude-redshift
relation. The theoretically expected apparent magnitude of the supernova at
redshift $z$ is
\begin{equation}
m_\text{th}(z) = 5 \log_{10}(H_0 D_L(z)) + \mathcal{M},
\end{equation}
where $\mathcal{M}$ is a nuisance parameter combining the intrinsic magnitude of
the supernova with the Hubble parameter $H_0$
\citep{Perlmutter_1999}. Therefore, each SNIa  constrains the luminosity
distance $D_L(z)$, with one overall nuisance parameter $\mathcal{M}$ to be
determined from the data as well.

There are several properties of supernovae that can change the magnitude of a 
supernova; these must be corrected for. The stretch (or broadness) of a 
supernova light curve is correlated with its brightness. Similarly, the color
of a supernova is also correlated with its brightness --- the broader
and bluer the supernova light curve, the brighter that supernova will be. We 
correct for these effects by writing the magnitude as \cite{Conley, Ruiz_Huterer}
\begin{equation}
m = m_\text{th} - \alpha_s\ (s-1) + \beta_\mathcal{C}\ \mathcal{C},
\end{equation}
where $s$ is the stretch and $\mathcal{C}$ the color of each SNIa, and $\alpha_s$ and 
$\beta_\mathcal{C}$ are additional, global nuisance parameters.

In addition to the statistical errors for each supernova measurement, we also
include the correlated systematic errors between each supernova measurement 
\cite{Conley, Ruiz_Huterer}. The covariance matrix resulting from these 
correlations  is also a function of $\alpha_s$ and $\beta_\mathcal{C}$.
Finally, we take into account host-galaxy effects in the value of $\mathcal{M}$
\cite{Conley, Shafer_Huterer} in our analysis. We allow two values of $\mathcal{M}$,
one for supernovae in lower-mass host galaxies and one for higher-mass galaxies.
These two $\mathcal{M}$'s are then marginalized over analytically. See Appendix C 
of \citet{Conley} for details.

\begin{figure}[t]
\includegraphics[width=0.45\textwidth]{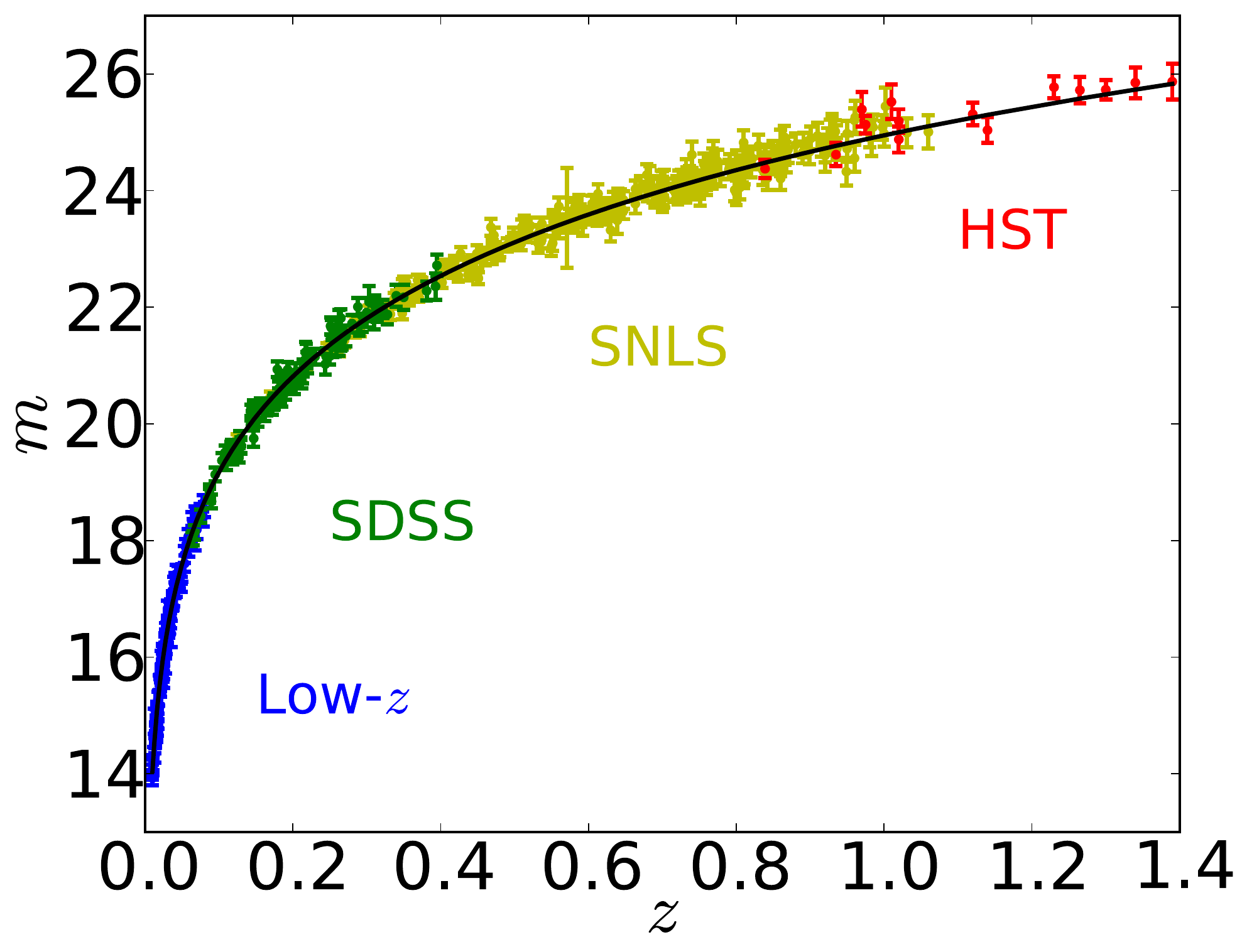}
\caption{Plot showing the set of 472 supernovae used in this work. Error bars are from
diagonal-only statistical errors. The black line shows the best-fit $\Lambda$CDM model 
with parameter values from column 2 of Table \ref{tab:results}.}
\label{fig:snedata}
\end{figure}

We use the Supernova Legacy Survey (SNLS) data compilation from \citet{Conley},
which contains 472 supernovae from various surveys,
including SNLS itself, the Sloan Digital Sky Survey (SDSS), some high redshift
supernovae  observed  by the Hubble Space Telescope (HST), and a selection of low-$z$
supernovae observed by various ground-based telescopes, collectively named
the ``Low-$z$'' sample. Supernova observations are summarized in Table \ref{tab:snedata}.

\begin{table}[t]
\begin{center}
\begin{tabular}{ c c c }
\hline \hline
\rule[-2mm]{0mm}{6mm} Source  & \,\,$N_{\mathrm{SN}}$\,\, & Redshift range \\ \hline 
\rule[-1mm]{0mm}{4mm} Low-$z$ & 123   & $0.01 - 0.1$ \\
\rule[-1mm]{0mm}{4mm} SDSS    & 93    & $0.06 - 0.4$ \\ 
\rule[-1mm]{0mm}{4mm} SNLS    & 242   & $0.08 - 1.05$\\ 
\rule[-1mm]{0mm}{4mm} HST     & 14    & $0.7 - 1.4$  \\ \hline
\end{tabular}
\end{center}
\caption{Summary of SNIa observations included in this analysis, showing the
  number of SN included from each survey and the approximate redshift
  ranges.}
\label{tab:snedata}
\end{table}

\subsection{CMB Peak Location}\label{sec:CMB}

The hot and cold spots of the cosmic microwave background (CMB) anisotropies
provide an excellent standard ruler: their angular separation, combined with
the sound horizon distance that is independently well determined (from the CMB
peaks' morphology), provides a single yet accurate measurement of the angular
diameter distance $D_A(z_*)$ to recombination. In addition to being very
high-redshift, this measurement of $D_A(z_*)$ is unique in that the physical
matter density $\Omega_M h^2$ is essentially fixed by the CMB peaks'
height. This is why the CMB peak location measurement traces out a very
complementary degeneracy direction in the $\Omega_M$--$w$ plane to
low-redshift measurements of distance \cite{Frieman:2002wi}.

For simplicity and clarity, we only use the geometrical measurement provided
by the CMB acoustic peaks' locations. The integrated Sachs-Wolfe (ISW) effect
of dark energy imprints on the CMB angular power spectrum on very large scales
adds very little to the information due to large cosmic variance. CMB is also
sensitive to the physics at the last-scattering surface \cite{Zahn:2002rr},
but recall that we decided to study the growth vs.\ geometry only in the late
universe, when dark energy becomes significant. Our use of the peaks'
  location only obviates the use the numerical  CMB codes
  that evaluate a full set of Einstein-Boltzmann equations,
  and speeds on this aspect of computation by a factor of $O(100)$.

Therefore, we use the aforementioned angular diameter distance to last
scattering with $\Omega_M h^2$ fixed, which is sometimes referred to as the
``shift parameter'' $R$, defined as
\begin{equation}
R = \sqrt{\Omega_m H_0^2}\, (1+z_*) D_A(z_*).
\label{eq:shiftpar}
\end{equation} 
To obtain a value of $R$, we use the Planck collaboration's Planck + WP
measurements of $r_*$ and $\theta_*$ \cite{Planck2013_XVI}; since $\theta_* =
r_*/D_A(z_*)$, we marginalize over these measurements assuming the $\Lambda$CDM
  cosmological model, as in \citep{Planck2013_XVI} to get a value for
$D_A(z_*)$. Combining this with the Planck values of $\Omega_M h^2$ and $z_*$,
we obtain 
\begin{equation}
R =1.7502 \pm 0.0073
\label{eq:shiftpar_meas}
\end{equation}
for their value of $z_* = 1090.48$. Being only sensitive to $\Omega_M$ and
$w$, $R$ presents a handy yet powerful constraint on the late universe.  When
using the CMB peak information alone, measurement of parameter $R$ in
Eq.~(\ref{eq:shiftpar_meas}) therefore provides complete information -- modulo
the aforementioned small ISW contribution -- about CMB's constraint on the late
universe.

Once we combine the CMB peaks information with that of other cosmological
probes and add the CMB early-universe prior (discussed further below in
Sec.~\ref{sec:param_space}), simply including the $R$ measurement would be
inconsistent as $R$ is necessarily correlated with the early universe
parameters, e.g.\ $\Omega_M h^2$. To do it correctly, we first extract the
$5\times 5$ covariance matrix from Planck which contains the $4\times 4$ early
universe prior shown in Table \ref{tab:covprior}, plus an additional row and
column corresponding to $R$. We than use the $5\times 5$ matrix as our early
universe prior that automatically and consistently includes the CMB peaks
information. Other probes are then added straightforwardly; see
Sec.~\ref{sec:like} for details.

\subsection{Baryon Acoustic Oscillations}\label{sec:BAO}

Baryonic acoustic oscillations (BAO) are features that arise from the
propagating sound waves in the early universe. The distance the sound wave can
travel between the Big Bang and decoupling -- the sound horizon -- imprints a
characteristic scale not only in the CMB fluctuations, but also in the
clustering two-point correlation function of galaxies. Roughly speaking, the
two-point correlation function is enhanced by $\sim 10\%$ at distances of
$\sim 100\hinvmpc$. This latter distance is, similarly to the CMB case,
well measured by the early-universe parameters ($\Omega_M h^2$ and $\Omega_B
h^2$ principally), but where we observe it is dependent on the expansion
history of the universe between the time that light from the galaxies is
emitted and today.

Specifically, for two galaxies at the same redshift separated by comoving
distance $r$ and seen with separation angle $\theta$, we have $\theta =
r/D_A(z)$ which enables measurement of the angular diameter distance given
known separation between galaxies. Similarly, two galaxies at the same angular
location but separated by redshift difference $\Delta z$ are separated by
comoving distance $r$, with the two quantities related via $\Delta z =r
H(z)$. The information from these transverse and radial sensitivities can be
conveniently combined into a single quantity, a generalized distance
$D_V(z_\text{eff})$ defined as \cite{Eisenstein2005}
\begin{equation}
D_V(z) \equiv \left( \frac{(1+z)^2 D_A^2(z) c z}{H(z)}\right)^{1/3}.
\end{equation}
The BAO surveys measure $r_s(z_d)/D_V(z_\text{eff})$ (or its inverse), where
$r_s(z_d)$ is the comoving sound horizon at the redshift of the baryon drag
epoch $z_d$,
\begin{equation}
r_s(z) = \frac{1}{\sqrt{3}}\int_0^{1/(1+z)} \frac{da^\prime}{a^{\prime 2} H(a^\prime) \sqrt{1 + 3\rho_b/4\rho_\gamma} }.
\label{eq:rs}
\end{equation}
In addition to the late-universe parameters, these BAO observable quantities
are only sensitive to the early-universe physics via a fixed single
combination, the sound horizon $r_s(z_d)$.

\begin{table}[t]
\begin{center}
\begin{tabular}{c c c c}
\hline \hline
\rule[-2mm]{0mm}{6mm} Survey & \,$z_\text{eff}$\, & \,Parameter\, & Measurement\\ \hline
\rule[-1mm]{0mm}{4mm} 6dFGS \cite{6dFGS}       & 0.106 & $r_s / D_V $ & $0.336 \pm 0.015 $ \\
\rule[-1mm]{0mm}{4mm} SDSS LRG \cite{SDSS_LRG} & 0.35  & $D_V / r_s $ & $8.88  \pm 0.17  $ \\
\rule[-1mm]{0mm}{4mm} BOSS CMASS \cite{BOSS}         & 0.57  & $D_V / r_s $ & $13.67 \pm 0.22  $ \\ \hline
\end{tabular}
\caption{BAO data measurements used here, together with the effective redshift
  for the corresponding galaxy sample.}
\label{tab:BAOdata}
\end{center}
\end{table}

It is important to note that the radiation term must be included in $H(a)$ in
Eq.~\eqref{eq:rs}.  The radiation energy density relative to critical is
$\Omega_r = \Omega_M a_\text{eq}$, where $a_\text{eq} = 1/(1 + z_\text{eq})$
is the scale factor at matter-radiation equality and
\begin{equation}
z_\text{eq} \approx 25000\ \Omega_M h^2 \left( \frac{T_\text{CMB}}{2.7 \text{K}}\right)^{-4}.
\end{equation}
The ratio of the baryonic density to the radiation density can be
approximated as
\begin{equation}
\frac{3\rho_b}{4\rho_\gamma} \approx 31500\ \Omega_B h^2 \left(
\frac{T_\text{CMB}}{2.7 \text{K}}\right)^{-4} a.
\end{equation}
We assume a value of $T_\text{CMB} = 2.7255 \text{K}$.

The redshift of the drag epoch can be approximated by the fitting formula \cite{Eisenstein1997}
\begin{equation}
z_d = \frac{1291 (\Omega_M h^2)^{0.251}}{1 + 0.659 (\Omega_M h^2)^{0.828}} \left[1 + b_1 (\Omega_B h^2)^{b_2}  \right]  ,
\end{equation}
where
\begin{align}
b_1 &= 0.313 (\Omega_M h^2)^{-0.419} \left[1 + 0.607 (\Omega_M h^2)^{0.674} \right], \\
b_2 &= 0.238 (\Omega_M h^2)^{0.223}. \nonumber
\end{align}

We use three sources of data for BAO constraints: the Six-degree-Field Galaxy
Survey (6dFGS) \cite{6dFGS}, the SDSS Luminous Red Galaxies (SDSS LRG)
\cite{SDSS_LRG}, and the SDSS-III DR9 Baryon Oscillation Spectroscopic Survey
(BOSS) \cite{BOSS}. These measurements and the corresponding redshift ranges
of their galaxy samples are summarized in Table \ref{tab:BAOdata}.

\subsection{Cluster Counts: MaxBCG}\label{sec:Clusters}

Counts of galaxy clusters are a particularly useful probe for this work, as
they probe both growth and geometry (for a review see
\citet{Allen:2011zs}). Cluster number density and its dependence on the
cosmological model are calibrated from N-body simulations; they are determined
by the growth of structure. On the other hand, the volume is purely a
geometric quantity that is straightforwardly calculated from first
principles. Product of the number density and volume gives the number
of clusters in some mass and redshift range, which can be compared to
measurements.

More specifically, the number of clusters within some mass and redshift range is
\begin{equation}
N = \int dM\ dz\  \frac{dn}{dM} \frac{dV}{dz} \psi(M)\phi(z)
\end{equation}
where $dn/dM$ is the halo mass function, $dV/dz$ is the comoving volume per
unit redshift, and $\psi(M)$ and $\phi(z)$ are the top-hat functions that
specify our binning in mass and redshift, that is, $\psi(M) = 1$ if $M$ is in
the mass bin of interest and 0 otherwise, and likewise for $\phi(z)$.

Here we use the measurements from the MaxBCG cluster catalog (\citet{Rozo}),
based on measurements from the Sloan Digital Sky Survey \cite{Koester:2007bg}.
A key proxy for measuring cluster masses is ``richness'', defined as
  the number of galaxies in $R_{200}$, the radius at which the average density
  of the cluster is 200 times that of the critical density of the universe.
The richness-mass relation has been calibrated using weak gravitational
lensing measurements from \citet{Johnston:2007}. For clarity and completeness,
we give further details of the \citet{Rozo} analysis that we adopt in Appendix
\ref{App:clusters}.

Cluster mass and redshift are not directly observable, but instead we rely on
cluster richness-mass relation and photometric redshift of cluster galaxy members,
respectively.  We  define $P(N_{200}|M)$ to be the probability that a
cluster of mass $M$ has a richness $N_{200}$, and $P(z_\text{photo}|z)$ to be
the probability that a cluster at redshift $z$ is observed with a photometric
redshift $z_\text{photo}$. We  redefine $\psi =
\psi(N_{200})$ and $\phi = \phi(z_\text{photo})$. The expected number of clusters then becomes
\begin{equation}
\langle N \rangle = \int dM\, dz\, \frac{dn}{dM} \frac{dV}{dz} \langle \psi|M\rangle \langle \phi|z\rangle
\label{eq:numexpect}
\end{equation}
where we introduce the probability weighting functions
\begin{align}
\langle\psi|M\rangle &= \int dN_{200}\, P(N_{200}|M)\psi(N_{200}), \label{eq:mprob}\\[0.1cm]
\langle\phi|z\rangle &= \int dz_\text{photo}\, P(z_\text{photo}|z)\phi(z_\text{photo}). \label{eq:zprob}
\end{align}
Here $P(z_\text{photo}|z)$ is modeled as a Gaussian distribution
as discussed in \citet{Rozo}. Meanwhile, $P(N_{200}|M)$ is modeled as
log-normal distribution, with the mean $\langle \ln N_{200} | M \rangle$
assumed to vary linearly with mass, resulting in two free parameters and an
unknown variance which is also treated as free parameter. These parameters
are marginalized in the analysis; see Appendix \ref{App:clusters} for details.

In a similar fashion, the expected total mass of clusters in a richness bin is
given by
\begin{equation}
\langle N\bar{M} \rangle = \beta\int dM\, dz \frac{dn}{dM} \frac{dV}{dz} \langle \psi|M\rangle \langle \phi|z\rangle.
\label{eq:massexpect}
\end{equation}
where another nuisance parameter $\beta$ is introduced to take into account
the uncertainty in the overall calibration of mass;
$\bar{M}_\text{obs}\rightarrow \beta \bar{M}_\text{obs}$. The comoving volume is simply
\begin{equation}
\frac{dV}{dz} = \Omega_\text{sky}\frac{r^2(z)}{H(z)}
\end{equation}
where $\Omega_\text{sky} = 2.254\text{ sr}$ is the solid angle covered by
SDSS and $r(z)$ is the comoving distance.

Finally, we use the Tinker mass function \cite{Tinker2008} for our halo mass
function $dn/dM$. The mass function requires the matter power spectrum as
input, and to speed up the code we calculate $P(k)$ semianalytically; for
that purpose we use the Eisenstein and Hu transfer function
\cite{Eisenstein1997}. We have checked that our calculation leads to
negligible differences in the results compared to one using CAMB's matter power
spectrum as input.

\subsection{Weak Lensing Shear: CFHTLens}\label{sec:WL}

Recent measurements by the Canada-France Hawaii Telescope Lensing Survey
(CFHTLenS) provide a very appealing test bed to apply our methodology and test
the consistency of the cosmological model, as weak lensing is  sensitive
to both growth and distance.

The CFHTLenS survey \cite{Erben_CFHTLens,Heymans_CFHTLens} covered 154 square
degrees over a period of five years in five wavebands ({\it ugriz}). The
resolved galaxy density is 17/arcmin$^2$. What is particularly appealing for
cosmological tests is that the survey is very deep (mean redshift $z_{\rm
  mean}\simeq 0.75$), implying that potentially strong constraints on the
temporal evolution of the effects of dark energy -- and, therefore, the growth
and geometry parameters -- can be achieved. A detailed analysis by the
CFHTLenS team made the shape measurements and obtained the photometric
redshift of galaxies, all the while dealing with a host of observational and
astrophysical systematic errors. The results are publicly available at the
survey web
site\footnote{\url{http://www.cfhtlens.org/astronomers/content-suitable-astronomers}}. We
use their {\tt blu\_sample} data, which were shown in \cite{Heymans_CFHTLens}
to have a negligible intrinsic alignment signal. The data are given in six
tomographic redshift bins, and presented at five different angles,
$\theta=\{1.73'', 3.75''. 8.13'', 17.6'', 37.9''\}$. The data is given for the
two 2-point correlation functions $\xi^+$ and $\xi^-$, defined as
\begin{equation}
\xi_{ij}^{\pm} = \frac{1}{2\pi}\int_0^\infty d\ell\, \ell\,
P^\kappa_{ij}(\ell)J^{\pm}(\ell\theta),
\end{equation}
where $\ell$ is the multipole, and
$J^{+}(x) \equiv J_0(x)$ and $J^{-}(x) \equiv J_4(x)$. Here $P^\kappa$ is the
weak lensing convergence power spectrum, that is, the two-point correlation
function of the convergence field on the sky, given as a function of the
multipole $\ell$. In the Limber approximation, which only includes
modes perpendicular to the line of sight and is an excellent approximation at
scales of interest, the convergence power is given as 
\begin{equation}
P^\kappa_{ij}(\ell) = \int dz\,\frac{r^2(z)}{H(z)} W_i(z) W_j(z)\,
P\left (k=\frac{\ell}{r(z)}\right),
\end{equation}
where $r(z)$ and $H(z)$ are the comoving distance and Hubble parameter
respectively, and the weight functions involve the distribution of galaxies
$dN/dz$ in each redshift bin
\begin{equation}
W_i(z) = \frac{3}{2}\Omega_M H_0^2 g_i(z) (1+z),
\end{equation}
where the weight function is given in terms of the radial distance $\chi= \int
dz/H(z)$,
\begin{eqnarray}
g_i(\chi(z)) &=& r(\chi) \int_{\chi}^\infty d\chi_s   n_i(\chi_s)
\,\frac{r(\chi_s-\chi)}{r(\chi_s)} \\[0.2cm]
&\longrightarrow & r(z) \int_{z}^\infty \frac{dz_s}{H(z_s)}  
n_i(z_s)\,\frac{r(z_s)-r(z)}{r(z_s)}.\nonumber 
\end{eqnarray}
Here the second line holds in the special case of a flat universe which we
adopt in the paper, and where $n(z_s)$ is the distribution of source galaxies in each
redshift bin, normalized to $n_i(z_s)dz_s=1$, and provided by CFHTLenS for each
tomographic bin (see Fig.~1 of \citet{Heymans_CFHTLens}). 

Finally, special attention is required to modeling the power spectrum $P(k)$,
given that scales probed are small --- consider, for example, that the
smallest angle $\theta=1.73''$, at the mean redshift of the survey $z\simeq 1$
spans $k\simeq 1 \hmpcinv$, which is in a regime of strongly nonlinear
clustering. It is imperative to have an accurate theoretical prediction for
the dark matter clustering at these scales which are a ``sweet spot'' for
sensitivity for weak lensing surveys \cite{Huterer_Takada}. Here we adopt an
updated version of the {\tt halofit} \cite{halofit} prescription for
nonlinear clustering given by \citet{Takahashi_2012}. This fit has the same
functional form as the original {\tt halofit}, but with updated parameter
values. The formula has been optimized for the dark energy
  equation of state $w\simeq -1$, justifying its use in this analysis. We
find that the Takahashi et al.\ prescription makes a non-negligible difference
relative to the original; for example, the best-fit $\sigma_8$ value, in a
simplified analysis we ran as a check, moves downwards by $\sim$0.03 relative
to the original {\tt halofit}, returning $\sigma_8\simeq 0.74$ (for a fixed
$\Omega_M=0.3$), in agreement with \citet{Heymans_CFHTLens}.

We also checked the robustness of the data assumptions by verifying that the
{\tt blue} and {\tt full} data sets from CFHTLens give very similar
constraints.

\begin{figure}[b]
\includegraphics[width=0.45\textwidth]{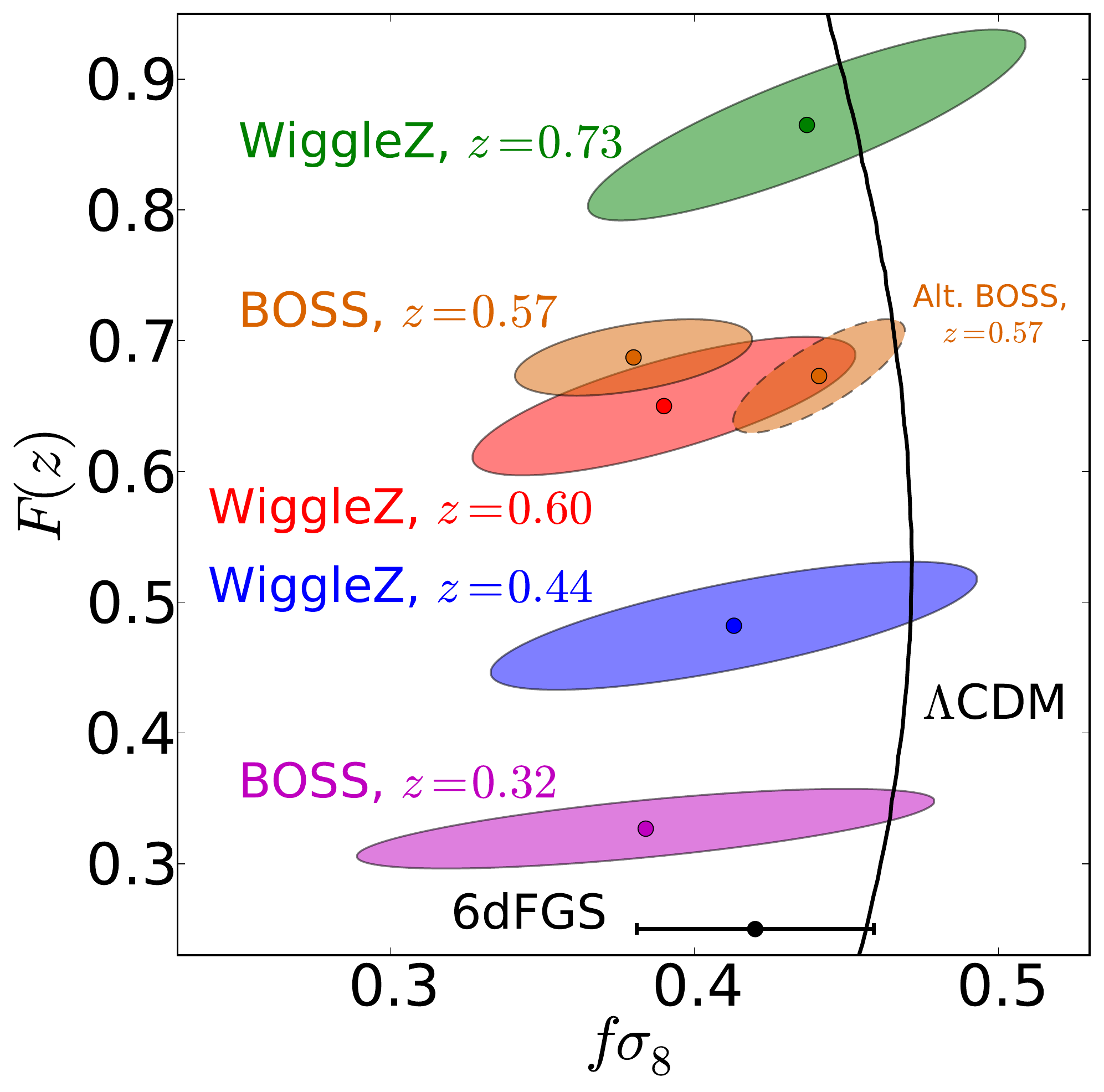}
\caption{RSD data used in our analysis, shown in the $f\sigma_8$--$F$ plane;
  more details can be found in Table \ref{tab:rsddata}. The black line shows
  the best-fit $\Lambda$CDM model with our best-fit parameter values given
  in the second column of Table \ref{tab:results}. The low-redshift 6dFGS measurement does
  not have an associated value for $F(z)$, and we therefore only show its
  horizontal error bar. The BOSS constraint on $F$ is obtained from the
  covariance of $H(z)$ and $D_A(z)$; see Appendix \ref{app:rsd-details} for
  details. The dashed error ellipse corresponds to an alternative RSD
  measurement at $z=0.57$ from \citet{Samushia:2013yga}; for details, see
  Sec.~\ref{sec:discussion}.}
\label{fig:rsd-data}
\end{figure}

\subsection{Redshift Space Distortions}\label{sec:RSD}

Redshift space distortions (RSD) refer to the effect of how density modes
affect velocity distribution of galaxies in their vicinity. Galaxies' peculiar
velocities are imprinted in galaxy redshift surveys in which recessional
velocity is used as the line-of-sight coordinate for galaxy positions, leading
to an apparent compression of radial clustering relative to transverse
clustering on large spatial scales (a few tens of Mpc). On smaller scales (a
few Mpc), one additionally observes the ``finger-of-God'' elongation
\cite{Jackson_FoG} due to nonlinear effects.  The spatial clustering of
galaxies is affected on scales corresponding to the size of the largest
objects (galaxy clusters) and larger, all the way up to $\sim 100$Mpc.
Measuring the clustering at these scales and at various redshifts provides
valuable information about the growth of structure across cosmic history.

RSD measurements are uniquely sensitive to the combination of cosmological
parameters $f(a) \sigma_8(a)$ (often just referred to as $f\sigma_8$) \cite{Song:2008qt}, where
$f (a)\equiv d\ln D/d\ln a$ and $D(a)$ is the linear growth factor.

In addition to pure growth information, however, we must take into account the
geometrical aspect of the RSD measurements, which comes about from the
breaking of underlying isotropy of galaxy clustering when observed in
redshift space. The effect is accurately captured by the parameter which
serves to compare clustering in the radial and tangential directions
\cite{Ballinger:1996cd,Matsubara:1996nf,Simpson:2009zj}, and which has been
motivated by the original analysis by \citet{Alcock:1979mp}
\begin{equation}
F(z) \equiv (1+z) H(z) D_A(z)/c
\label{eq:F}
\end{equation}
where $H(z)$ is the Hubble parameter and $D_A(z)$ is the angular
distance. Intuitively, the comoving diameter a spherical object (or, more
  generally, a feature in the clustering of galaxies) $ds$ at redshift $z$ is
  related to its angular size on the sky $\Delta\theta$ by
  $ds=D_A(z)\Delta\theta$. The diameter of the feature can also be related to its redshift
  extent $\Delta z$ via $ds=c\Delta z/[(1+z)H(z)]$.  By comparing the angular
  and redshift dimensions of the feature (i.e.\ measuring $\Delta\theta/\Delta
  z$) one can then determine the parameter combination given in
  Eq.~(\ref{eq:F}). Alternatively, the effect is captured by the separate but
correlated measurements of $H(z)$ and $D_A(z)$. These parameters all measure
geometric effects and thus grant RSD the ability to test both geometry and
growth.

We use a compilation of measurements of $f\sigma_8$, $F(z)$, $H(z)$, and $D_A(z)$
from a number of spectroscopic surveys; these are summarized in Table 
\ref{tab:rsddata} and illustrated in Fig. \ref{fig:rsd-data}.

\begin{table}[ht]
\begin{center}
\begin{tabular}{c c c c}
\hline \hline
$z$ & Parameter & Measurement (diag) & Survey\\ \hline
$0.067$ & $f\sigma_8$ & $0.42 \pm 0.06$ & 6dFGS \cite{Beutler:2012px}\\
$0.32$  & $   H(z)  $ & $78.1 \pm 7.1$  & BOSS Low-z \cite{Chuang:2013wga}\\
$0.32$  & $  D_A(z) $ & $950  \pm 61$   & BOSS Low-z \cite{Chuang:2013wga}\\
$0.32$  & $f\sigma_8$ & $0.38 \pm 0.10$ & BOSS Low-z \cite{Chuang:2013wga}\\
$0.44$  & $   F(z)  $ & $0.48 \pm 0.05$ & WiggleZ \cite{Blake:2012pj}\\
$0.44$  & $f\sigma_8$ & $0.41 \pm 0.08$ & WiggleZ \cite{Blake:2012pj}\\
$0.57$  & $   H(z)  $ & $97.1 \pm 5.5$  & BOSS CMASS \cite{Chuang:2013wga}\\
$0.57$  & $  D_A(z) $ & $1351 \pm 60$   & BOSS CMASS \cite{Chuang:2013wga}\\
$0.57$  & $f\sigma_8$ & $0.38 \pm 0.04$ & BOSS CMASS \cite{Chuang:2013wga}\\
$0.60$  & $   F(z)  $ & $0.65 \pm 0.05$ & WiggleZ \cite{Blake:2012pj}\\
$0.60$  & $f\sigma_8$ & $0.39 \pm 0.06$ & WiggleZ \cite{Blake:2012pj}\\
$0.73$  & $   F(z)  $ & $0.87 \pm 0.07$ & WiggleZ \cite{Blake:2012pj}\\
$0.73$  & $f\sigma_8$ & $0.44 \pm 0.07$ & WiggleZ \cite{Blake:2012pj}\\\hline

\end{tabular}
\caption{RSD measurements from various surveys. Each line shows the effective
  redshift associated with the data point, the measured parameter, the value
  of that parameter with associated diagonal error, and the data point's
  associated survey.  Measurements from the same survey are correlated;
  \citep{Beutler:2012px,Chuang:2013wga,Blake:2012pj}; for brevity we show the
  diagonal errors (i.e.\ square roots of parameter variances) here and the full
  covariance matrices in Appendix \ref{app:rsd-details}.}
\label{tab:rsddata}
\end{center}
\end{table}

\section{Parameters and analysis}\label{sec:analysis}

\subsection{Parameter space}\label{sec:param_space}

We adopt the following set of fundamental cosmological parameters
\begin{equation}
\vec{\textbf{p}}^\text{fund} = \lbrace \Omega_M, \Omega_M h^2, \Omega_B h^2, w, 10^9A, n_s \rbrace
\label{eq:par_fund}
\end{equation}
where $\Omega_M$ and $\Omega_B$ are the energy densities in matter and
baryons relative to critical density, $w$ is the equation of state of dark
energy, $A$ is the amplitude of the primordial curvature power spectrum on
scale of 0.05 Mpc$^{-1}$, and $n_s$ is the scalar spectral index of curvature
perturbations.  We also include the nuisance
parameters
\begin{equation}
\vec{\textbf{p}}^\text{nuis} = \lbrace \alpha_s, \beta_\mathcal{C}, \langle \ln N|M_1 \rangle,
\langle \ln N|M_2 \rangle, \sigma_{NM}, \beta \rbrace,
\label{eq:par_nuis}
\end{equation}
where $\alpha_s$ and $\beta_\mathcal{C}$ are the supernovae nuisance
parameters, while the others enter the cluster count analysis.  Our analysis
also produces constraints on several derived parameters,
\begin{equation}
\vec{\textbf{p}}^\text{deriv} = \lbrace \sigma_8, h;\sigma_{MN} \rbrace.
\label{eq:par_deriv}
\end{equation}
Here, $\sigma_{MN}$ is the scatter of the richness for a given mass
(opposed to $\sigma_{NM}$, which is the scatter of the mass for a given richness),
and is considered a derived nuisance parameter.

Throughout we assume a constant equation of state parameter $w$ for analyses,
as well as a flat universe ($\Omega_K = 0$). The latter assumption
  effectively assumes standard inflation, and also has a very practical benefit
  of improving the convergence of the parameter constraints. At any rate, in
  this paper we are interested in testing the consistency of the dark energy
  sector, which is typically unrelated to the flatness of the universe.  In
addition, we set the sum of neutrino masses to $m_\nu = 0.06$ eV, which is
consistent with atmospheric and solar data on neutrino flavor oscillations and
a normal hierarchy between individual mass eigenstates \cite{Beringer:1900zz}.
Note that, in our extended tests in Sec.~\ref{sec:discussion}, we also vary
the neutrino mass $m_\nu$. The number of neutrino species is held fixed at
$N_\nu=3.046$ throughout the analysis, as predicted by the standard model.

We adopt priors on $\Omega_M$, $\sigma_{NM}$, $\beta$, and $\sigma_{MN}$ from
\citet{Rozo}. In addition, we add very weak top-hat priors on $h$, $w$ and
$n_s$.  See Table \ref{tab:params} for details.

We also impose a multidimensional Gaussian prior based on Planck constraints
on $\Omega_M h^2$, $\Omega_B h^2$, $10^9A$, and $n_s$; we term this the
early-universe prior (``EU'' for short in our plots). While we would have
ideally liked to run our analyses without this prior, we find that the MCMC
runs without the prior have difficulty converging in the large parameter space
with split geometry and growth late-universe parameters.  The early-universe
prior correlation matrix is calculated from Planck $\Lambda$CDM (+ lowl) MCMC
chains \cite{Planck2013_XVI}; see Table \ref{tab:covprior}. The square roots
of the diagonal entries of the full covariance matrix prior -- the
unmarginalized errors of the prior -- are shown in Table \ref{tab:params}. We
apply this full prior covariance to RSD, WL and clusters, and the overall
combined constraint.  In the case of BAO, we apply only information coming
from the $2\times 2$ subset of this matrix containing $\Omega_M h^2$ and
$\Omega_B h^2$, corresponding to the sound horizon (``SH'' in our plots). The
Planck prior changes very little if one assumes the underlying Planck $w$CDM
model instead of $\Lambda$CDM, as has been verified explicitly by the authors,
implying that it should represent the early-universe information with the
sufficient accuracy even when the late-universe parameters have been split.

\begin{table}[th]
\footnotetext[1]{\label{foot:covprior}These errors are the diagonal parts of
  the full covariance matrix prior. See table \ref{tab:covprior} for further
  details and the full correlation matrix.}
\begin{center}
\begin{tabular}{c c c c}
\hline \hline
Parameter & Priors & Geometry & Growth\\ \hline
$\Omega_M$ & [0.05, 0.95] & \checkmark & \checkmark\\
$\Omega_M h^2$ & $0.1423 \pm 0.0029$\textsuperscript{\ref{foot:covprior}} &  & \checkmark\\
$\Omega_B h^2$ & $0.02207 \pm 0.00033$\textsuperscript{\ref{foot:covprior}}&  & \checkmark\\
$w$ & [-2, 0] & \checkmark & \checkmark\\
$10^9A$ & $2.215\pm 0.16$\textsuperscript{\ref{foot:covprior}}&  & \checkmark\\
$n_s$ & [0.9, 1.1], $0.9616 \pm 0.0094$\textsuperscript{\ref{foot:covprior}}&  & \checkmark\\\hline
$\sigma_8$ & --- &  \multicolumn{2}{c}{derived par.} \\
$h$ & [0.5, 1.0] &  \multicolumn{2}{c}{derived par.}\\\hline
$\alpha_s$                    & --- & \multicolumn{2}{c}{nuisance par.} \\
$\beta_\mathcal{C}$            & --- & \multicolumn{2}{c}{nuisance par.} \\
$\langle \ln N | M_1 \rangle$ & --- & \multicolumn{2}{c}{nuisance par.}\\
$\langle \ln N | M_2 \rangle$ & --- & \multicolumn{2}{c}{nuisance par.}\\
$\sigma_{NM}$           & [0.1, 1.5] & \multicolumn{2}{c}{nuisance par.} \\
$\beta$ & [0.5, 1.5], $1.0 \pm 0.06$ & \multicolumn{2}{c}{nuisance par.} \\\hline
$\sigma_{MN}$ & $0.45 \pm 0.1$ &  \multicolumn{2}{c}{der.\ nuis.\ par.} \\\hline

\end{tabular}
\caption{Parameters used in our analysis. The first seven parameters lying
  above the horizontal line are the fundamental quantities that we varied in
  the Markov chains. The next two parameters are derived from the fundamental
  parameters. Those in the final sections are nuisance parameters, again
  separated into fundamental (six) and derived (one). In the 'Priors' column,
  notation $[a,b]$ indicates a flat prior between the end points $a$ and $b$,
  while $c \pm d$ indicates a Gaussian prior with mean $c$ and standard
  deviation $d$. For the basic set of cosmological parameters (i.e.\ the first
  six above), we include information about whether they enter the geometry or
  growth in the final two columns. If a parameter is found in both columns, it
  is necessarily a split parameter.}
\label{tab:params}
\end{center}
\end{table}

\subsection{Likelihood}\label{sec:like}

We assume that the likelihood is Gaussian in suitably chosen meta-parameters
for each cosmological probe. We assign the individual likelihoods as follows:
\begin{itemize}
\item SNIa: the data vector consists of SN magnitudes, and we calculate the full
  off-diagonal covariance matrix that takes into account errors in magnitude,
  stretch factor, color, redshift, and gravitational lensing. See Appendix C
  of \citet{Conley} for details.
\item CMB peak location: the data vector consists of the single measurement of
  the ``shift parameter'' $R$; see Eq.~(\ref{eq:shiftpar}). In the
  combined-probe analysis, we account for the correlation of $R$ and the
  early-universe parameters, as explained near the end of Sec.~\ref{sec:CMB}.
\item BAO: data vector and corresponding (diagonal) errors are quantities given in
  Table \ref{tab:BAOdata}. Because the SDSS and BOSS CMASS samples cover different
  redshift ranges, and the two are in the northern hemisphere while 6dFGS is
  in the south, it is a good approximation to ignore correlations between
  these three surveys. 
\item Clusters: following \citet{Rozo}, we utilize both the number counts, and
  number-weighted mass counts in richness; details are explained in Appendix
  \ref{App:clusters}.
\item Weak lensing (WL): data vector are the correlation functions $\xi_{ij}^{\pm}(\theta)$
  given for six redshift bins (so $i\leq j\leq 6$) and for measurements
  at five values of $\theta$. The total length of the vector is therefore
  $2\times (6\times 7/2)\times 5 = 210$. The $210\times 210$ covariance
  matrix, calculated using numerical simulations, is provided by the CFHTLens
  team \cite{Heymans_CFHTLens}.
\item RSD: data vector and corresponding (diagonal) errors are quantities given in
  Table \ref{tab:rsddata}. The correlation matrices for the off-diagonal errors 
  between data points can be found in Tables \ref{tab:rsd-corr_boss} and 
  \ref{tab:rsd-corr_wigglez} in Appendix \ref{app:rsd-details}.
\end{itemize}
\begin{table}[th]
\begin{center}
\begin{tabular}{c|cccc} \hline \hline
 & $\Omega_M h^2$ & $\Omega_B h^2$ & $10^9A$  & $n_s$ \\\hline
$\Omega_M h^2$ & $1.00$  & $-0.62$ & $-0.51$ & $-0.84$ \\
$\Omega_B h^2$ &   ---   & $1.00$  & $0.56$  & $0.70$ \\
$10^9A$ &   ---   &   ---   & $1.00$  & $0.65$ \\
$n_s$          &   ---   &   ---   &   ---   & $1.00$ \\\hline
\end{tabular}
\caption{Correlation matrix corresponding to our early-universe prior (labeled
  as ``EU'' in our plots).  The correlation matrix is calculated from Planck
  $\Lambda$CDM (+ lowl) MCMC chains \cite{Planck2013_XVI}. The square roots of
  the diagonal entries of the full covariance matrix prior -- the
  unmarginalized errors of the prior -- are shown in Table
  \ref{tab:params}. We apply this full prior covariance to RSD, WL and
  clusters, and the overall combined constraint. In the case of BAO, we apply
  only information coming from the $2\times 2$ subset of this matrix
  containing $\Omega_M h^2$ and $\Omega_B h^2$, corresponding to the sound
  horizon (``SH'' in our plots).}
\label{tab:covprior}
\end{center}
\end{table}

\begin{figure*}[th]
\includegraphics[width=0.437\textwidth]{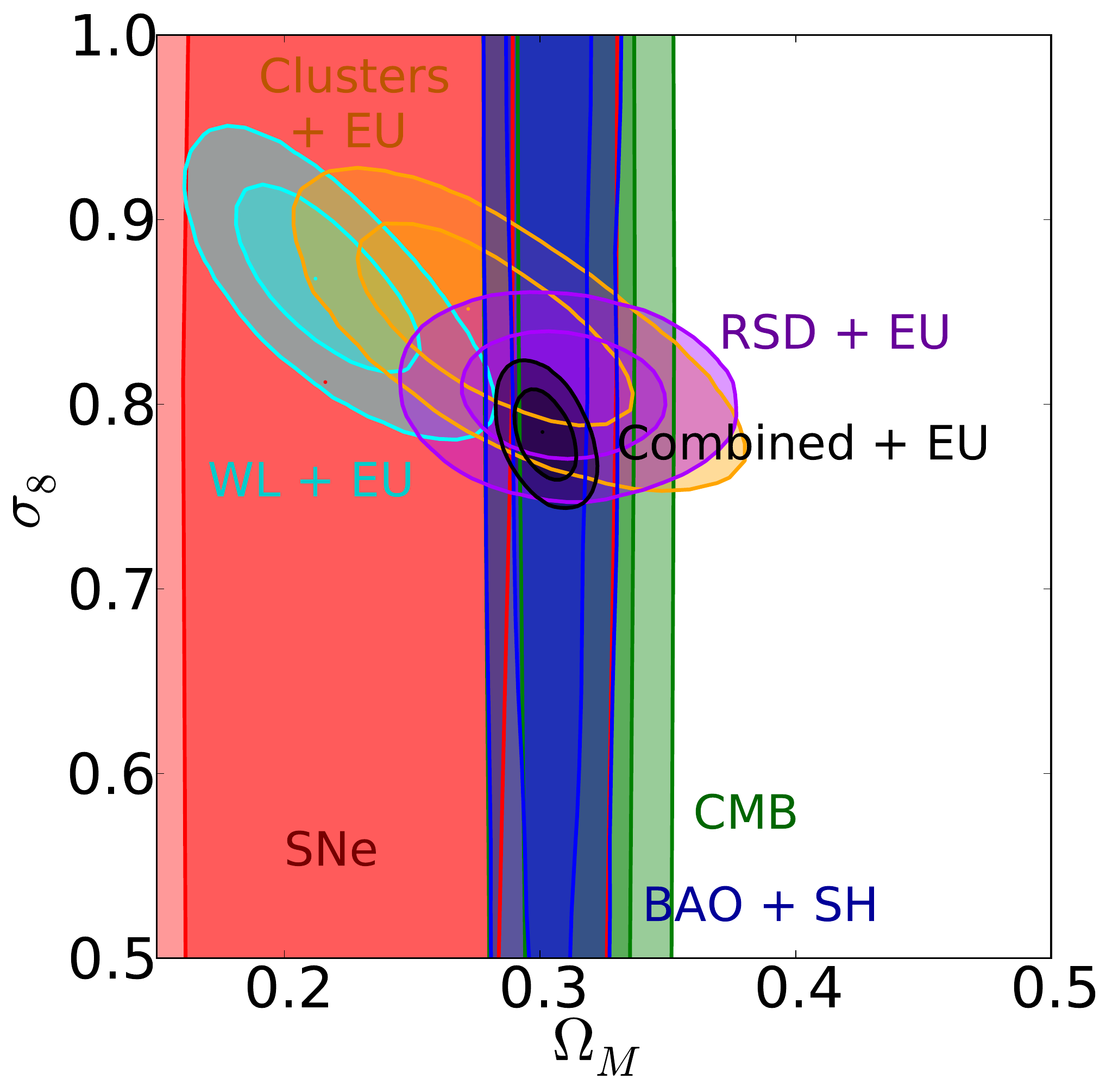}
\includegraphics[width=0.45\textwidth]{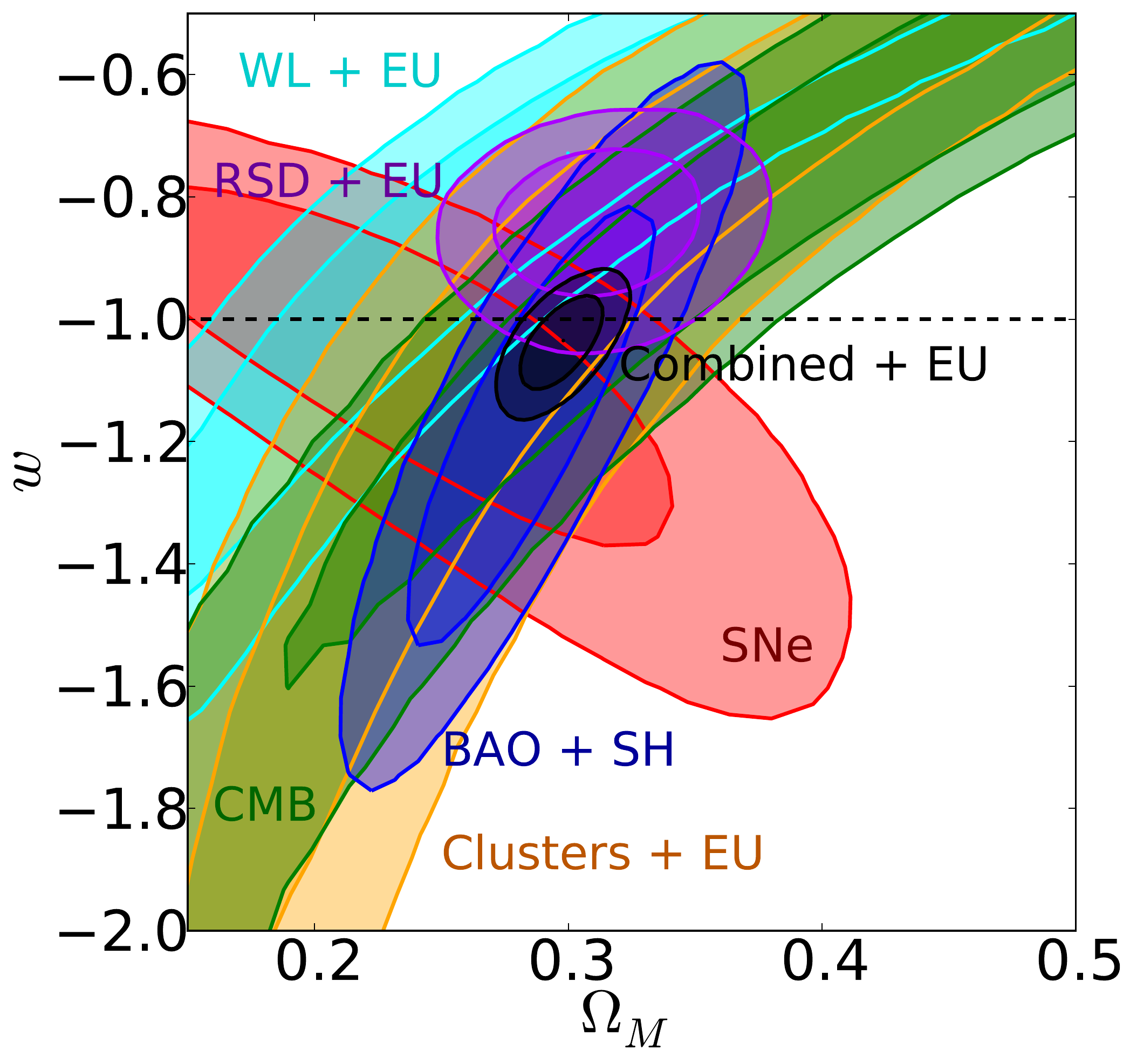}
\caption{Fiducial constraints from cosmological probes {\it before} the
  geometry-growth parameter split. We show the 68\% and 95\% confidence
  constraints in the $\Omega_M$--$\sigma_8$ plane assuming $w=-1$ held
  constant (left panel) and in the $\Omega_M$-$w$ plane (right panel). In the
  labels, ``EU" refers to our early universe prior, while ``SH" refers to the
  sound horizon prior; see Table \ref{tab:covprior} for relevant details.}
\label{fig:unsplit}
\end{figure*}

\begin{figure*}[th]
\includegraphics[width=0.80\textwidth]{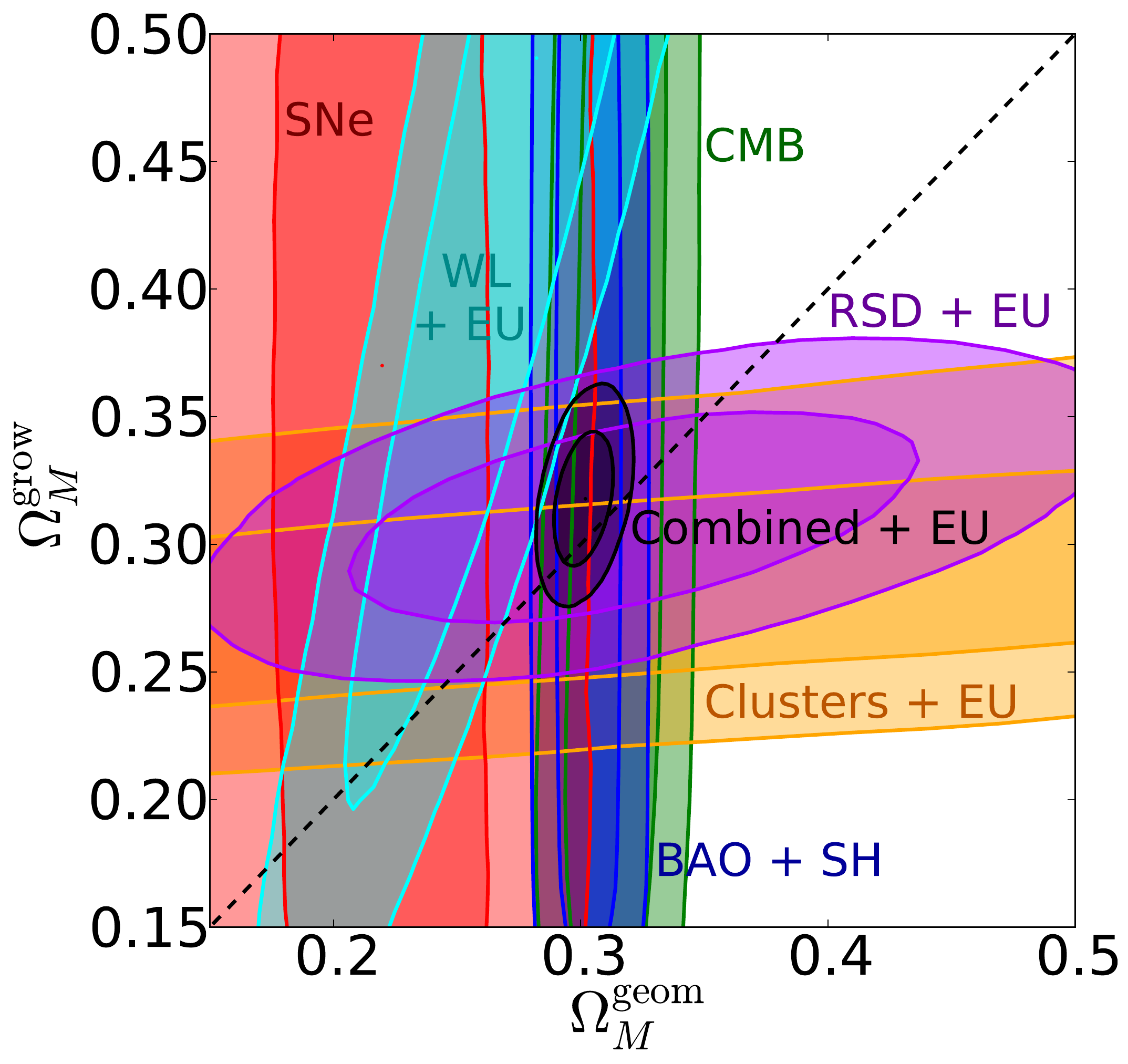}
\caption{ 68\% and 95\% confidence constraints in the split $\Omega_M$ plane
  with the equation of state held constant at the $\Lambda$CDM value
  ($w^\text{geom} = w^\text{grow}=-1$). As in Fig. \ref{fig:unsplit}, 
  ``EU" refers to our early universe prior, while ``SH" refers to the
  sound horizon prior.}
\label{fig:om-split}
\end{figure*}

\begin{figure*}[th]
\includegraphics[width=0.8\textwidth]{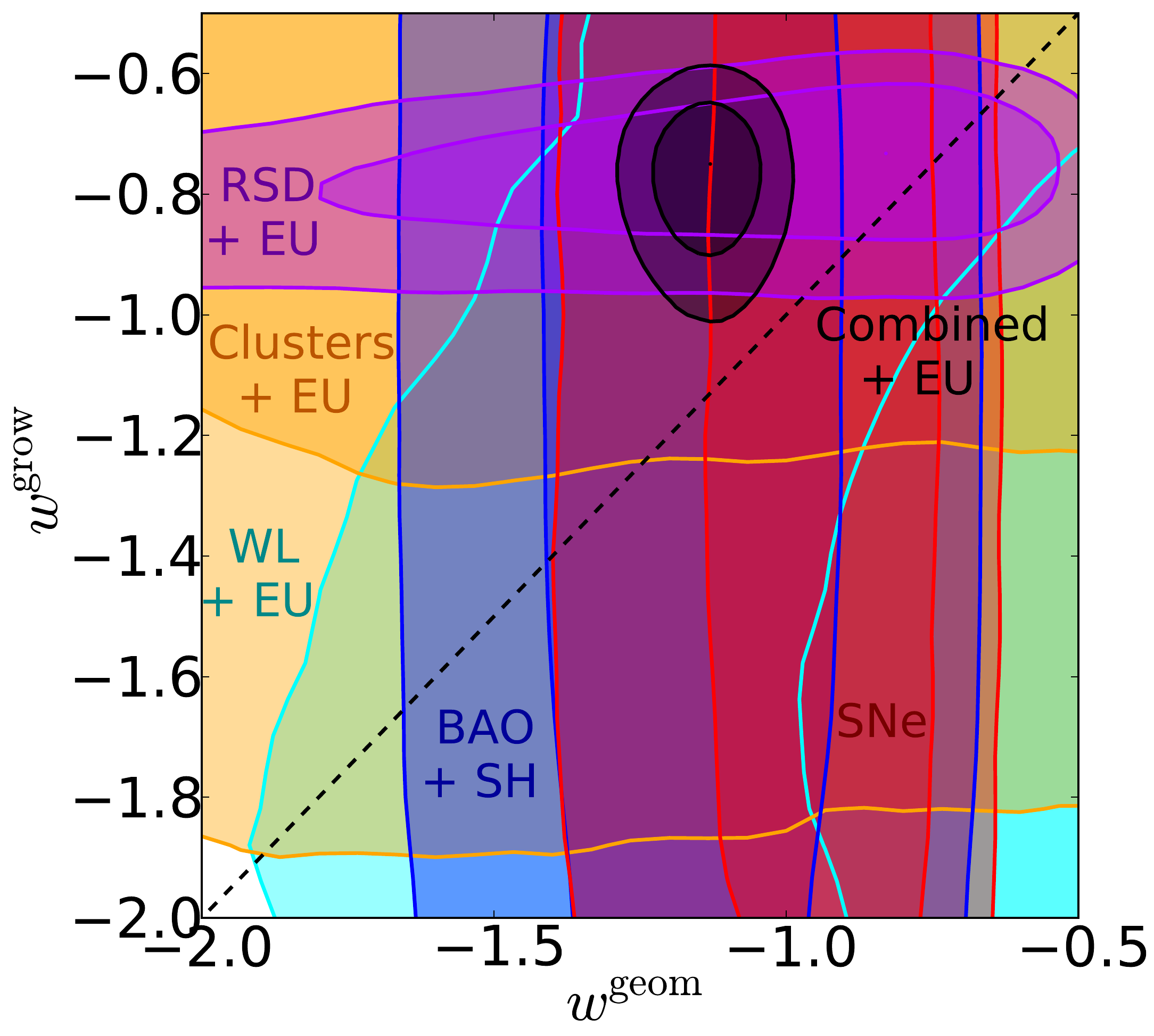}
\caption{68\% and 95\% confidence constraints in the split $w$ plane. Note
  that the combined $2-\sigma$ contour does not pass through the
  $w^\text{geom} = w^\text{grow}$ line. As before, ``EU" refers to our early
  universe prior, while ``SH" refers to the sound horizon prior. Individual CMB results
  have been omitted due to the poor constraints they provide in this plane,
  but they are included in the combined constraint. See text for details.}
\label{fig:w-split}
\end{figure*}

The likelihood of the combined cosmological probes is given by the product of
individual likelihoods:
\begin{equation}
\hspace{-0.2cm}
\mathcal{L} =
\mathcal{L}_\text{SNIa}\,
\mathcal{L}_\text{CMBpeak}\,
\mathcal{L}_\text{BAO}\,
\mathcal{L}_\text{cluster}\,
\mathcal{L}_\text{WL}\,
\mathcal{L}_\text{RSD} \,
\mathcal{L}_\text{prior}.
\end{equation}
The assumption that the individual likelihoods are independent may well be
questioned, but it is in practice well justified by the nature of the data sets
that we combine. CMB peak location is decoupled from other probes, as it is a
much higher-redshift measurement. Similarly, cluster counts are a 1-point
correlation function, and as such only weakly coupled to clustering. Weak
lensing is expected to be  slightly correlated with SNIa, as the latter
are also weakly lensed, but the effect is very small for current data.

Perhaps the biggest worry is potential correlation between the BAO and RSD,
since these use the same spatial scales (e.g.\ 32-100 Mpc for the BOSS CMASS
sample) and, in the case of both Wigglez and BOSS, the same galaxies. This
correlation occurs because the RSD are partially sensitive to the
Alcock-Paczynski parameter combination $F(z)\propto H(z)D_A(z)$; this in turn
{\it may} be slightly degenerate with BAO measurements, depending on the
treatment of the broadband clustering power in the BAO analysis.  Direct
estimates indicate that the correlation between the RSD and BAO measured
quantities are at the 10\% level (e.g.\ Table 2 of \citet{Blake:2012pj} and
Tables 2, 4 and 6 in \citet{Chuang:2013wga}). Therefore, simply multiplying
the BAO and RSD likelihoods is justified.

At face value, the Gaussian assumption for the likelihoods might seem risky
and unrealistic. Certainly, the {\it exact} likelihood in any given probe will
not be precisely Gaussian, even if evaluated in parameters that are
well measured by the cosmological probes (e.g.\ the apparent magnitudes of
SNIa). Nevertheless, in addition to making the problem vastly more tractable,
the assumption of Gaussianity seems to be well justified at this stage: for
cosmological models that fit the data well, tails of the distribution are not
as important. Had our analysis been oriented toward ruling out $w$CDM --
using, for example, Bayesian model-selection techniques -- then the analysis
would have perhaps warranted a much more careful accounting of the likelihood. This, in
turn, would have necessitated a vastly more complex data challenge -- for example,
fitting theoretical models to the observed galaxy clustering power spectrum,
as opposed to the convenient quantity $D_V(z)$. In this work, instead, we
follow a large body of literature in simplifying our likelihood as Gaussian in
the derived parameters since it is expected to be a very good approximation to
the truth.

\subsection{Parameter constraints}

We use a Markov Chain Monte Carlo (MCMC) algorithm to place constraints on cosmological
parameters. The MCMC algorithm estimates the posterior distribution of the 
cosmological, derived, and nuisance parameters by sampling the parameter space and 
evaluating the likelihood of each model with the data sets provided. Given the
likelihood $\mathcal{L}(\textbf{x}|\textbf{p})$ of the data set $\textbf{x}$
for the parameters $\textbf{p}$, the posterior distribution is obtained using 
Bayes' Theorem
\begin{equation}
\mathcal{P}(\textbf{p}|\textbf{x}) = \frac{\mathcal{L}(\textbf{x}|\textbf{p})\mathcal{P}(\textbf{p})}
{\int d\textbf{p}\mathcal{L}(\textbf{x}|\textbf{p})\mathcal{P}(\textbf{p})}
\end{equation}
where $\mathcal{P}(\textbf{p})$ is the prior probability density. The MCMC
algorithm produces the posterior probability in the parameter space including
the parameter mean values, covariances, and confidence intervals.

We analyze our models using an MCMC code that one of us (E.\ R.) developed
specifically for this purpose. We initially generate an optimized parameter
covariance matrix calculated using several shorter MCMC runs to optimize the
MCMC step size and direction and to minimize the overall runtime. The initial
10\% of the chains are thrown out, and the resulting chains are analyzed for
convergence using the Gelman-Rubin criteria \cite{GelmanRubin}, with a
conservative convergence requirement for the convergence parameter of $r <
1.03$ across a minimum of six chains for each case. Additionally, the step
sizes in parameters are optimized so that they have an acceptance rate of
$\sim$35\%. The resulting chains are then binned and smoothed with a Gaussian
filter for plotting.

\section{Results}\label{sec:results}

\subsection{Unsplit case}

Before splitting the late-universe parameters into those sensitive to geometry
and growth, we first show the fiducial constraints to make sure they are in
reasonably good agreement with similar recent constraints in the
literature. The left panel of Fig.~\ref{fig:unsplit} shows constraints on the
$\Omega_M-\sigma_8$ plane assuming $w=-1$, while the right panel shows the
constraints in the $\Omega_M-w$ plane. Note that these plots include
marginalization over four other cosmological parameters ($\Omega_M h^2,
\Omega_B h^2, 10^9A$, and $n_s$), in addition to several SNIa and cluster
nuisance parameters; see Eqs.~(\ref{eq:par_fund}) and (\ref{eq:par_nuis}). We
can already see the complementarity of the various cosmological probes: SNIa,
BAO and the CMB distance are sensitive only to geometry, so they measure
$\Omega_M$ and $w$ quite well, but are not sensitive to $\sigma_8$. In
contrast, WL, RSD and, to a smaller extent, cluster counts constrain (in the
case of $w=-1$) the characteristic combinations
\begin{equation}
  \begin{aligned}
    (\Omega_M/0.3)^{0.28}\sigma_8&=0.799 \pm 0.018\quad \mbox{(WL)},\\[0.1cm]
    (\Omega_M/0.3)^{0.04}\sigma_8&=0.809 \pm 0.022\quad \mbox{(RSD)},\\[0.1cm]
    (\Omega_M/0.3)^{0.27}\sigma_8&=0.837 \pm 0.021\quad \mbox{(clusters)}.
  \end{aligned}
\end{equation}
To obtain these best-constrained combinations of $\Omega_M$ and $\sigma_8$, we
simply varied the power $\alpha$ until the error in $(\Omega_M/0.3)^\alpha
\sigma_8$ was minimized.

Note that WL constraints favor a somewhat lower value of $\Omega_M$ and a
higher value of $\sigma_8$ than those favored by the combination of other
data sets. This has been noted and extensively explored in
\citet{MacCrann:2014wfa} who discuss possible reasons for this parameter
tension. Given that weak lensing is currently less mature than most of the
other cosmological probes, and the fact that WL only weakly contributes to our
principal constraints to be discussed below, we do not discuss this point
further.

The final combined constraints on $\Omega_M$ and $w$ are
\begin{equation}
  \begin{aligned}
    \Omega_M &= 0.299 \pm 0.010 \\[0.02cm]
           w &= -1.03 \pm 0.05
  \end{aligned}
  \qquad\mbox{(unsplit case)}
\label{eq:Om_w_unsplit_constraints}
\end{equation}
Constraints on all other parameters can be found in the third column of Table
\ref{tab:results}. For completeness,we also show constraints on the unsplit
case with $w=-1$ held fixed in the second column of the same Table.

We next study constraints when the late-universe parameters are split into
geometry and growth components.

\begin{table*}[ht]
\begin{center}
\begin{tabular}{ccccc} 
\hline \hline
Parameter & Unsplit, $w=-1$ & Unsplit, $w$ free & Split, $w=-1$ & Split, $w$ free \\ \hline
$ \Omega_M\left \{ \begin{array}{cl}
\Omega_M^\mathrm{geom} \\
\Omega_M^\mathrm{grow} \\
\end{array}\right .$ & $0.303 \pm 0.008 $ & $  0.299\pm 0.010 $ & $\begin{array}{cl}
0.302 \pm 0.008 \\
0.321 \pm 0.017 \\
\end{array}$ & $\begin{array}{cl}
0.283 \pm 0.011 \\
0.311 \pm 0.017 \\
\end{array}$ \\
$\Omega_M h^2$ & $  0.140 \pm 0.001 $ & $  0.141 \pm 0.002 $ & $  0.140 \pm 0.001 $ & $  0.142 \pm 0.002 $ \\
$\Omega_B h^2$ & $  0.0221 \pm 0.0002 $ & $  0.0220 \pm 0.0003 $ & $  0.0221 \pm 0.0002 $ & $  0.0221 \pm 0.0003 $ \\
$ w\left \{ \begin{array}{cl}
w^\mathrm{geom} \\
w^\mathrm{grow} \\
\end{array}\right . $ & -----  & $ -1.03 \pm 0.05 $ & $\begin{array}{cl}
\text{-----} \\
\text{-----} \\
\end{array}$ & $\begin{array}{cl}
-1.13 \pm 0.06 \\
-0.77 \pm 0.08 \\
\end{array}$ \\
$10^{9}A$ & $1.95 \pm 0.09 $ & $1.91 \pm 0.10 $ & $  1.96 \pm 0.09 $ & $  2.17 \pm 0.13 $ \\
$n_s$ & $  0.961 \pm 0.005 $ & $  0.959 \pm 0.006 $ & $  0.962 \pm 0.005 $ & $  0.961 \pm 0.006 $ \\\hline
$\sigma_8$ & $  0.786 \pm 0.015 $ & $  0.788\pm 0.016 $ & $  0.782 \pm 0.016 $ & $  0.771 \pm 0.017 $ \\
$h$ & $  0.680 \pm 0.006 $ & $  0.687 \pm 0.012 $ & $  0.661 \pm 0.017 $ & $  0.677 \pm 0.018 $ \\\hline
$\alpha_s$ & $  1.44 \pm 0.11 $ & $  1.44 \pm 0.11 $ & $  1.44 \pm 0.11 $ & $  1.44 \pm 0.11 $ \\
$\beta_c$ & $  3.26 \pm 0.11 $ & $  3.26 \pm 0.11 $ & $  3.26 \pm 0.11 $ & $  3.27 \pm 0.11 $ \\
$\ln(N|M_1)$ & $  2.36 \pm 0.06 $ & $  2.37 \pm 0.06 $ & $  2.29 \pm 0.08 $ & $  2.33 \pm 0.08 $ \\
$\ln(N|M_2)$ & $  4.15 \pm 0.09 $ & $  4.16 \pm 0.09 $ & $  4.09 \pm 0.11 $ & $  4.15 \pm 0.11 $ \\
$\sigma_{NM}$ & $  0.359 \pm 0.057 $ & $  0.357 \pm 0.057 $ & $  0.378 \pm 0.059 $ & $  0.367 \pm 0.060 $ \\
$\beta$ & $  1.041 \pm 0.050 $ & $  1.045 \pm 0.051 $ & $  1.018 \pm 0.054 $ & $  1.036 \pm 0.055 $ \\\hline
$\sigma_{MN}$ & $  0.462 \pm 0.081 $ & $  0.459 \pm 0.082 $ & $  0.486 \pm 0.085 $ & $  0.464 \pm 0.084 $ \\\hline

\end{tabular}
\caption{ 
  Constraints on the cosmological parameters from the combined
  probes. The second column shows constraints in the unsplit $\Lambda$CDM (so
  $w=-1$) model, while the third column also shows the standard unsplit case
  but allows $w$ to vary.  The fourth and fifth columns are our main results, and
  show the split-parameter cases where $\Omega_M$ is split and
  $w^\mathrm{geom} =w^\mathrm{grow} =-1$ is fixed (fourth column), and finally
  where both $\Omega_M$ and $w$ are split and allowed to vary (fifth
  column). In cases of parameters that can be split, the constraints are given
  either on the unsplit parameter (vertically centered number) or separate
  constraints on the geometry and growth split parameters (vertically offset
  pair of numbers).  }
\label{tab:results}
\end{center}
\end{table*}

\subsection{Split case: \texorpdfstring{$\Omega_M$}{Om} alone}

We now carry out the first of our analyses where the late-universe,
dark-energy parameters have been split into those governing geometry and
growth. Recall, the parameter split has been described at length in
Sec.~\ref{sec:probes}, and summarized in Table \ref{tab:summary}.

Fixing $w^\mathrm{geom} = w^\mathrm{grow} =-1$, we first split the matter
density alone into two separate parameters, $\Omega_M^{\rm grow}$ and
$\Omega_M^{\rm geom}$. In addition to these two parameters, we assume the usual set of four
additional fundamental early-universe parameters $\lbrace \Omega_M h^2,
\Omega_B h^2, 10^9A, n_s \rbrace$, plus the nuisance parameters.  Constraints
are shown in Fig.~\ref{fig:om-split} and in the fourth column of Table
\ref{tab:results}. Here we learn the first interesting lessons in how surveys
complement in measuring growth and distance.

Some trends are fully as expected: CMB distance and BAO are sensitive exclusively
to the geometry, and both prefer $\Omega_M^{\rm geom}\simeq 0.30$; recall that
BAO requires the help of the sound horizon prior, otherwise its constraints
become much weaker. We do not add any priors to Type Ia supernovae, which are
able to constrain $\Omega_M^{\rm geom}$, preferring however somewhat lower
values but with errors large enough to encompass the value of 0.3 at
2-$\sigma$. On the other hand RSD, combined with the early-universe prior, is
sensitive to both geometry and growth, though it constrains either only weakly.

The first small surprise is that clusters are {\it much} more sensitive to
growth than geometry, despite the fact that they probe both (recall the
summary in Table \ref{tab:summary}).  This is excellent news for consistency
tests of $w$CDM, since growth is typically more weakly probed than geometry
and ``needs more help''. The cluster constraint, combined with the
early-universe prior, is broadly consistent with $\Omega_M^{\rm grow}\simeq
0.25$-$0.30$. Finally, weak lensing constrains both geometry and growth about
equally well, but the overall constraint is rather weak and consistent with a
wide range of values of the two $\Omega_M$s.

On the whole, Fig.~\ref{fig:om-split} shows an impressive complementarity
between the different cosmological probes in how they constrain geometry and
growth. It also shows the huge progress in the field since similar constraints
imposed by \citet{Wang_split} seven years ago. Because the constraints are
mutually consistent, it is reasonable to combine them; the fully marginalized
constraints on the matter energy density relative to critical is
\begin{equation}
  \begin{aligned}
    \Omega_M^{\rm geom} &= 0.302 \pm 0.008 \\[0.02cm]
    \Omega_M^{\rm grow} &= 0.321 \pm 0.017 
  \end{aligned}
  \qquad\mbox{($\Omega_M$ split, $w\equiv -1$)}
\label{eq:Om_split_constraints}
\end{equation}
Clearly, in this $w = -1$ split case the geometry and growth constraints are
perfectly consistent with each other. The geometry constraint is stronger, as
expected.

\subsection{Split case: \texorpdfstring{$\Omega_M$}{Om} and \texorpdfstring{$w$}{w}}

A much more challenging task is to constrain the geometry and growth
components of the dark energy equation of state, since in that case one also
has to split the matter density and therefore deals with the dark energy
sector parameter space consisting of four parameters: $\Omega_M^{\rm geom},
\Omega_M^{\rm grow}, w^{\rm geom}$ and $w^{\rm grow}$. Before we show the
constraints, let us emphasize that, despite their relatively weak {\it
  individual} constraints on the equation of state, all of the cosmological
probes are invaluable since in combination they help break degeneracies in the
full $\sim 10$-dimensional parameter space and lead to excellent combined
constraints.

\begin{figure*}[t]
\includegraphics[width=0.45\textwidth]{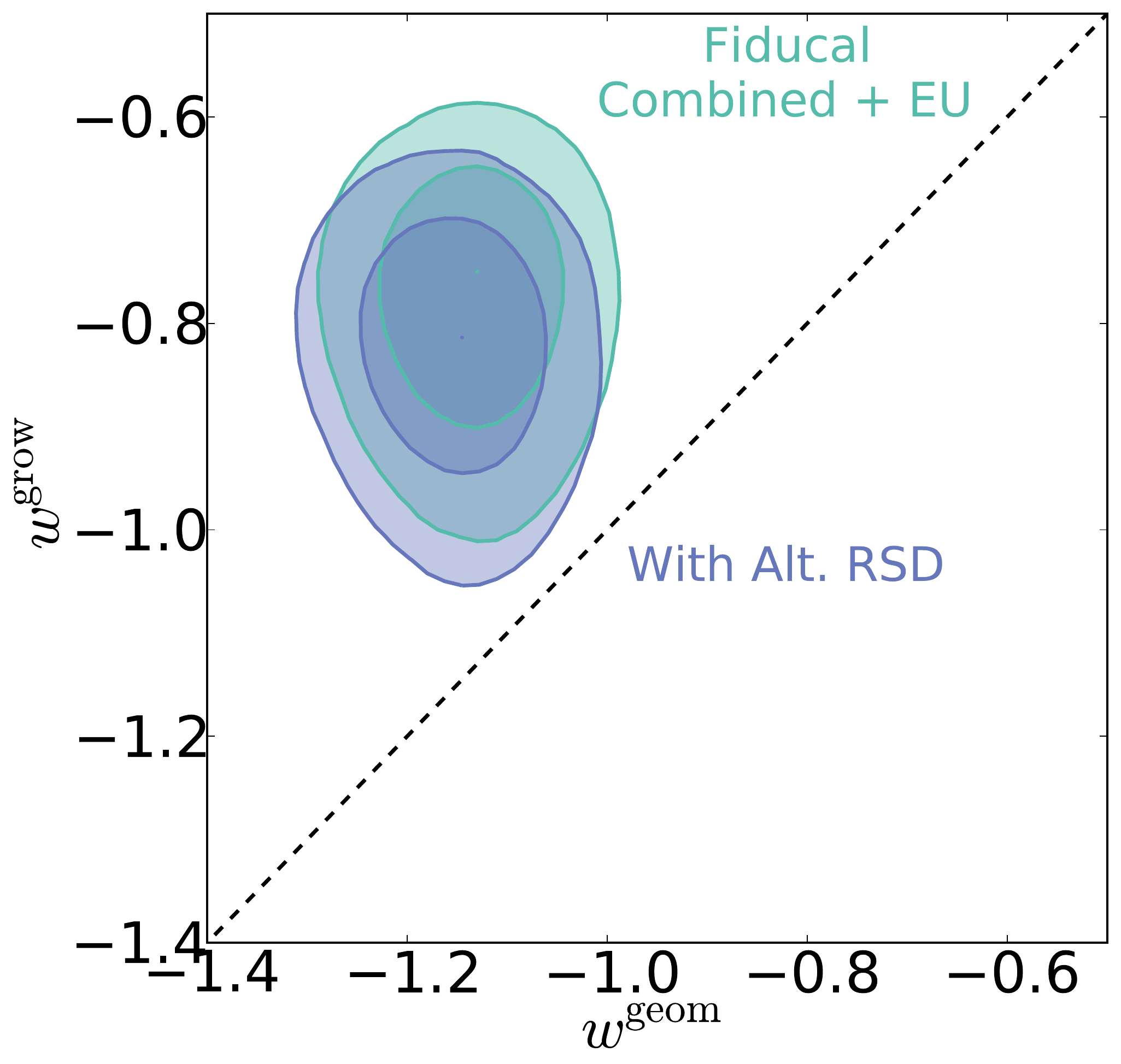}
\includegraphics[width=0.45\textwidth]{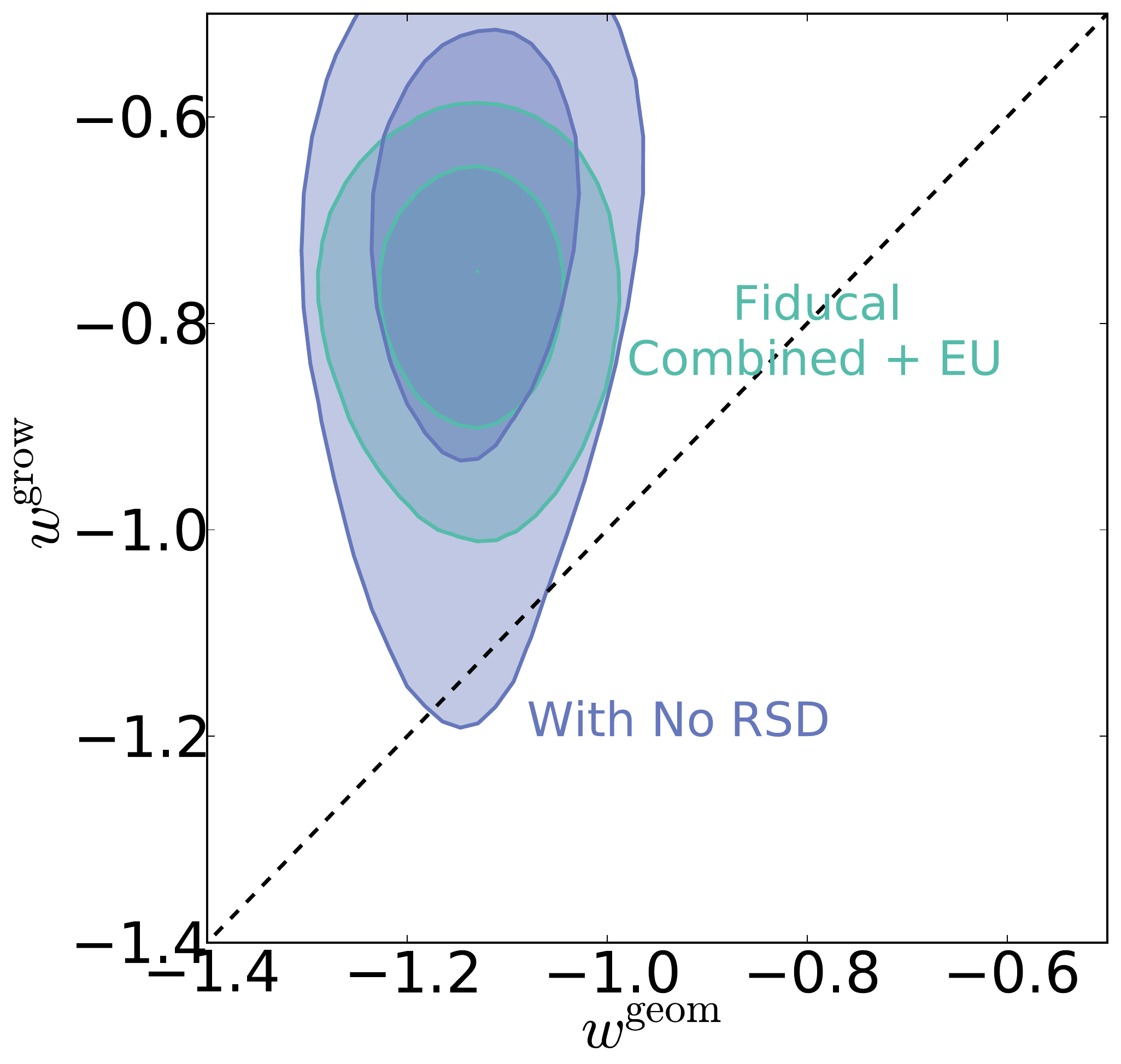}
\caption{ Dependence of our results on the RSD data and their analyses.
  Left panel: Combined constraints for the case where we replace the $z=0.57$
  RSD measurement from \cite{Chuang:2013wga} with the alternative BOSS
  measurement that uses the same raw data but a different analysis
  \cite{Samushia:2013yga}; see Fig.~\ref{fig:rsd-data}. The combined
  constraints are now only slightly less discrepant with the $w^{\rm geom} = w^{\rm
    grow}$ line. Right panel: Combined constraints, but
  without the RSD data employed. The combined contour is now larger in
  the growth direction; however it is still somewhat discrepant with the
  $w^{\rm geom} = w^{\rm grow}$ line, though less so than with the RSD data
  included. See text for details.}
\label{fig:w-split-no-rsd}
\end{figure*}

In Fig.~\ref{fig:w-split}, we show constraints on $ w^\text{geom}$ and
$w^\text{grow}$, marginalized (for each probe) over $\lbrace \Omega_M^{\rm
  geom}, \Omega_M^{\rm grow}, \Omega_M h^2, \Omega_B h^2, 10^9A, n_s \rbrace$,
plus the nuisance parameters as before. As in the previous case when only the
matter density parameter was split, we find largely expected directions probed
in this plane. However, because we now fully marginalize over the matter
density parameters $\Omega_M^{\rm geom}$ and $\Omega_M^{\rm grow}$, the
constraints on the equation of state are necessarily weaker. Nevertheless, BAO
and SNIa still do an admirable job in constraining the geometric $w$. The CMB
distance, being a single quantity, is subject to degeneracy between
$\Omega_M^{\rm geom}$ and $w^{\rm geom}$ and, by itself, provides no
constraint on either parameter alone. Finally WL and clusters also weakly
constrain either equation of state parameters due to partial degeneracies. All
of the aforementioned probes are broadly consistent with the $\Lambda$CDM
value $w^\mathrm{geom} =w^\mathrm{grow} =-1$.  In addition, we want to check
that our constraints are comparable to those obtained previously.  To that
effect, we get constraints using only the combined CMB and weak lensing, and
find that these are similar to comparible constraints obtained
\citet{Wang_split} and shown in Fig. 3 of that work.

The one significant outlier are the RSD; they alone, combined with the Planck
early-universe prior, precisely constrain the growth equation of state, but
with the value
\begin{equation}
w^{\rm grow, RSD}=-0.760 \pm 0.085, 
\end{equation}
which is clearly far from the $\Lambda$CDM value of $-1$.

The RSD data clearly pull the combined constraints away from the $w^{\rm
  geom}=w^{\rm grow}$ line, as a simple visual inspection of
Fig.~\ref{fig:w-split} shows. The fully marginalized combined constraints from
all cosmological probes, including the discrepant RSD, are
\begin{equation}
  \begin{aligned}
    \Omega_M^{\rm geom} &= 0.283 \pm 0.011 \\[0.02cm]
    \Omega_M^{\rm grow} &= 0.311 \pm 0.017 \\[0.02cm]
           w^{\rm geom} &= -1.13 \pm 0.06 \\[0.02cm]
           w^{\rm grow} &= -0.77 \pm 0.08 
  \end{aligned}
  \qquad\mbox{($\Omega_M$ and $w$ both split)}
\label{eq:w_split_constraints}
\end{equation}
and those on all other parameters can be found in the last column of Table
\ref{tab:results}. Note also that the overall goodness of fit with or without
RSD is satisfactory: with RSD $\chi^2/{\rm dof} = 728/699 = 1.04$, while when the
redshift space distortions are removed, $\chi^2/{\rm dof} = 719/686 = 1.05$.

We can easily quantify the significance of the pull away from the $w^{\rm
  geom}=w^{\rm grow}$ line by calculating the fraction of the likelihood for $w^{\rm
  geom}>w^{\rm grow}$, which is the p value defined as
\begin{equation}
p = \frac{\int_{w^\text{geom} > w^\text{grow}} 
dw^\text{geom} dw^\text{grow} \mathcal{L}(w^\text{geom}, w^\text{grow})}
{\int dw^\text{geom} dw^\text{grow} \mathcal{L}(w^\text{geom}, w^\text{grow})}.
\label{eq:p-value}
\end{equation}
The $p$-value is 0.0010 for the combined constraints,
corresponding\footnote{To convert this p value to ``sigmas'', we assumed the
  p value represents one tail of a two-sided Gaussian distribution: we would
  have been equally surprised to obtain the opposite result, namely $w^{\rm
    geom}>w^{\rm grow}$, and so this more conservative number of sigmas seems
  appropriate.} to an inconsistency with $w$CDM at $3.3\sigma$.

\section{Discussion}\label{sec:discussion}

Let us consider possible reasons for the pull of redshift-space distortions
toward $w^{\rm grow}>-1$. This result is qualitatively not new: a number of
recent investigations have already been established that the RSD data are in
some conflict with $\Lambda$CDM, suggesting less growth at
  recent times than predicted by the standard model \cite{Macaulay:2013swa}.
For example, \citet{Beutler:2013yhm} have measured a $>2$-$\sigma$ tension in
measurements of the growth index $\gamma= 0.772^{+0.124}_{-0.097}$ relative to
the $\Lambda$CDM (and, for that matter, also $w$CDM) prediction $\gamma\simeq
0.55$. Similarly, \citet{Samushia:2013yga}, using DR11 CMASS sample, and the
more precise results by \citet{Reid:2014iaa} that utilized smaller spatial
scales by doing extensive halo occupation distribution modeling, have obtained
similar results, indicating that growth is suppressed relative to $\Lambda$CDM
prediction at approximately the 2-$\sigma$ level. Moreover,
\citet{Beutler:2014yhv} find a $\sim$2.5$\sigma$ evidence for nonzero neutrino
mass, again a signature of the hints of the departure from the standard model.
Finally, \citet{Salvatelli:2014zta} utilize the combined cosmological
  probes (including the RSD) in the context of a model where vacuum energy 
  interacts with dark matter, and interpret the results as detection of
  nonzero interactions between dark matter and dark energy --- another
  possible interpretation of the departure from the standard $\Lambda$CDM
  model. 

Degeneracy with optical depth may play an important role here: our RSD
measurement is combined with the early-universe prior, whose crucial input is
the measurement of the optical depth to reionization $\tau$ which has been
most accurately measured by WMAP's polarization data. The higher the $\tau$,
the higher the primordial fluctuation amplitude $A$ or, roughly equivalently,
amplitude of mass fluctuations $\sigma_8$ at low redshift, and thus the larger
the discrepancy. Recall from Fig.~\ref{fig:rsd-data} that all RSD data, except
perhaps the higher-redshift WiggleZ measurement, pull toward low values of
$f\sigma_8$ relative to those predicted by the standard model. Therefore, the
anomalous RSD results {\it may} perhaps partly be explained by a high
WMAP-polarization estimate of $\tau$. Forthcoming Planck polarization
  measurements will provide more accurate constraints on the optical depth and
should clarify this issue.

\begin{figure*}[t]
\includegraphics[width=0.46215\textwidth]{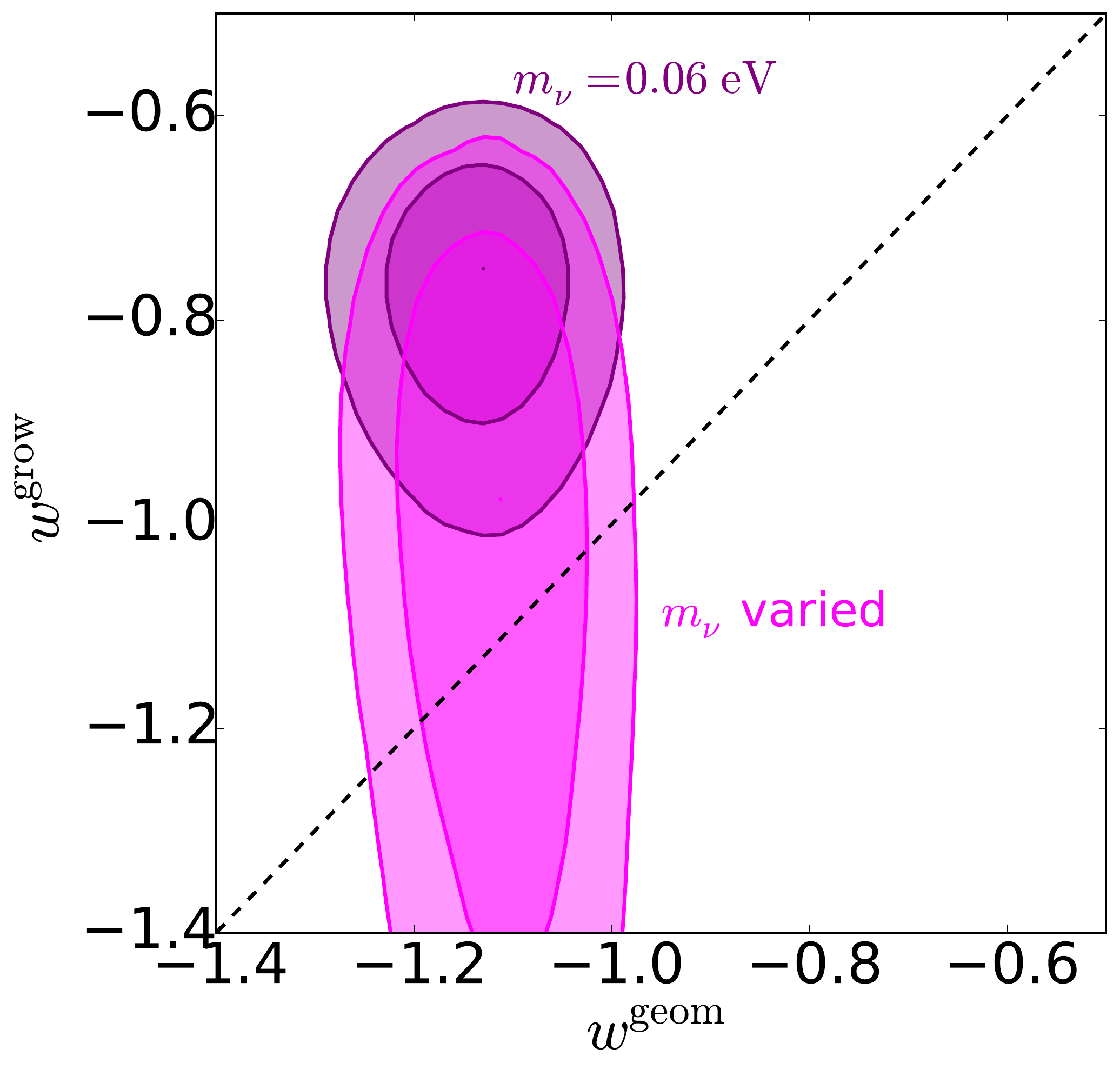}
\includegraphics[width=0.40\textwidth]{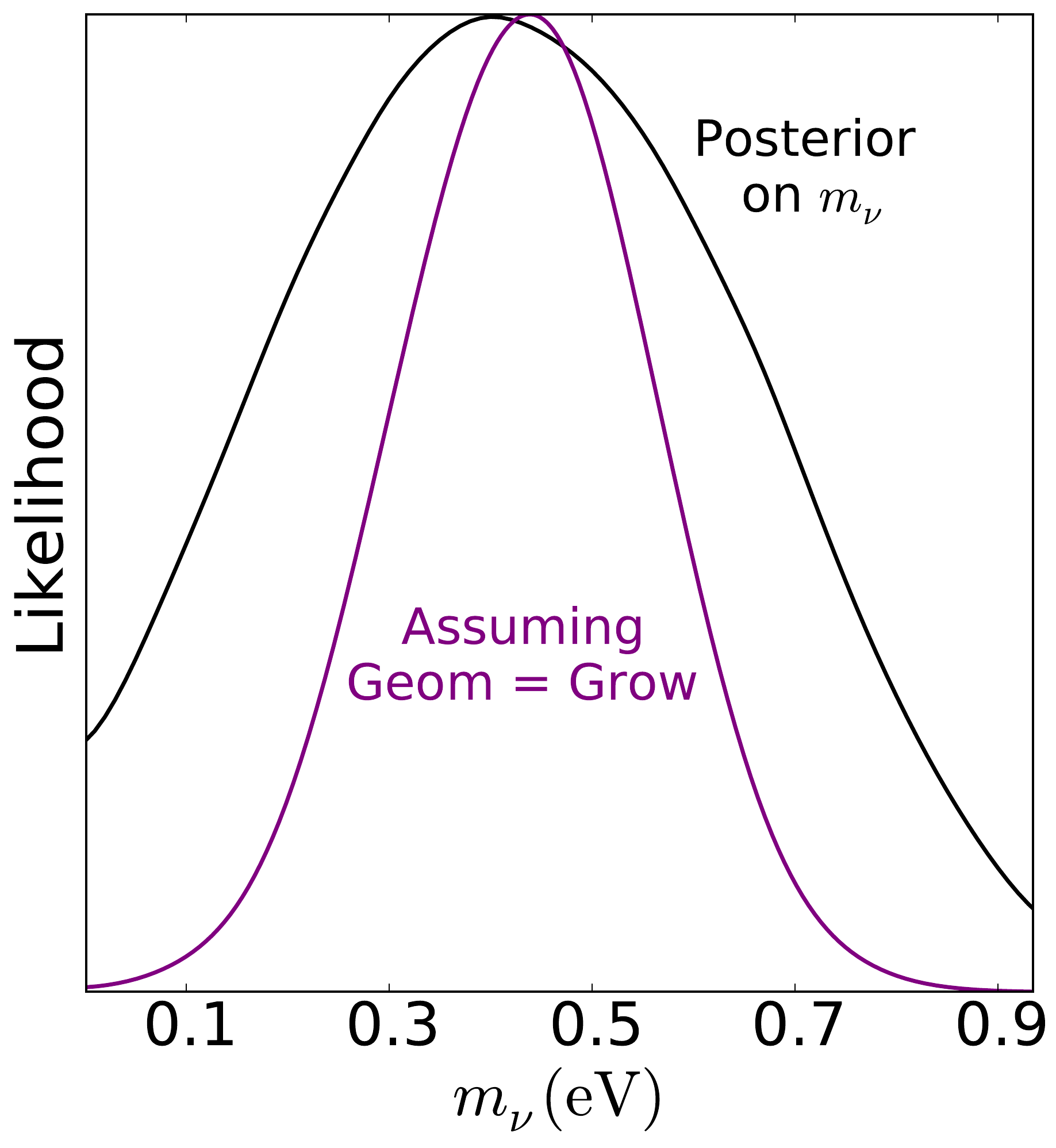}
\caption{Left panel: The effects on the combined constraints when the sum of
  the neutrino masses $m_\nu$ is allowed to vary, compared to our fiducial
  assumption of holding it fixed at $0.06$ eV. The constraints are now fully
  consistent with the $w^{\rm geom} = w^{\rm grow}$ line. Right panel:
  Posterior likelihood on $m_\nu$ for when $\Omega_M$ and $w$ are split (wider
  curve), and when growth = geometry correspondence ($\Omega_M^{\rm
      geom} = \Omega_M^{\rm grow}$ and $w^{\rm geom} = w^{\rm grow}$) is
    enforced (narrower curve). In both cases a value of $m_\nu \simeq
  0.45$ eV is preferred; see text for details.}
\label{fig:mnu-vary}
\end{figure*}

Perhaps of most interest is investigating how our results depend on the choice
of RSD analyses. Even within BOSS, different analyses make different
assumptions and give somewhat different results; this is clearly shown for the
$z=0.57$ measurements shown in Fig.~\ref{fig:rsd-data}. We do our best to
avoid the {\it a posteriori} bias of hand-picking analyses that give results
that are closer, or further away, from the concordance $\Lambda$CDM model. To
that extent, we keep our original choice of the RSD data from
Fig.~\ref{fig:rsd-data} and Table \ref{tab:rsddata} as fiducial but, as an
alternative, choose to investigate what happens in the combined analysis when
the measurement at $z=0.57$, which clearly is most responsible for the
discrepancy with the standard model, is replaced by the alternative analysis
of the same data \cite{Samushia:2013yga}. That alternative determination of
$(F, f\sigma_8)$ at $z=0.57$ is less discrepant with the $\Lambda$CDM
model; see Fig.~\ref{fig:rsd-data}. The results are shown in the left panel of
Fig.~\ref{fig:w-split-no-rsd}. Clearly, the combined constraints (RSD +
everything else) are now slightly closer to the geometry=growth line, but the
p value is still small (0.0020), indicating a 3.1-$\sigma$ discrepancy with the
standard geometry=growth assumption.  The constraints on cosmological
parameters with the alternate RSD $z=0.57$ measurement from BOSS are
\begin{equation}
  \begin{aligned}
    \Omega^{\rm geom} &= 0.279 \pm 0.011 \\[0.02cm]
    \Omega^{\rm grow} &= 0.319 \pm 0.021 \\[0.02cm]
         w^{\rm geom} &= -1.14 \pm 0.06 \\[0.02cm]
         w^{\rm grow} &= -0.81 \pm 0.08
  \end{aligned}
  \qquad\mbox{(w/ alternate RSD)}.
  \label{eq:alt-rsd-constraints}
\end{equation}
The goodness-of-fit for this case is also satisfactory, $\chi^2/$dof $=
724/699 = 1.04$.  

The RSD results are therefore reasonably stable with respect to the choice of
data. However, while the data in the RSD analyses that we employed typically
include information from large scale (roughly $10$-$30\hinvmpc\lesssim
\sqrt{r_\parallel^2+r_\perp^2} \lesssim 150$-$200\hinvmpc$) --- scales
considered well modeled by theory --- some analyses are subject to
contributions from shorter scales perpendicular to the line of sight (small
$r_\perp$), making those measurements subject to increased theory systematics
\cite{Song:2014nba}. Therefore, it is prudent to be cautious in interpreting
the RSD observations at this early stage.

We next investigate the implications of completely removing the RSD in the combined
constraints in the right panel of Fig.~\ref{fig:w-split-no-rsd}. In this case,
the combined constraints are more consistent with the geometry=growth
expectations, though the p value is still somewhat small at 0.0204, corresponding
to a discrepancy of 2.3$\sigma$. 
As mentioned earlier, the goodness-of-fit is entirely
satisfactory both with and without the RSD data. Clearly, RSD currently
provide by far the strongest constraint on the growth of structure.

It is also interesting to study the effect of the neutrino mass.
So far, cosmology has provided rather stringent upper limits to the sum of neutrino
masses, roughly $m_\nu\lesssim 0.3$ eV \citep[e.g.][]{Seljak:2006bg}. Recently
several papers have claimed evidence for the positive neutrino mass in order
to alleviate the discrepancy between the RSD data and the standard
$\Lambda$CDM model \cite{Beutler:2014yhv}, or the twin tensions between the
local measurements of the expansion history and Planck data
\cite{Hou:2012xq,Dvorkin:2014lea,Bocquet:2014lmj,Costanzi:2014tna}, and Planck and BICEP2 constraints on the
amplitude of gravitational waves \cite{Dvorkin:2014lea,Archidiacono:2014apa}.

To test the effect of neutrino mass sum on our combined constraints (including
RSD), we allow it to vary within the range $ m_\nu\in [0, 1]$ eV.  We
compare the combined results to our fiducial case of fixing the mass sum to
$m_\nu = 0.06$ eV, the results of which can be seen in the left panel of
Fig.~\ref{fig:mnu-vary}. Allowing the combined masses of neutrinos to vary
results in a significant increase in the range of values allowed by the
combined data, and the constraints become fully consistent with the
growth=geometry expectation:
\begin{equation}
  \begin{aligned}
    \Omega_M^{\rm geom} &= 0.289 \pm 0.012 \\[0.02cm]
    \Omega_M^{\rm grow} &= 0.319 \pm 0.018 \\[0.02cm]
           w^{\rm geom} &= -1.11 \pm 0.06 \\[0.02cm]
           w^{\rm grow} &= -1.10 \pm 0.28 
  \end{aligned}
  \quad\mbox{($m_\nu$ marginalized over)}.
\label{eq:mnu-varied_constraints}
\end{equation}

Neutrino mass therefore relieves tension between geometry and growth. It is
then of particular interest to report what neutrino mass sum is favored by the
data. The posterior probability on $m_\nu$ is shown in the right panel of
Fig. \ref{fig:mnu-vary}. In the case where both $\Omega_M$ and $w$ are split,
$m_\nu = 0.45 \pm 0.21$ eV, higher than our fiducial, normal-hierarchy value
(which assumes the massless lightest-mass eigenstate) of $m_\nu = 0.06$ eV by
$\sim{2}$-$\sigma$.  As a further test, we place constraints on $m_\nu$ in the
case of unsplit parameters (i.e.\ enforcing $\Omega_M^{\rm geom} =
\Omega_M^{\rm grow}$ and $w^{\rm geom} = w^{\rm grow}$), obtaining $m_\nu =
0.45 \pm 0.12$ eV. Our results are in good agreement with
\citet{Beutler:2014yhv} who favor similar neutrino mass, $m_\nu = 0.36 \pm
0.10$ eV, using the combined BAO+RSD+Planck data.

From Fig.~\ref{fig:w-split} and Eq.~(\ref{eq:w_split_constraints}) we see that
the {\it geometric} equation of state is also somewhat incompatible with the
$\Lambda$CDM value, since the combined data mildly prefer a value $w^{\rm
  geom}= -1.13 \pm 0.06$. We find that most of the pull toward such negative
values is provided by the BAO. The fact that $w^{\rm grow}>-1$ while $w^{\rm
  geom}<-1$ clearly exacerbates the disagreement between geometry and growth,
leading to the 3.3$\sigma$ incompatibility calculated above; growth however
clearly exhibits the more pronounced tension with the standard value.

Finally, we investigate whether there is something about the Planck
early-universe prior that pushes the combined constraints away from the
standard assumption that geometry=growth. To that effect, we replace the
Planck prior in Table \ref{tab:covprior} with the equivalent based on WMAP
nine-year data \citep{Hinshaw:2012aka}. Runs with this prior indicate that
$w^{\rm geom}=-1.13 \pm 0.06$, $w^{\rm grow} = -0.78 \pm 0.08$, with $w^{\rm
  grow} > w^{\rm geom}$ now favored at $3.1\sigma$ (p value=0.0017). These
constraints with WMAP9 are very similar to those obtained with Planck, so
differences between the two CMB probes' measurements are not responsible for
the tensions we observe.

\section{Conclusions}\label{sec:conclusion}

In this paper we have carried out a general, weakly model-dependent test of
the consistency of the $w$CDM cosmological model using current
cosmological data from Type Ia supernovae, CMB peak location, baryon acoustic
oscillations, redshift space distortions, cluster counts, and weak lensing. We
split each late-universe parameter that describes the effects of dark energy
into two parameters, one that comes from observed quantities that are governed
by geometry of the cosmological model, and one that is determined by the
growth of structure. Assuming flat universe, we first assume
the dark energy equation of state of $-1$ and constrain the parameters
determining the matter density relative to critical, $\Omega_M^{\rm geom}$ and
$\Omega_M^{\rm grow}$. We then consider the case when, in addition to the
matter density, the equation of state of dark energy can vary and hence
$w^{\rm geom}$ and $w^{\rm grow}$ can be constrained. We marginalize over five
additional early-universe parameters including the neutrino mass, plus several nuisance
parameters that are specific to individual cosmological probes. As a 
check, we show constraints projected on popular parameter combinations
$(\Omega_M, \sigma_8)$ and $(\Omega_M, w)$ in Fig.~\ref{fig:unsplit}.

The main results --- constraints on the geometry and growth components of
$\Omega_M$ and $w$ --- are shown in Figs.~\ref{fig:om-split} and
\ref{fig:w-split}, respectively. The complementarity of various probes is
impressive; this is especially visually evident in the $\Omega_M^{\rm
  geom}-\Omega_M^{\rm grow}$ plane in Fig.~\ref{fig:om-split} which shows that
SNIa, BAO and CMB peak location determine distance; the remaining
  three probes are sensitive to both geometry and growth -- RSD and cluster
  counts are largely sensitive to growth, while weak lensing mostly constrains
  the geometry. 
The overall goodness of fit is satisfactory, and the
constraints on the late-universe parameters of interest, given in
Eqs.~(\ref{eq:Om_split_constraints}) and (\ref{eq:w_split_constraints}) and
summarized in Table \ref{tab:summary}, are very tight.

One surprise are the redshift-space distortions, which are in a $\simeq
3$-$\sigma$ conflict with $w$CDM. The RSD prefer less growth at late times
than in the standard model; this can visually be seen in the RSD data ---
Fig.~\ref{fig:rsd-data} shows preference for a lower $f\sigma_8$ than in the
standard Planck $\Lambda$CDM model.  The tension is most clearly seen in the
$w$-split plane, Fig.~\ref{fig:w-split}, which shows that RSD alone prefers
$w^{\rm grow, RSD}=-0.760\pm 0.085$, and in fact pulls the {\it combined}
constraint from all probes to $w^{\rm grow}=-0.77\pm 0.08$. We quantify the
tension with $w$CDM to be 3.3$\sigma$ (p value of $w^{\rm geom}\geq w^{\rm
  grow}$ is 0.0010). This tension brought about with current RSD measurements
has already been noticed and discussed in the literature. In the Discussion
section, we demonstrate that the discrepancy remains at the still-significant
3.1$\sigma$ level once the most discrepant RSD measurement is replaced by one
from an alternative analysis. The discrepancy may be resolved with a higher
value of the sum of the neutrino masses than what is expected in the normal
hierarchy between the mass eigenstates with the lightest eigenstate being
massless, $m_\nu = 0.45 \pm 0.12$ eV; see Fig. \ref{fig:mnu-vary}.  However,
systematics may play a role in resolving the discrepancy; more work in this
area is needed to determine which of these effects is responsible.

On the whole, our results demonstrate very explicitly how the diverse
cosmological probes complement each other and not just break degeneracy in the
multidimensional parameter space, but also effectively specialize in
constraining geometry, growth, or both.  The resulting combined constraints on
the geometry and growth are impressively tight. The next generation of surveys
--- Stage III and IV in the language of the Dark Energy Task Force --- are
sure to improve them further.

Over the past few years, as the cosmological constraints improved, we and
others hoped that nature will be kind enough to provide hints for departure
from the standard $\Lambda$CDM model in order to help reveal the dynamics of
dark energy. We already see those hints, and it will be interesting to see
whether they are cracks in the cosmic egg\footnote{As expressed by Michael
  Turner, Aspen, summer 2014.} or perhaps systematics in data and observations.

\section*{Acknowledgments}
We thank Chris Blake, Catherine Heymans, Eric Linder, Will Percival, Martin White, and
especially Daniel Shafer for many useful conversations and comments. We are
supported by the DOE grant under contract DE-FG02-95ER40899 and NSF under
contract AST-0807564. DH thanks the Aspen Center for Physics, supported by NSF
Grant \#1066293, for hospitality during the completion of this work.

\bibliography{banana_split}

\begin{thebibliography}{74}%
\makeatletter
\providecommand \@ifxundefined [1]{%
 \@ifx{#1\undefined}
}%
\providecommand \@ifnum [1]{%
 \ifnum #1\expandafter \@firstoftwo
 \else \expandafter \@secondoftwo
 \fi
}%
\providecommand \@ifx [1]{%
 \ifx #1\expandafter \@firstoftwo
 \else \expandafter \@secondoftwo
 \fi
}%
\providecommand \natexlab [1]{#1}%
\providecommand \enquote  [1]{``#1''}%
\providecommand \bibnamefont  [1]{#1}%
\providecommand \bibfnamefont [1]{#1}%
\providecommand \citenamefont [1]{#1}%
\providecommand \href@noop [0]{\@secondoftwo}%
\providecommand \href [0]{\begingroup \@sanitize@url \@href}%
\providecommand \@href[1]{\@@startlink{#1}\@@href}%
\providecommand \@@href[1]{\endgroup#1\@@endlink}%
\providecommand \@sanitize@url [0]{\catcode `\\12\catcode `\$12\catcode
  `\&12\catcode `\#12\catcode `\^12\catcode `\_12\catcode `\%12\relax}%
\providecommand \@@startlink[1]{}%
\providecommand \@@endlink[0]{}%
\providecommand \url  [0]{\begingroup\@sanitize@url \@url }%
\providecommand \@url [1]{\endgroup\@href {#1}{\urlprefix }}%
\providecommand \urlprefix  [0]{URL }%
\providecommand \Eprint [0]{\href }%
\providecommand \doibase [0]{http://dx.doi.org/}%
\providecommand \selectlanguage [0]{\@gobble}%
\providecommand \bibinfo  [0]{\@secondoftwo}%
\providecommand \bibfield  [0]{\@secondoftwo}%
\providecommand \translation [1]{[#1]}%
\providecommand \BibitemOpen [0]{}%
\providecommand \bibitemStop [0]{}%
\providecommand \bibitemNoStop [0]{.\EOS\space}%
\providecommand \EOS [0]{\spacefactor3000\relax}%
\providecommand \BibitemShut  [1]{\csname bibitem#1\endcsname}%
\let\auto@bib@innerbib\@empty
\bibitem [{\citenamefont {Riess}\ \emph {et~al.}(1998)\citenamefont {Riess}
  \emph {et~al.}}]{Riess_1998}%
  \BibitemOpen
  \bibfield  {author} {\bibinfo {author} {\bibfnamefont {A.~G.}\ \bibnamefont
  {Riess}} \emph {et~al.},\ }\href@noop {} {\bibfield  {journal} {\bibinfo
  {journal} {Astron. J.}\ }\textbf {\bibinfo {volume} {116}},\ \bibinfo {pages}
  {1009} (\bibinfo {year} {1998})},\ \Eprint
  {http://arxiv.org/abs/astro-ph/9805201} {astro-ph/9805201} \BibitemShut
  {NoStop}%
\bibitem [{\citenamefont {Perlmutter}\ \emph {et~al.}(1999)\citenamefont
  {Perlmutter} \emph {et~al.}}]{Perlmutter_1999}%
  \BibitemOpen
  \bibfield  {author} {\bibinfo {author} {\bibfnamefont {S.}~\bibnamefont
  {Perlmutter}} \emph {et~al.},\ }\href@noop {} {\bibfield  {journal} {\bibinfo
   {journal} {Astrophys. J.}\ }\textbf {\bibinfo {volume} {517}},\ \bibinfo
  {pages} {565} (\bibinfo {year} {1999})},\ \Eprint
  {http://arxiv.org/abs/astro-ph/9812133} {astro-ph/9812133} \BibitemShut
  {NoStop}%
\bibitem [{\citenamefont {Frieman}\ \emph {et~al.}(2008)\citenamefont
  {Frieman}, \citenamefont {Turner},\ and\ \citenamefont
  {Huterer}}]{Frieman:2008sn}%
  \BibitemOpen
  \bibfield  {author} {\bibinfo {author} {\bibfnamefont {J.}~\bibnamefont
  {Frieman}}, \bibinfo {author} {\bibfnamefont {M.}~\bibnamefont {Turner}}, \
  and\ \bibinfo {author} {\bibfnamefont {D.}~\bibnamefont {Huterer}},\ }\href
  {\doibase 10.1146/annurev.astro.46.060407.145243} {\bibfield  {journal}
  {\bibinfo  {journal} {Ann.Rev.Astron.Astrophys.}\ }\textbf {\bibinfo {volume}
  {46}},\ \bibinfo {pages} {385} (\bibinfo {year} {2008})},\ \Eprint
  {http://arxiv.org/abs/0803.0982} {arXiv:0803.0982 [astro-ph]} \BibitemShut
  {NoStop}%
\bibitem [{\citenamefont {Joyce}\ \emph {et~al.}(2015)\citenamefont {Joyce},
  \citenamefont {Jain}, \citenamefont {Khoury},\ and\ \citenamefont
  {Trodden}}]{Joyce:2014kja}%
  \BibitemOpen
  \bibfield  {author} {\bibinfo {author} {\bibfnamefont {A.}~\bibnamefont
  {Joyce}}, \bibinfo {author} {\bibfnamefont {B.}~\bibnamefont {Jain}},
  \bibinfo {author} {\bibfnamefont {J.}~\bibnamefont {Khoury}}, \ and\ \bibinfo
  {author} {\bibfnamefont {M.}~\bibnamefont {Trodden}},\ }\href {\doibase
  10.1016/j.physrep.2014.12.002} {\bibfield  {journal} {\bibinfo  {journal}
  {Phys.Rept.}\ }\textbf {\bibinfo {volume} {568}},\ \bibinfo {pages} {1}
  (\bibinfo {year} {2015})},\ \Eprint {http://arxiv.org/abs/1407.0059}
  {arXiv:1407.0059 [astro-ph.CO]} \BibitemShut {NoStop}%
\bibitem [{\citenamefont {Ishak}\ \emph {et~al.}(2006)\citenamefont {Ishak},
  \citenamefont {Upadhye},\ and\ \citenamefont {Spergel}}]{Ishak:2005zs}%
  \BibitemOpen
  \bibfield  {author} {\bibinfo {author} {\bibfnamefont {M.}~\bibnamefont
  {Ishak}}, \bibinfo {author} {\bibfnamefont {A.}~\bibnamefont {Upadhye}}, \
  and\ \bibinfo {author} {\bibfnamefont {D.~N.}\ \bibnamefont {Spergel}},\
  }\href {\doibase 10.1103/PhysRevD.74.043513} {\bibfield  {journal} {\bibinfo
  {journal} {Phys.Rev.}\ }\textbf {\bibinfo {volume} {D74}},\ \bibinfo {pages}
  {043513} (\bibinfo {year} {2006})},\ \Eprint
  {http://arxiv.org/abs/astro-ph/0507184} {arXiv:astro-ph/0507184 [astro-ph]}
  \BibitemShut {NoStop}%
\bibitem [{\citenamefont {Zhan}\ \emph {et~al.}(2009)\citenamefont {Zhan},
  \citenamefont {Knox},\ and\ \citenamefont {Tyson}}]{Zhan:2008jh}%
  \BibitemOpen
  \bibfield  {author} {\bibinfo {author} {\bibfnamefont {H.}~\bibnamefont
  {Zhan}}, \bibinfo {author} {\bibfnamefont {L.}~\bibnamefont {Knox}}, \ and\
  \bibinfo {author} {\bibfnamefont {J.~A.}\ \bibnamefont {Tyson}},\ }\href
  {\doibase 10.1088/0004-637X/690/1/923} {\bibfield  {journal} {\bibinfo
  {journal} {Astrophys.J.}\ }\textbf {\bibinfo {volume} {690}},\ \bibinfo
  {pages} {923} (\bibinfo {year} {2009})},\ \Eprint
  {http://arxiv.org/abs/0806.0937} {arXiv:0806.0937 [astro-ph]} \BibitemShut
  {NoStop}%
\bibitem [{\citenamefont {Mortonson}\ \emph {et~al.}(2009)\citenamefont
  {Mortonson}, \citenamefont {Hu},\ and\ \citenamefont
  {Huterer}}]{Mortonson:2008qy}%
  \BibitemOpen
  \bibfield  {author} {\bibinfo {author} {\bibfnamefont {M.~J.}\ \bibnamefont
  {Mortonson}}, \bibinfo {author} {\bibfnamefont {W.}~\bibnamefont {Hu}}, \
  and\ \bibinfo {author} {\bibfnamefont {D.}~\bibnamefont {Huterer}},\ }\href
  {\doibase 10.1103/PhysRevD.79.023004} {\bibfield  {journal} {\bibinfo
  {journal} {Phys.Rev.}\ }\textbf {\bibinfo {volume} {D79}},\ \bibinfo {pages}
  {023004} (\bibinfo {year} {2009})},\ \Eprint {http://arxiv.org/abs/0810.1744}
  {arXiv:0810.1744 [astro-ph]} \BibitemShut {NoStop}%
\bibitem [{\citenamefont {Mortonson}\ \emph {et~al.}(2010)\citenamefont
  {Mortonson}, \citenamefont {Hu},\ and\ \citenamefont
  {Huterer}}]{Mortonson:2009hk}%
  \BibitemOpen
  \bibfield  {author} {\bibinfo {author} {\bibfnamefont {M.~J.}\ \bibnamefont
  {Mortonson}}, \bibinfo {author} {\bibfnamefont {W.}~\bibnamefont {Hu}}, \
  and\ \bibinfo {author} {\bibfnamefont {D.}~\bibnamefont {Huterer}},\ }\href
  {\doibase 10.1103/PhysRevD.81.063007} {\bibfield  {journal} {\bibinfo
  {journal} {Phys.Rev.}\ }\textbf {\bibinfo {volume} {D81}},\ \bibinfo {pages}
  {063007} (\bibinfo {year} {2010})},\ \Eprint {http://arxiv.org/abs/0912.3816}
  {arXiv:0912.3816 [astro-ph.CO]} \BibitemShut {NoStop}%
\bibitem [{\citenamefont {Acquaviva}\ and\ \citenamefont
  {Gawiser}(2010)}]{Acquaviva:2010vr}%
  \BibitemOpen
  \bibfield  {author} {\bibinfo {author} {\bibfnamefont {V.}~\bibnamefont
  {Acquaviva}}\ and\ \bibinfo {author} {\bibfnamefont {E.}~\bibnamefont
  {Gawiser}},\ }\href {\doibase 10.1103/PhysRevD.82.082001} {\bibfield
  {journal} {\bibinfo  {journal} {Phys.Rev.}\ }\textbf {\bibinfo {volume}
  {D82}},\ \bibinfo {pages} {082001} (\bibinfo {year} {2010})},\ \Eprint
  {http://arxiv.org/abs/1008.3392} {arXiv:1008.3392 [astro-ph.CO]} \BibitemShut
  {NoStop}%
\bibitem [{\citenamefont {Vanderveld}\ \emph {et~al.}(2012)\citenamefont
  {Vanderveld}, \citenamefont {Mortonson}, \citenamefont {Hu},\ and\
  \citenamefont {Eifler}}]{Vanderveld:2012ec}%
  \BibitemOpen
  \bibfield  {author} {\bibinfo {author} {\bibfnamefont {R.~A.}\ \bibnamefont
  {Vanderveld}}, \bibinfo {author} {\bibfnamefont {M.~J.}\ \bibnamefont
  {Mortonson}}, \bibinfo {author} {\bibfnamefont {W.}~\bibnamefont {Hu}}, \
  and\ \bibinfo {author} {\bibfnamefont {T.}~\bibnamefont {Eifler}},\ }\href
  {\doibase 10.1103/PhysRevD.85.103518} {\bibfield  {journal} {\bibinfo
  {journal} {Phys.Rev.}\ }\textbf {\bibinfo {volume} {D85}},\ \bibinfo {pages}
  {103518} (\bibinfo {year} {2012})},\ \Eprint {http://arxiv.org/abs/1203.3195}
  {arXiv:1203.3195 [astro-ph.CO]} \BibitemShut {NoStop}%
\bibitem [{\citenamefont {Zhao}\ \emph {et~al.}(2010)\citenamefont {Zhao},
  \citenamefont {Giannantonio}, \citenamefont {Pogosian}, \citenamefont
  {Silvestri}, \citenamefont {Bacon} \emph {et~al.}}]{Zhao:2010dz}%
  \BibitemOpen
  \bibfield  {author} {\bibinfo {author} {\bibfnamefont {G.-B.}\ \bibnamefont
  {Zhao}}, \bibinfo {author} {\bibfnamefont {T.}~\bibnamefont {Giannantonio}},
  \bibinfo {author} {\bibfnamefont {L.}~\bibnamefont {Pogosian}}, \bibinfo
  {author} {\bibfnamefont {A.}~\bibnamefont {Silvestri}}, \bibinfo {author}
  {\bibfnamefont {D.~J.}\ \bibnamefont {Bacon}},  \emph {et~al.},\ }\href
  {\doibase 10.1103/PhysRevD.81.103510} {\bibfield  {journal} {\bibinfo
  {journal} {Phys.Rev.}\ }\textbf {\bibinfo {volume} {D81}},\ \bibinfo {pages}
  {103510} (\bibinfo {year} {2010})},\ \Eprint {http://arxiv.org/abs/1003.0001}
  {arXiv:1003.0001 [astro-ph.CO]} \BibitemShut {NoStop}%
\bibitem [{\citenamefont {Bean}\ and\ \citenamefont
  {Tangmatitham}(2010)}]{Bean:2010zq}%
  \BibitemOpen
  \bibfield  {author} {\bibinfo {author} {\bibfnamefont {R.}~\bibnamefont
  {Bean}}\ and\ \bibinfo {author} {\bibfnamefont {M.}~\bibnamefont
  {Tangmatitham}},\ }\href {\doibase 10.1103/PhysRevD.81.083534} {\bibfield
  {journal} {\bibinfo  {journal} {Phys.Rev.}\ }\textbf {\bibinfo {volume}
  {D81}},\ \bibinfo {pages} {083534} (\bibinfo {year} {2010})},\ \Eprint
  {http://arxiv.org/abs/1002.4197} {arXiv:1002.4197 [astro-ph.CO]} \BibitemShut
  {NoStop}%
\bibitem [{\citenamefont {Zhao}\ \emph {et~al.}(2012)\citenamefont {Zhao},
  \citenamefont {Li}, \citenamefont {Linder}, \citenamefont {Koyama},
  \citenamefont {Bacon} \emph {et~al.}}]{Zhao:2011te}%
  \BibitemOpen
  \bibfield  {author} {\bibinfo {author} {\bibfnamefont {G.-B.}\ \bibnamefont
  {Zhao}}, \bibinfo {author} {\bibfnamefont {H.}~\bibnamefont {Li}}, \bibinfo
  {author} {\bibfnamefont {E.~V.}\ \bibnamefont {Linder}}, \bibinfo {author}
  {\bibfnamefont {K.}~\bibnamefont {Koyama}}, \bibinfo {author} {\bibfnamefont
  {D.~J.}\ \bibnamefont {Bacon}},  \emph {et~al.},\ }\href {\doibase
  10.1103/PhysRevD.85.123546} {\bibfield  {journal} {\bibinfo  {journal}
  {Phys.Rev.}\ }\textbf {\bibinfo {volume} {D85}},\ \bibinfo {pages} {123546}
  (\bibinfo {year} {2012})},\ \Eprint {http://arxiv.org/abs/1109.1846}
  {arXiv:1109.1846 [astro-ph.CO]} \BibitemShut {NoStop}%
\bibitem [{\citenamefont {Hojjati}\ \emph {et~al.}(2012)\citenamefont
  {Hojjati}, \citenamefont {Zhao}, \citenamefont {Pogosian}, \citenamefont
  {Silvestri}, \citenamefont {Crittenden} \emph {et~al.}}]{Hojjati:2011xd}%
  \BibitemOpen
  \bibfield  {author} {\bibinfo {author} {\bibfnamefont {A.}~\bibnamefont
  {Hojjati}}, \bibinfo {author} {\bibfnamefont {G.-B.}\ \bibnamefont {Zhao}},
  \bibinfo {author} {\bibfnamefont {L.}~\bibnamefont {Pogosian}}, \bibinfo
  {author} {\bibfnamefont {A.}~\bibnamefont {Silvestri}}, \bibinfo {author}
  {\bibfnamefont {R.}~\bibnamefont {Crittenden}},  \emph {et~al.},\ }\href
  {\doibase 10.1103/PhysRevD.85.043508} {\bibfield  {journal} {\bibinfo
  {journal} {Phys.Rev.}\ }\textbf {\bibinfo {volume} {D85}},\ \bibinfo {pages}
  {043508} (\bibinfo {year} {2012})},\ \Eprint {http://arxiv.org/abs/1111.3960}
  {arXiv:1111.3960 [astro-ph.CO]} \BibitemShut {NoStop}%
\bibitem [{\citenamefont {Dossett}\ \emph
  {et~al.}(2011{\natexlab{a}})\citenamefont {Dossett}, \citenamefont
  {Moldenhauer},\ and\ \citenamefont {Ishak}}]{Dossett:2011zp}%
  \BibitemOpen
  \bibfield  {author} {\bibinfo {author} {\bibfnamefont {J.}~\bibnamefont
  {Dossett}}, \bibinfo {author} {\bibfnamefont {J.}~\bibnamefont
  {Moldenhauer}}, \ and\ \bibinfo {author} {\bibfnamefont {M.}~\bibnamefont
  {Ishak}},\ }\href {\doibase 10.1103/PhysRevD.84.023012} {\bibfield  {journal}
  {\bibinfo  {journal} {Phys.Rev.}\ }\textbf {\bibinfo {volume} {D84}},\
  \bibinfo {pages} {023012} (\bibinfo {year} {2011}{\natexlab{a}})},\ \Eprint
  {http://arxiv.org/abs/1103.1195} {arXiv:1103.1195 [astro-ph.CO]} \BibitemShut
  {NoStop}%
\bibitem [{\citenamefont {Dossett}\ \emph
  {et~al.}(2011{\natexlab{b}})\citenamefont {Dossett}, \citenamefont {Ishak},\
  and\ \citenamefont {Moldenhauer}}]{Dossett:2011tn}%
  \BibitemOpen
  \bibfield  {author} {\bibinfo {author} {\bibfnamefont {J.~N.}\ \bibnamefont
  {Dossett}}, \bibinfo {author} {\bibfnamefont {M.}~\bibnamefont {Ishak}}, \
  and\ \bibinfo {author} {\bibfnamefont {J.}~\bibnamefont {Moldenhauer}},\
  }\href {\doibase 10.1103/PhysRevD.84.123001} {\bibfield  {journal} {\bibinfo
  {journal} {Phys.Rev.}\ }\textbf {\bibinfo {volume} {D84}},\ \bibinfo {pages}
  {123001} (\bibinfo {year} {2011}{\natexlab{b}})},\ \Eprint
  {http://arxiv.org/abs/1109.4583} {arXiv:1109.4583 [astro-ph.CO]} \BibitemShut
  {NoStop}%
\bibitem [{\citenamefont {Silvestri}\ \emph {et~al.}(2013)\citenamefont
  {Silvestri}, \citenamefont {Pogosian},\ and\ \citenamefont
  {Buniy}}]{Silvestri:2013ne}%
  \BibitemOpen
  \bibfield  {author} {\bibinfo {author} {\bibfnamefont {A.}~\bibnamefont
  {Silvestri}}, \bibinfo {author} {\bibfnamefont {L.}~\bibnamefont {Pogosian}},
  \ and\ \bibinfo {author} {\bibfnamefont {R.~V.}\ \bibnamefont {Buniy}},\
  }\href {\doibase 10.1103/PhysRevD.87.104015} {\bibfield  {journal} {\bibinfo
  {journal} {Phys.Rev.}\ }\textbf {\bibinfo {volume} {D87}},\ \bibinfo {pages}
  {104015} (\bibinfo {year} {2013})},\ \Eprint {http://arxiv.org/abs/1302.1193}
  {arXiv:1302.1193 [astro-ph.CO]} \BibitemShut {NoStop}%
\bibitem [{\citenamefont {Wang}\ \emph {et~al.}(2007)\citenamefont {Wang},
  \citenamefont {Hui}, \citenamefont {May},\ and\ \citenamefont
  {Haiman}}]{Wang_split}%
  \BibitemOpen
  \bibfield  {author} {\bibinfo {author} {\bibfnamefont {S.}~\bibnamefont
  {Wang}}, \bibinfo {author} {\bibfnamefont {L.}~\bibnamefont {Hui}}, \bibinfo
  {author} {\bibfnamefont {M.}~\bibnamefont {May}}, \ and\ \bibinfo {author}
  {\bibfnamefont {Z.}~\bibnamefont {Haiman}},\ }\href {\doibase
  10.1103/PhysRevD.76.063503} {\bibfield  {journal} {\bibinfo  {journal}
  {Phys.Rev.}\ }\textbf {\bibinfo {volume} {D76}},\ \bibinfo {pages} {063503}
  (\bibinfo {year} {2007})},\ \Eprint {http://arxiv.org/abs/0705.0165}
  {arXiv:0705.0165 [astro-ph]} \BibitemShut {NoStop}%
\bibitem [{\citenamefont {Zhang}\ \emph {et~al.}(2005)\citenamefont {Zhang},
  \citenamefont {Hui},\ and\ \citenamefont {Stebbins}}]{Zhang:2003ii}%
  \BibitemOpen
  \bibfield  {author} {\bibinfo {author} {\bibfnamefont {J.}~\bibnamefont
  {Zhang}}, \bibinfo {author} {\bibfnamefont {L.}~\bibnamefont {Hui}}, \ and\
  \bibinfo {author} {\bibfnamefont {A.}~\bibnamefont {Stebbins}},\ }\href
  {\doibase 10.1086/497676} {\bibfield  {journal} {\bibinfo  {journal}
  {Astrophys.J.}\ }\textbf {\bibinfo {volume} {635}},\ \bibinfo {pages} {806}
  (\bibinfo {year} {2005})},\ \Eprint {http://arxiv.org/abs/astro-ph/0312348}
  {arXiv:astro-ph/0312348 [astro-ph]} \BibitemShut {NoStop}%
\bibitem [{\citenamefont {Chu}\ and\ \citenamefont {Knox}(2005)}]{Chu:2004qx}%
  \BibitemOpen
  \bibfield  {author} {\bibinfo {author} {\bibfnamefont {M.}~\bibnamefont
  {Chu}}\ and\ \bibinfo {author} {\bibfnamefont {L.}~\bibnamefont {Knox}},\
  }\href {\doibase 10.1086/427064} {\bibfield  {journal} {\bibinfo  {journal}
  {Astrophys.J.}\ }\textbf {\bibinfo {volume} {620}},\ \bibinfo {pages} {1}
  (\bibinfo {year} {2005})},\ \Eprint {http://arxiv.org/abs/astro-ph/0407198}
  {arXiv:astro-ph/0407198 [astro-ph]} \BibitemShut {NoStop}%
\bibitem [{\citenamefont {{Abate}}\ and\ \citenamefont
  {{Lahav}}(2008)}]{Abate:2008au}%
  \BibitemOpen
  \bibfield  {author} {\bibinfo {author} {\bibfnamefont {A.}~\bibnamefont
  {{Abate}}}\ and\ \bibinfo {author} {\bibfnamefont {O.}~\bibnamefont
  {{Lahav}}},\ }\href {\doibase 10.1111/j.1745-3933.2008.00519.x} {\bibfield
  {journal} {\bibinfo  {journal} {\mnras}\ }\textbf {\bibinfo {volume} {389}},\
  \bibinfo {pages} {L47} (\bibinfo {year} {2008})},\ \Eprint
  {http://arxiv.org/abs/0805.3160} {arXiv:0805.3160} \BibitemShut {NoStop}%
\bibitem [{\citenamefont {Krauss}\ and\ \citenamefont
  {Turner}(1995)}]{Krauss_Turner_95}%
  \BibitemOpen
  \bibfield  {author} {\bibinfo {author} {\bibfnamefont {L.~M.}\ \bibnamefont
  {Krauss}}\ and\ \bibinfo {author} {\bibfnamefont {M.~S.}\ \bibnamefont
  {Turner}},\ }\href {\doibase 10.1007/BF02108229} {\bibfield  {journal}
  {\bibinfo  {journal} {Gen.Rel.Grav.}\ }\textbf {\bibinfo {volume} {27}},\
  \bibinfo {pages} {1137} (\bibinfo {year} {1995})},\ \Eprint
  {http://arxiv.org/abs/astro-ph/9504003} {arXiv:astro-ph/9504003 [astro-ph]}
  \BibitemShut {NoStop}%
\bibitem [{\citenamefont {Scolnic}\ \emph {et~al.}(2014)\citenamefont
  {Scolnic}, \citenamefont {Rest}, \citenamefont {Riess}, \citenamefont
  {Huber}, \citenamefont {Foley} \emph {et~al.}}]{Scolnic:2013efb}%
  \BibitemOpen
  \bibfield  {author} {\bibinfo {author} {\bibfnamefont {D.}~\bibnamefont
  {Scolnic}}, \bibinfo {author} {\bibfnamefont {A.}~\bibnamefont {Rest}},
  \bibinfo {author} {\bibfnamefont {A.}~\bibnamefont {Riess}}, \bibinfo
  {author} {\bibfnamefont {M.}~\bibnamefont {Huber}}, \bibinfo {author}
  {\bibfnamefont {R.}~\bibnamefont {Foley}},  \emph {et~al.},\ }\href {\doibase
  10.1088/0004-637X/795/1/45} {\bibfield  {journal} {\bibinfo  {journal}
  {Astrophys.J.}\ }\textbf {\bibinfo {volume} {795}},\ \bibinfo {pages} {45}
  (\bibinfo {year} {2014})},\ \Eprint {http://arxiv.org/abs/1310.3824}
  {arXiv:1310.3824 [astro-ph.CO]} \BibitemShut {NoStop}%
\bibitem [{\citenamefont {Cheng}\ and\ \citenamefont
  {Huang}(2014)}]{Cheng:2013csa}%
  \BibitemOpen
  \bibfield  {author} {\bibinfo {author} {\bibfnamefont {C.}~\bibnamefont
  {Cheng}}\ and\ \bibinfo {author} {\bibfnamefont {Q.-G.}\ \bibnamefont
  {Huang}},\ }\href {\doibase 10.1103/PhysRevD.89.043003} {\bibfield  {journal}
  {\bibinfo  {journal} {Phys.Rev.}\ }\textbf {\bibinfo {volume} {D89}},\
  \bibinfo {pages} {043003} (\bibinfo {year} {2014})},\ \Eprint
  {http://arxiv.org/abs/1306.4091} {arXiv:1306.4091 [astro-ph.CO]} \BibitemShut
  {NoStop}%
\bibitem [{\citenamefont {Xia}\ \emph {et~al.}(2013)\citenamefont {Xia},
  \citenamefont {Li},\ and\ \citenamefont {Zhang}}]{Xia:2013dea}%
  \BibitemOpen
  \bibfield  {author} {\bibinfo {author} {\bibfnamefont {J.-Q.}\ \bibnamefont
  {Xia}}, \bibinfo {author} {\bibfnamefont {H.}~\bibnamefont {Li}}, \ and\
  \bibinfo {author} {\bibfnamefont {X.}~\bibnamefont {Zhang}},\ }\href
  {\doibase 10.1103/PhysRevD.88.063501} {\bibfield  {journal} {\bibinfo
  {journal} {Phys.Rev.}\ }\textbf {\bibinfo {volume} {D88}},\ \bibinfo {pages}
  {063501} (\bibinfo {year} {2013})},\ \Eprint {http://arxiv.org/abs/1308.0188}
  {arXiv:1308.0188 [astro-ph.CO]} \BibitemShut {NoStop}%
\bibitem [{\citenamefont {Shafer}\ and\ \citenamefont
  {Huterer}(2014)}]{Shafer_Huterer}%
  \BibitemOpen
  \bibfield  {author} {\bibinfo {author} {\bibfnamefont {D.~L.}\ \bibnamefont
  {Shafer}}\ and\ \bibinfo {author} {\bibfnamefont {D.}~\bibnamefont
  {Huterer}},\ }\href {\doibase 10.1103/PhysRevD.89.063510} {\bibfield
  {journal} {\bibinfo  {journal} {Phys.Rev.}\ }\textbf {\bibinfo {volume}
  {D89}},\ \bibinfo {pages} {063510} (\bibinfo {year} {2014})},\ \Eprint
  {http://arxiv.org/abs/1312.1688} {arXiv:1312.1688 [astro-ph.CO]} \BibitemShut
  {NoStop}%
\bibitem [{\citenamefont {{Linder}}(2005)}]{Linder_gamma}%
  \BibitemOpen
  \bibfield  {author} {\bibinfo {author} {\bibfnamefont {E.~V.}\ \bibnamefont
  {{Linder}}},\ }\href {\doibase 10.1103/PhysRevD.72.043529} {\bibfield
  {journal} {\bibinfo  {journal} {Phys. Rev.}\ }\textbf {\bibinfo {volume}
  {D72}},\ \bibinfo {pages} {043529} (\bibinfo {year} {2005})},\ \Eprint
  {http://arxiv.org/abs/astro-ph/0507263} {astro-ph/0507263} \BibitemShut
  {NoStop}%
\bibitem [{\citenamefont {Conley}\ \emph {et~al.}(2011)\citenamefont {Conley}
  \emph {et~al.}}]{Conley}%
  \BibitemOpen
  \bibfield  {author} {\bibinfo {author} {\bibfnamefont {A.}~\bibnamefont
  {Conley}} \emph {et~al.} (\bibinfo {collaboration} {SNLS Collaboration}),\
  }\href {\doibase 10.1088/0067-0049/192/1/1} {\bibfield  {journal} {\bibinfo
  {journal} {Astrophys.J.Suppl.}\ }\textbf {\bibinfo {volume} {192}},\ \bibinfo
  {pages} {1} (\bibinfo {year} {2011})},\ \Eprint
  {http://arxiv.org/abs/1104.1443} {arXiv:1104.1443 [astro-ph.CO]} \BibitemShut
  {NoStop}%
\bibitem [{\citenamefont {Ruiz}\ \emph {et~al.}(2012)\citenamefont {Ruiz},
  \citenamefont {Shafer}, \citenamefont {Huterer},\ and\ \citenamefont
  {Conley}}]{Ruiz_Huterer}%
  \BibitemOpen
  \bibfield  {author} {\bibinfo {author} {\bibfnamefont {E.~J.}\ \bibnamefont
  {Ruiz}}, \bibinfo {author} {\bibfnamefont {D.~L.}\ \bibnamefont {Shafer}},
  \bibinfo {author} {\bibfnamefont {D.}~\bibnamefont {Huterer}}, \ and\
  \bibinfo {author} {\bibfnamefont {A.}~\bibnamefont {Conley}},\ }\href
  {\doibase 10.1103/PhysRevD.86.103004} {\bibfield  {journal} {\bibinfo
  {journal} {Phys.Rev.}\ }\textbf {\bibinfo {volume} {D86}},\ \bibinfo {pages}
  {103004} (\bibinfo {year} {2012})},\ \Eprint {http://arxiv.org/abs/1207.4781}
  {arXiv:1207.4781 [astro-ph.CO]} \BibitemShut {NoStop}%
\bibitem [{\citenamefont {Frieman}\ \emph {et~al.}(2003)\citenamefont
  {Frieman}, \citenamefont {Huterer}, \citenamefont {Linder},\ and\
  \citenamefont {Turner}}]{Frieman:2002wi}%
  \BibitemOpen
  \bibfield  {author} {\bibinfo {author} {\bibfnamefont {J.~A.}\ \bibnamefont
  {Frieman}}, \bibinfo {author} {\bibfnamefont {D.}~\bibnamefont {Huterer}},
  \bibinfo {author} {\bibfnamefont {E.~V.}\ \bibnamefont {Linder}}, \ and\
  \bibinfo {author} {\bibfnamefont {M.~S.}\ \bibnamefont {Turner}},\ }\href
  {\doibase 10.1103/PhysRevD.67.083505} {\bibfield  {journal} {\bibinfo
  {journal} {Phys.Rev.}\ }\textbf {\bibinfo {volume} {D67}},\ \bibinfo {pages}
  {083505} (\bibinfo {year} {2003})},\ \Eprint
  {http://arxiv.org/abs/astro-ph/0208100} {arXiv:astro-ph/0208100 [astro-ph]}
  \BibitemShut {NoStop}%
\bibitem [{\citenamefont {Zahn}\ and\ \citenamefont
  {Zaldarriaga}(2003)}]{Zahn:2002rr}%
  \BibitemOpen
  \bibfield  {author} {\bibinfo {author} {\bibfnamefont {O.}~\bibnamefont
  {Zahn}}\ and\ \bibinfo {author} {\bibfnamefont {M.}~\bibnamefont
  {Zaldarriaga}},\ }\href {\doibase 10.1103/PhysRevD.67.063002} {\bibfield
  {journal} {\bibinfo  {journal} {Phys.Rev.}\ }\textbf {\bibinfo {volume}
  {D67}},\ \bibinfo {pages} {063002} (\bibinfo {year} {2003})},\ \Eprint
  {http://arxiv.org/abs/astro-ph/0212360} {arXiv:astro-ph/0212360 [astro-ph]}
  \BibitemShut {NoStop}%
\bibitem [{\citenamefont {Ade}\ \emph {et~al.}(2014)\citenamefont {Ade} \emph
  {et~al.}}]{Planck2013_XVI}%
  \BibitemOpen
  \bibfield  {author} {\bibinfo {author} {\bibfnamefont {P.}~\bibnamefont
  {Ade}} \emph {et~al.} (\bibinfo {collaboration} {Planck Collaboration}),\
  }\href {\doibase 10.1051/0004-6361/201321591} {\bibfield  {journal} {\bibinfo
   {journal} {Astron.Astrophys.}\ } (\bibinfo {year} {2014}),\
  10.1051/0004-6361/201321591},\ \Eprint {http://arxiv.org/abs/1303.5076}
  {arXiv:1303.5076 [astro-ph.CO]} \BibitemShut {NoStop}%
\bibitem [{\citenamefont {Eisenstein}\ \emph {et~al.}(2005)\citenamefont
  {Eisenstein} \emph {et~al.}}]{Eisenstein2005}%
  \BibitemOpen
  \bibfield  {author} {\bibinfo {author} {\bibfnamefont {D.~J.}\ \bibnamefont
  {Eisenstein}} \emph {et~al.} (\bibinfo {collaboration} {SDSS
  Collaboration}),\ }\href {\doibase 10.1086/466512} {\bibfield  {journal}
  {\bibinfo  {journal} {Astrophys.J.}\ }\textbf {\bibinfo {volume} {633}},\
  \bibinfo {pages} {560} (\bibinfo {year} {2005})},\ \Eprint
  {http://arxiv.org/abs/astro-ph/0501171} {arXiv:astro-ph/0501171 [astro-ph]}
  \BibitemShut {NoStop}%
\bibitem [{\citenamefont {Beutler}\ \emph {et~al.}(2011)\citenamefont
  {Beutler}, \citenamefont {Blake}, \citenamefont {Colless}, \citenamefont
  {Jones}, \citenamefont {Staveley-Smith} \emph {et~al.}}]{6dFGS}%
  \BibitemOpen
  \bibfield  {author} {\bibinfo {author} {\bibfnamefont {F.}~\bibnamefont
  {Beutler}}, \bibinfo {author} {\bibfnamefont {C.}~\bibnamefont {Blake}},
  \bibinfo {author} {\bibfnamefont {M.}~\bibnamefont {Colless}}, \bibinfo
  {author} {\bibfnamefont {D.~H.}\ \bibnamefont {Jones}}, \bibinfo {author}
  {\bibfnamefont {L.}~\bibnamefont {Staveley-Smith}},  \emph {et~al.},\ }\href
  {\doibase 10.1111/j.1365-2966.2011.19250.x} {\bibfield  {journal} {\bibinfo
  {journal} {Mon.Not.Roy.Astron.Soc.}\ }\textbf {\bibinfo {volume} {416}},\
  \bibinfo {pages} {3017} (\bibinfo {year} {2011})},\ \Eprint
  {http://arxiv.org/abs/1106.3366} {arXiv:1106.3366 [astro-ph.CO]} \BibitemShut
  {NoStop}%
\bibitem [{\citenamefont {Padmanabhan}\ \emph {et~al.}(2012)\citenamefont
  {Padmanabhan}, \citenamefont {Xu}, \citenamefont {Eisenstein}, \citenamefont
  {Scalzo}, \citenamefont {Cuesta} \emph {et~al.}}]{SDSS_LRG}%
  \BibitemOpen
  \bibfield  {author} {\bibinfo {author} {\bibfnamefont {N.}~\bibnamefont
  {Padmanabhan}}, \bibinfo {author} {\bibfnamefont {X.}~\bibnamefont {Xu}},
  \bibinfo {author} {\bibfnamefont {D.~J.}\ \bibnamefont {Eisenstein}},
  \bibinfo {author} {\bibfnamefont {R.}~\bibnamefont {Scalzo}}, \bibinfo
  {author} {\bibfnamefont {A.~J.}\ \bibnamefont {Cuesta}},  \emph {et~al.},\
  }\href {\doibase 10.1111/j.1365-2966.2012.21888.x} {\bibfield  {journal}
  {\bibinfo  {journal} {Mon.Not.Roy.Astron.Soc.}\ }\textbf {\bibinfo {volume}
  {427}},\ \bibinfo {pages} {2132} (\bibinfo {year} {2012})},\ \Eprint
  {http://arxiv.org/abs/1202.0090} {arXiv:1202.0090 [astro-ph.CO]} \BibitemShut
  {NoStop}%
\bibitem [{\citenamefont {Anderson}\ \emph {et~al.}(2012)\citenamefont
  {Anderson}, \citenamefont {Aubourg}, \citenamefont {Bailey}, \citenamefont
  {Bizyaev}, \citenamefont {Blanton} \emph {et~al.}}]{BOSS}%
  \BibitemOpen
  \bibfield  {author} {\bibinfo {author} {\bibfnamefont {L.}~\bibnamefont
  {Anderson}}, \bibinfo {author} {\bibfnamefont {E.}~\bibnamefont {Aubourg}},
  \bibinfo {author} {\bibfnamefont {S.}~\bibnamefont {Bailey}}, \bibinfo
  {author} {\bibfnamefont {D.}~\bibnamefont {Bizyaev}}, \bibinfo {author}
  {\bibfnamefont {M.}~\bibnamefont {Blanton}},  \emph {et~al.},\ }\href
  {\doibase 10.1111/j.1365-2966.2012.22066.x} {\bibfield  {journal} {\bibinfo
  {journal} {Mon.Not.Roy.Astron.Soc.}\ }\textbf {\bibinfo {volume} {427}},\
  \bibinfo {pages} {3435} (\bibinfo {year} {2012})},\ \Eprint
  {http://arxiv.org/abs/1203.6594} {arXiv:1203.6594 [astro-ph.CO]} \BibitemShut
  {NoStop}%
\bibitem [{\citenamefont {Eisenstein}\ and\ \citenamefont
  {Hu}(1999)}]{Eisenstein1997}%
  \BibitemOpen
  \bibfield  {author} {\bibinfo {author} {\bibfnamefont {D.~J.}\ \bibnamefont
  {Eisenstein}}\ and\ \bibinfo {author} {\bibfnamefont {W.}~\bibnamefont
  {Hu}},\ }\href {\doibase 10.1086/306640} {\bibfield  {journal} {\bibinfo
  {journal} {Astrophys.J.}\ }\textbf {\bibinfo {volume} {511}},\ \bibinfo
  {pages} {5} (\bibinfo {year} {1999})},\ \Eprint
  {http://arxiv.org/abs/astro-ph/9710252} {arXiv:astro-ph/9710252 [astro-ph]}
  \BibitemShut {NoStop}%
\bibitem [{\citenamefont {Allen}\ \emph {et~al.}(2011)\citenamefont {Allen},
  \citenamefont {Evrard},\ and\ \citenamefont {Mantz}}]{Allen:2011zs}%
  \BibitemOpen
  \bibfield  {author} {\bibinfo {author} {\bibfnamefont {S.~W.}\ \bibnamefont
  {Allen}}, \bibinfo {author} {\bibfnamefont {A.~E.}\ \bibnamefont {Evrard}}, \
  and\ \bibinfo {author} {\bibfnamefont {A.~B.}\ \bibnamefont {Mantz}},\ }\href
  {\doibase 10.1146/annurev-astro-081710-102514} {\bibfield  {journal}
  {\bibinfo  {journal} {Ann.Rev.Astron.Astrophys.}\ }\textbf {\bibinfo {volume}
  {49}},\ \bibinfo {pages} {409} (\bibinfo {year} {2011})},\ \Eprint
  {http://arxiv.org/abs/1103.4829} {arXiv:1103.4829 [astro-ph.CO]} \BibitemShut
  {NoStop}%
\bibitem [{\citenamefont {Rozo}\ \emph {et~al.}(2010)\citenamefont {Rozo} \emph
  {et~al.}}]{Rozo}%
  \BibitemOpen
  \bibfield  {author} {\bibinfo {author} {\bibfnamefont {E.}~\bibnamefont
  {Rozo}} \emph {et~al.} (\bibinfo {collaboration} {DSDD Collaboration}),\
  }\href {\doibase 10.1088/0004-637X/708/1/645} {\bibfield  {journal} {\bibinfo
   {journal} {Astrophys.J.}\ }\textbf {\bibinfo {volume} {708}},\ \bibinfo
  {pages} {645} (\bibinfo {year} {2010})},\ \Eprint
  {http://arxiv.org/abs/0902.3702} {arXiv:0902.3702 [astro-ph.CO]} \BibitemShut
  {NoStop}%
\bibitem [{\citenamefont {Koester}\ \emph {et~al.}(2007)\citenamefont {Koester}
  \emph {et~al.}}]{Koester:2007bg}%
  \BibitemOpen
  \bibfield  {author} {\bibinfo {author} {\bibfnamefont {B.}~\bibnamefont
  {Koester}} \emph {et~al.} (\bibinfo {collaboration} {SDSS Collaboration}),\
  }\href {\doibase 10.1086/509599} {\bibfield  {journal} {\bibinfo  {journal}
  {Astrophys.J.}\ }\textbf {\bibinfo {volume} {660}},\ \bibinfo {pages} {239}
  (\bibinfo {year} {2007})},\ \Eprint {http://arxiv.org/abs/astro-ph/0701265}
  {arXiv:astro-ph/0701265 [astro-ph]} \BibitemShut {NoStop}%
\bibitem [{\citenamefont {Johnston}\ \emph {et~al.}(2007)\citenamefont
  {Johnston} \emph {et~al.}}]{Johnston:2007}%
  \BibitemOpen
  \bibfield  {author} {\bibinfo {author} {\bibfnamefont {D.~E.}\ \bibnamefont
  {Johnston}} \emph {et~al.} (\bibinfo {collaboration} {SDSS Collaboration}),\
  }\href@noop {} {\  (\bibinfo {year} {2007})},\ \Eprint
  {http://arxiv.org/abs/0709.1159} {arXiv:0709.1159 [astro-ph]} \BibitemShut
  {NoStop}%
\bibitem [{\citenamefont {Tinker}\ \emph {et~al.}(2008)\citenamefont {Tinker},
  \citenamefont {Kravtsov}, \citenamefont {Klypin}, \citenamefont {Abazajian},
  \citenamefont {Warren} \emph {et~al.}}]{Tinker2008}%
  \BibitemOpen
  \bibfield  {author} {\bibinfo {author} {\bibfnamefont {J.~L.}\ \bibnamefont
  {Tinker}}, \bibinfo {author} {\bibfnamefont {A.~V.}\ \bibnamefont
  {Kravtsov}}, \bibinfo {author} {\bibfnamefont {A.}~\bibnamefont {Klypin}},
  \bibinfo {author} {\bibfnamefont {K.}~\bibnamefont {Abazajian}}, \bibinfo
  {author} {\bibfnamefont {M.~S.}\ \bibnamefont {Warren}},  \emph {et~al.},\
  }\href {\doibase 10.1086/591439} {\bibfield  {journal} {\bibinfo  {journal}
  {Astrophys.J.}\ }\textbf {\bibinfo {volume} {688}},\ \bibinfo {pages} {709}
  (\bibinfo {year} {2008})},\ \Eprint {http://arxiv.org/abs/0803.2706}
  {arXiv:0803.2706 [astro-ph]} \BibitemShut {NoStop}%
\bibitem [{\citenamefont {{Erben}}\ \emph {et~al.}(2013)\citenamefont
  {{Erben}}, \citenamefont {{Hildebrandt}}, \citenamefont {{Miller}},
  \citenamefont {{van Waerbeke}}, \citenamefont {{Heymans}} \emph
  {et~al.}}]{Erben_CFHTLens}%
  \BibitemOpen
  \bibfield  {author} {\bibinfo {author} {\bibfnamefont {T.}~\bibnamefont
  {{Erben}}}, \bibinfo {author} {\bibfnamefont {H.}~\bibnamefont
  {{Hildebrandt}}}, \bibinfo {author} {\bibfnamefont {L.}~\bibnamefont
  {{Miller}}}, \bibinfo {author} {\bibfnamefont {L.}~\bibnamefont {{van
  Waerbeke}}}, \bibinfo {author} {\bibnamefont {{Heymans}}},  \emph {et~al.},\
  }\href {\doibase 10.1093/mnras/stt928} {\bibfield  {journal} {\bibinfo
  {journal} {\mnras}\ }\textbf {\bibinfo {volume} {433}},\ \bibinfo {pages}
  {2545} (\bibinfo {year} {2013})},\ \Eprint {http://arxiv.org/abs/1210.8156}
  {arXiv:1210.8156 [astro-ph.CO]} \BibitemShut {NoStop}%
\bibitem [{\citenamefont {{Heymans}}\ \emph {et~al.}(2012)\citenamefont
  {{Heymans}}, \citenamefont {{Van Waerbeke}}, \citenamefont {{Miller}},
  \citenamefont {{Erben}}, \citenamefont {{Hildebrandt}} \emph
  {et~al.}}]{Heymans_CFHTLens}%
  \BibitemOpen
  \bibfield  {author} {\bibinfo {author} {\bibfnamefont {C.}~\bibnamefont
  {{Heymans}}}, \bibinfo {author} {\bibfnamefont {L.}~\bibnamefont {{Van
  Waerbeke}}}, \bibinfo {author} {\bibfnamefont {L.}~\bibnamefont {{Miller}}},
  \bibinfo {author} {\bibfnamefont {T.}~\bibnamefont {{Erben}}}, \bibinfo
  {author} {\bibnamefont {{Hildebrandt}}},  \emph {et~al.},\ }\href {\doibase
  10.1111/j.1365-2966.2012.21952.x} {\bibfield  {journal} {\bibinfo  {journal}
  {\mnras}\ }\textbf {\bibinfo {volume} {427}},\ \bibinfo {pages} {146}
  (\bibinfo {year} {2012})},\ \Eprint {http://arxiv.org/abs/1210.0032}
  {arXiv:1210.0032 [astro-ph.CO]} \BibitemShut {NoStop}%
\bibitem [{\citenamefont {{Huterer}}\ and\ \citenamefont
  {{Takada}}(2005)}]{Huterer_Takada}%
  \BibitemOpen
  \bibfield  {author} {\bibinfo {author} {\bibfnamefont {D.}~\bibnamefont
  {{Huterer}}}\ and\ \bibinfo {author} {\bibfnamefont {M.}~\bibnamefont
  {{Takada}}},\ }\href {\doibase 10.1016/j.astropartphys.2005.02.006}
  {\bibfield  {journal} {\bibinfo  {journal} {Astroparticle Physics}\ }\textbf
  {\bibinfo {volume} {23}},\ \bibinfo {pages} {369} (\bibinfo {year} {2005})},\
  \Eprint {http://arxiv.org/abs/astro-ph/0412142} {astro-ph/0412142}
  \BibitemShut {NoStop}%
\bibitem [{\citenamefont {{Smith}}\ \emph {et~al.}(2003)\citenamefont
  {{Smith}}, \citenamefont {{Peacock}}, \citenamefont {{Jenkins}},
  \citenamefont {{White}}, \citenamefont {{Frenk}}, \citenamefont {{Pearce}},
  \citenamefont {{Thomas}}, \citenamefont {{Efstathiou}},\ and\ \citenamefont
  {{Couchman}}}]{halofit}%
  \BibitemOpen
  \bibfield  {author} {\bibinfo {author} {\bibfnamefont {R.~E.}\ \bibnamefont
  {{Smith}}}, \bibinfo {author} {\bibfnamefont {J.~A.}\ \bibnamefont
  {{Peacock}}}, \bibinfo {author} {\bibfnamefont {A.}~\bibnamefont
  {{Jenkins}}}, \bibinfo {author} {\bibfnamefont {S.~D.~M.}\ \bibnamefont
  {{White}}}, \bibinfo {author} {\bibfnamefont {C.~S.}\ \bibnamefont
  {{Frenk}}}, \bibinfo {author} {\bibfnamefont {F.~R.}\ \bibnamefont
  {{Pearce}}}, \bibinfo {author} {\bibfnamefont {P.~A.}\ \bibnamefont
  {{Thomas}}}, \bibinfo {author} {\bibfnamefont {G.}~\bibnamefont
  {{Efstathiou}}}, \ and\ \bibinfo {author} {\bibfnamefont {H.~M.~P.}\
  \bibnamefont {{Couchman}}},\ }\href {\doibase
  10.1046/j.1365-8711.2003.06503.x} {\bibfield  {journal} {\bibinfo  {journal}
  {\mnras}\ }\textbf {\bibinfo {volume} {341}},\ \bibinfo {pages} {1311}
  (\bibinfo {year} {2003})},\ \Eprint {http://arxiv.org/abs/astro-ph/0207664}
  {astro-ph/0207664} \BibitemShut {NoStop}%
\bibitem [{\citenamefont {Takahashi}\ \emph {et~al.}(2012)\citenamefont
  {Takahashi}, \citenamefont {Sato}, \citenamefont {Nishimichi}, \citenamefont
  {Taruya},\ and\ \citenamefont {Oguri}}]{Takahashi_2012}%
  \BibitemOpen
  \bibfield  {author} {\bibinfo {author} {\bibfnamefont {R.}~\bibnamefont
  {Takahashi}}, \bibinfo {author} {\bibfnamefont {M.}~\bibnamefont {Sato}},
  \bibinfo {author} {\bibfnamefont {T.}~\bibnamefont {Nishimichi}}, \bibinfo
  {author} {\bibfnamefont {A.}~\bibnamefont {Taruya}}, \ and\ \bibinfo {author}
  {\bibfnamefont {M.}~\bibnamefont {Oguri}},\ }\href {\doibase
  10.1088/0004-637X/761/2/152} {\bibfield  {journal} {\bibinfo  {journal}
  {Astrophys.J.}\ }\textbf {\bibinfo {volume} {761}},\ \bibinfo {pages} {152}
  (\bibinfo {year} {2012})},\ \Eprint {http://arxiv.org/abs/1208.2701}
  {arXiv:1208.2701 [astro-ph.CO]} \BibitemShut {NoStop}%
\bibitem [{\citenamefont {Samushia}\ \emph {et~al.}(2014)\citenamefont
  {Samushia}, \citenamefont {Reid}, \citenamefont {White}, \citenamefont
  {Percival}, \citenamefont {Cuesta} \emph {et~al.}}]{Samushia:2013yga}%
  \BibitemOpen
  \bibfield  {author} {\bibinfo {author} {\bibfnamefont {L.}~\bibnamefont
  {Samushia}}, \bibinfo {author} {\bibfnamefont {B.~A.}\ \bibnamefont {Reid}},
  \bibinfo {author} {\bibfnamefont {M.}~\bibnamefont {White}}, \bibinfo
  {author} {\bibfnamefont {W.~J.}\ \bibnamefont {Percival}}, \bibinfo {author}
  {\bibfnamefont {A.~J.}\ \bibnamefont {Cuesta}},  \emph {et~al.},\ }\href
  {\doibase 10.1093/mnras/stu197} {\bibfield  {journal} {\bibinfo  {journal}
  {Mon.Not.Roy.Astron.Soc.}\ }\textbf {\bibinfo {volume} {439}},\ \bibinfo
  {pages} {3504} (\bibinfo {year} {2014})},\ \Eprint
  {http://arxiv.org/abs/1312.4899} {arXiv:1312.4899 [astro-ph.CO]} \BibitemShut
  {NoStop}%
\bibitem [{\citenamefont {{Jackson}}(1972)}]{Jackson_FoG}%
  \BibitemOpen
  \bibfield  {author} {\bibinfo {author} {\bibfnamefont {J.~C.}\ \bibnamefont
  {{Jackson}}},\ }\href@noop {} {\bibfield  {journal} {\bibinfo  {journal}
  {\mnras}\ }\textbf {\bibinfo {volume} {156}},\ \bibinfo {pages} {1P}
  (\bibinfo {year} {1972})}\BibitemShut {NoStop}%
\bibitem [{\citenamefont {Song}\ and\ \citenamefont
  {Percival}(2009)}]{Song:2008qt}%
  \BibitemOpen
  \bibfield  {author} {\bibinfo {author} {\bibfnamefont {Y.-S.}\ \bibnamefont
  {Song}}\ and\ \bibinfo {author} {\bibfnamefont {W.~J.}\ \bibnamefont
  {Percival}},\ }\href {\doibase 10.1088/1475-7516/2009/10/004} {\bibfield
  {journal} {\bibinfo  {journal} {JCAP}\ }\textbf {\bibinfo {volume} {0910}},\
  \bibinfo {pages} {004} (\bibinfo {year} {2009})},\ \Eprint
  {http://arxiv.org/abs/0807.0810} {arXiv:0807.0810 [astro-ph]} \BibitemShut
  {NoStop}%
\bibitem [{\citenamefont {Ballinger}\ \emph {et~al.}(1996)\citenamefont
  {Ballinger}, \citenamefont {Peacock},\ and\ \citenamefont
  {Heavens}}]{Ballinger:1996cd}%
  \BibitemOpen
  \bibfield  {author} {\bibinfo {author} {\bibfnamefont {W.}~\bibnamefont
  {Ballinger}}, \bibinfo {author} {\bibfnamefont {J.}~\bibnamefont {Peacock}},
  \ and\ \bibinfo {author} {\bibfnamefont {A.}~\bibnamefont {Heavens}},\ }\href
  {\doibase 10.1093/mnras/282.3.877} {\bibfield  {journal} {\bibinfo  {journal}
  {Mon.Not.Roy.Astron.Soc.}\ }\textbf {\bibinfo {volume} {282}},\ \bibinfo
  {pages} {877} (\bibinfo {year} {1996})},\ \Eprint
  {http://arxiv.org/abs/astro-ph/9605017} {arXiv:astro-ph/9605017 [astro-ph]}
  \BibitemShut {NoStop}%
\bibitem [{\citenamefont {Matsubara}\ and\ \citenamefont
  {Suto}(1996)}]{Matsubara:1996nf}%
  \BibitemOpen
  \bibfield  {author} {\bibinfo {author} {\bibfnamefont {T.}~\bibnamefont
  {Matsubara}}\ and\ \bibinfo {author} {\bibfnamefont {Y.}~\bibnamefont
  {Suto}},\ }\href {\doibase 10.1086/310290} {\bibfield  {journal} {\bibinfo
  {journal} {Astrophys.J.}\ }\textbf {\bibinfo {volume} {470}},\ \bibinfo
  {pages} {L1} (\bibinfo {year} {1996})},\ \Eprint
  {http://arxiv.org/abs/astro-ph/9604142} {arXiv:astro-ph/9604142 [astro-ph]}
  \BibitemShut {NoStop}%
\bibitem [{\citenamefont {Simpson}\ and\ \citenamefont
  {Peacock}(2010)}]{Simpson:2009zj}%
  \BibitemOpen
  \bibfield  {author} {\bibinfo {author} {\bibfnamefont {F.}~\bibnamefont
  {Simpson}}\ and\ \bibinfo {author} {\bibfnamefont {J.~A.}\ \bibnamefont
  {Peacock}},\ }\href {\doibase 10.1103/PhysRevD.81.043512} {\bibfield
  {journal} {\bibinfo  {journal} {Phys.Rev.}\ }\textbf {\bibinfo {volume}
  {D81}},\ \bibinfo {pages} {043512} (\bibinfo {year} {2010})},\ \Eprint
  {http://arxiv.org/abs/0910.3834} {arXiv:0910.3834 [astro-ph.CO]} \BibitemShut
  {NoStop}%
\bibitem [{\citenamefont {Alcock}\ and\ \citenamefont
  {Paczynski}(1979)}]{Alcock:1979mp}%
  \BibitemOpen
  \bibfield  {author} {\bibinfo {author} {\bibfnamefont {C.}~\bibnamefont
  {Alcock}}\ and\ \bibinfo {author} {\bibfnamefont {B.}~\bibnamefont
  {Paczynski}},\ }\href {\doibase 10.1038/281358a0} {\bibfield  {journal}
  {\bibinfo  {journal} {Nature}\ }\textbf {\bibinfo {volume} {281}},\ \bibinfo
  {pages} {358} (\bibinfo {year} {1979})}\BibitemShut {NoStop}%
\bibitem [{\citenamefont {Beutler}\ \emph {et~al.}(2012)\citenamefont
  {Beutler}, \citenamefont {Blake}, \citenamefont {Colless}, \citenamefont
  {Jones}, \citenamefont {Staveley-Smith} \emph {et~al.}}]{Beutler:2012px}%
  \BibitemOpen
  \bibfield  {author} {\bibinfo {author} {\bibfnamefont {F.}~\bibnamefont
  {Beutler}}, \bibinfo {author} {\bibfnamefont {C.}~\bibnamefont {Blake}},
  \bibinfo {author} {\bibfnamefont {M.}~\bibnamefont {Colless}}, \bibinfo
  {author} {\bibfnamefont {D.~H.}\ \bibnamefont {Jones}}, \bibinfo {author}
  {\bibfnamefont {L.}~\bibnamefont {Staveley-Smith}},  \emph {et~al.},\ }\href
  {\doibase 10.1111/j.1365-2966.2012.21136.x} {\bibfield  {journal} {\bibinfo
  {journal} {Mon.Not.Roy.Astron.Soc.}\ }\textbf {\bibinfo {volume} {423}},\
  \bibinfo {pages} {3430} (\bibinfo {year} {2012})},\ \Eprint
  {http://arxiv.org/abs/1204.4725} {arXiv:1204.4725 [astro-ph.CO]} \BibitemShut
  {NoStop}%
\bibitem [{\citenamefont {Chuang}\ \emph {et~al.}(2013)\citenamefont {Chuang},
  \citenamefont {Prada}, \citenamefont {Beutler}, \citenamefont {Eisenstein},
  \citenamefont {Escoffier} \emph {et~al.}}]{Chuang:2013wga}%
  \BibitemOpen
  \bibfield  {author} {\bibinfo {author} {\bibfnamefont {C.-H.}\ \bibnamefont
  {Chuang}}, \bibinfo {author} {\bibfnamefont {F.}~\bibnamefont {Prada}},
  \bibinfo {author} {\bibfnamefont {F.}~\bibnamefont {Beutler}}, \bibinfo
  {author} {\bibfnamefont {D.~J.}\ \bibnamefont {Eisenstein}}, \bibinfo
  {author} {\bibfnamefont {S.}~\bibnamefont {Escoffier}},  \emph {et~al.},\
  }\href@noop {} {\  (\bibinfo {year} {2013})},\ \Eprint
  {http://arxiv.org/abs/1312.4889} {arXiv:1312.4889 [astro-ph.CO]} \BibitemShut
  {NoStop}%
\bibitem [{\citenamefont {Blake}\ \emph {et~al.}(2012)\citenamefont {Blake},
  \citenamefont {Brough}, \citenamefont {Colless}, \citenamefont {Contreras},
  \citenamefont {Couch} \emph {et~al.}}]{Blake:2012pj}%
  \BibitemOpen
  \bibfield  {author} {\bibinfo {author} {\bibfnamefont {C.}~\bibnamefont
  {Blake}}, \bibinfo {author} {\bibfnamefont {S.}~\bibnamefont {Brough}},
  \bibinfo {author} {\bibfnamefont {M.}~\bibnamefont {Colless}}, \bibinfo
  {author} {\bibfnamefont {C.}~\bibnamefont {Contreras}}, \bibinfo {author}
  {\bibfnamefont {W.}~\bibnamefont {Couch}},  \emph {et~al.},\ }\href {\doibase
  10.1111/j.1365-2966.2012.21473.x} {\bibfield  {journal} {\bibinfo  {journal}
  {Mon.Not.Roy.Astron.Soc.}\ }\textbf {\bibinfo {volume} {425}},\ \bibinfo
  {pages} {405} (\bibinfo {year} {2012})},\ \Eprint
  {http://arxiv.org/abs/1204.3674} {arXiv:1204.3674 [astro-ph.CO]} \BibitemShut
  {NoStop}%
\bibitem [{\citenamefont {Beringer}\ \emph {et~al.}(2012)\citenamefont
  {Beringer} \emph {et~al.}}]{Beringer:1900zz}%
  \BibitemOpen
  \bibfield  {author} {\bibinfo {author} {\bibfnamefont {J.}~\bibnamefont
  {Beringer}} \emph {et~al.} (\bibinfo {collaboration} {Particle Data Group}),\
  }\href {\doibase 10.1103/PhysRevD.86.010001} {\bibfield  {journal} {\bibinfo
  {journal} {Phys.Rev.}\ }\textbf {\bibinfo {volume} {D86}},\ \bibinfo {pages}
  {010001} (\bibinfo {year} {2012})}\BibitemShut {NoStop}%
\bibitem [{\citenamefont {Gelman}\ and\ \citenamefont
  {Rubin}(1992)}]{GelmanRubin}%
  \BibitemOpen
  \bibfield  {author} {\bibinfo {author} {\bibfnamefont {A.}~\bibnamefont
  {Gelman}}\ and\ \bibinfo {author} {\bibfnamefont {D.}~\bibnamefont {Rubin}},\
  }\href@noop {} {\bibfield  {journal} {\bibinfo  {journal} {Statistical
  Science}\ }\textbf {\bibinfo {volume} {7}},\ \bibinfo {pages} {457} (\bibinfo
  {year} {1992})},\ \bibinfo {note}
  {\url{http://www.stat.columbia.edu/~gelman/research/published/itsim.pdf}}\BibitemShut
  {NoStop}%
\bibitem [{\citenamefont {MacCrann}\ \emph {et~al.}(2014)\citenamefont
  {MacCrann}, \citenamefont {Zuntz}, \citenamefont {Bridle}, \citenamefont
  {Jain},\ and\ \citenamefont {Becker}}]{MacCrann:2014wfa}%
  \BibitemOpen
  \bibfield  {author} {\bibinfo {author} {\bibfnamefont {N.}~\bibnamefont
  {MacCrann}}, \bibinfo {author} {\bibfnamefont {J.}~\bibnamefont {Zuntz}},
  \bibinfo {author} {\bibfnamefont {S.}~\bibnamefont {Bridle}}, \bibinfo
  {author} {\bibfnamefont {B.}~\bibnamefont {Jain}}, \ and\ \bibinfo {author}
  {\bibfnamefont {M.~R.}\ \bibnamefont {Becker}},\ }\href@noop {} {\  (\bibinfo
  {year} {2014})},\ \Eprint {http://arxiv.org/abs/1408.4742} {arXiv:1408.4742
  [astro-ph.CO]} \BibitemShut {NoStop}%
\bibitem [{\citenamefont {Macaulay}\ \emph {et~al.}(2013)\citenamefont
  {Macaulay}, \citenamefont {Wehus},\ and\ \citenamefont
  {Eriksen}}]{Macaulay:2013swa}%
  \BibitemOpen
  \bibfield  {author} {\bibinfo {author} {\bibfnamefont {E.}~\bibnamefont
  {Macaulay}}, \bibinfo {author} {\bibfnamefont {I.~K.}\ \bibnamefont {Wehus}},
  \ and\ \bibinfo {author} {\bibfnamefont {H.~K.}\ \bibnamefont {Eriksen}},\
  }\href {\doibase 10.1103/PhysRevLett.111.161301} {\bibfield  {journal}
  {\bibinfo  {journal} {Phys.Rev.Lett.}\ }\textbf {\bibinfo {volume} {111}},\
  \bibinfo {pages} {161301} (\bibinfo {year} {2013})},\ \Eprint
  {http://arxiv.org/abs/1303.6583} {arXiv:1303.6583 [astro-ph.CO]} \BibitemShut
  {NoStop}%
\bibitem [{\citenamefont {{Beutler}}\ \emph {et~al.}(2014)\citenamefont
  {{Beutler}}, \citenamefont {{Saito}}, \citenamefont {{Seo}}, \citenamefont
  {{Brinkmann}}, \citenamefont {{Dawson}} \emph {et~al.}}]{Beutler:2013yhm}%
  \BibitemOpen
  \bibfield  {author} {\bibinfo {author} {\bibfnamefont {F.}~\bibnamefont
  {{Beutler}}}, \bibinfo {author} {\bibfnamefont {S.}~\bibnamefont {{Saito}}},
  \bibinfo {author} {\bibfnamefont {H.-J.}\ \bibnamefont {{Seo}}}, \bibinfo
  {author} {\bibfnamefont {J.}~\bibnamefont {{Brinkmann}}}, \bibinfo {author}
  {\bibfnamefont {K.~S.}\ \bibnamefont {{Dawson}}},  \emph {et~al.},\ }\href
  {\doibase 10.1093/mnras/stu1051} {\bibfield  {journal} {\bibinfo  {journal}
  {\mnras}\ }\textbf {\bibinfo {volume} {443}},\ \bibinfo {pages} {1065}
  (\bibinfo {year} {2014})},\ \Eprint {http://arxiv.org/abs/1312.4611}
  {arXiv:1312.4611} \BibitemShut {NoStop}%
\bibitem [{\citenamefont {{Reid}}\ \emph {et~al.}(2014)\citenamefont {{Reid}},
  \citenamefont {{Seo}}, \citenamefont {{Leauthaud}}, \citenamefont
  {{Tinker}},\ and\ \citenamefont {{White}}}]{Reid:2014iaa}%
  \BibitemOpen
  \bibfield  {author} {\bibinfo {author} {\bibfnamefont {B.~A.}\ \bibnamefont
  {{Reid}}}, \bibinfo {author} {\bibfnamefont {H.-J.}\ \bibnamefont {{Seo}}},
  \bibinfo {author} {\bibfnamefont {A.}~\bibnamefont {{Leauthaud}}}, \bibinfo
  {author} {\bibfnamefont {J.~L.}\ \bibnamefont {{Tinker}}}, \ and\ \bibinfo
  {author} {\bibfnamefont {M.}~\bibnamefont {{White}}},\ }\href {\doibase
  10.1093/mnras/stu1391} {\bibfield  {journal} {\bibinfo  {journal} {\mnras}\
  }\textbf {\bibinfo {volume} {444}},\ \bibinfo {pages} {476} (\bibinfo {year}
  {2014})},\ \Eprint {http://arxiv.org/abs/1404.3742} {arXiv:1404.3742}
  \BibitemShut {NoStop}%
\bibitem [{\citenamefont {Beutler}\ \emph {et~al.}(2014)\citenamefont {Beutler}
  \emph {et~al.}}]{Beutler:2014yhv}%
  \BibitemOpen
  \bibfield  {author} {\bibinfo {author} {\bibfnamefont {F.}~\bibnamefont
  {Beutler}} \emph {et~al.} (\bibinfo {collaboration} {BOSS Collaboration}),\
  }\href {\doibase 10.1093/mnras/stu1702} {\bibfield  {journal} {\bibinfo
  {journal} {Mon.Not.Roy.Astron.Soc.}\ }\textbf {\bibinfo {volume} {444}},\
  \bibinfo {pages} {3501} (\bibinfo {year} {2014})},\ \Eprint
  {http://arxiv.org/abs/1403.4599} {arXiv:1403.4599 [astro-ph.CO]} \BibitemShut
  {NoStop}%
\bibitem [{\citenamefont {Salvatelli}\ \emph {et~al.}(2014)\citenamefont
  {Salvatelli}, \citenamefont {Said}, \citenamefont {Bruni}, \citenamefont
  {Melchiorri},\ and\ \citenamefont {Wands}}]{Salvatelli:2014zta}%
  \BibitemOpen
  \bibfield  {author} {\bibinfo {author} {\bibfnamefont {V.}~\bibnamefont
  {Salvatelli}}, \bibinfo {author} {\bibfnamefont {N.}~\bibnamefont {Said}},
  \bibinfo {author} {\bibfnamefont {M.}~\bibnamefont {Bruni}}, \bibinfo
  {author} {\bibfnamefont {A.}~\bibnamefont {Melchiorri}}, \ and\ \bibinfo
  {author} {\bibfnamefont {D.}~\bibnamefont {Wands}},\ }\href {\doibase
  10.1103/PhysRevLett.113.181301} {\bibfield  {journal} {\bibinfo  {journal}
  {Phys.Rev.Lett.}\ }\textbf {\bibinfo {volume} {113}},\ \bibinfo {pages}
  {181301} (\bibinfo {year} {2014})},\ \Eprint {http://arxiv.org/abs/1406.7297}
  {arXiv:1406.7297 [astro-ph.CO]} \BibitemShut {NoStop}%
\bibitem [{\citenamefont {Song}\ \emph {et~al.}(2014)\citenamefont {Song},
  \citenamefont {Sabiu}, \citenamefont {Okumura}, \citenamefont {Oh},\ and\
  \citenamefont {Linder}}]{Song:2014nba}%
  \BibitemOpen
  \bibfield  {author} {\bibinfo {author} {\bibfnamefont {Y.-S.}\ \bibnamefont
  {Song}}, \bibinfo {author} {\bibfnamefont {C.~G.}\ \bibnamefont {Sabiu}},
  \bibinfo {author} {\bibfnamefont {T.}~\bibnamefont {Okumura}}, \bibinfo
  {author} {\bibfnamefont {M.}~\bibnamefont {Oh}}, \ and\ \bibinfo {author}
  {\bibfnamefont {E.~V.}\ \bibnamefont {Linder}},\ }\href {\doibase
  10.1088/1475-7516/2014/12/005} {\bibfield  {journal} {\bibinfo  {journal}
  {JCAP}\ }\textbf {\bibinfo {volume} {1412}},\ \bibinfo {pages} {005}
  (\bibinfo {year} {2014})},\ \Eprint {http://arxiv.org/abs/1407.2257}
  {arXiv:1407.2257 [astro-ph.CO]} \BibitemShut {NoStop}%
\bibitem [{\citenamefont {Seljak}\ \emph {et~al.}(2006)\citenamefont {Seljak},
  \citenamefont {Slosar},\ and\ \citenamefont {McDonald}}]{Seljak:2006bg}%
  \BibitemOpen
  \bibfield  {author} {\bibinfo {author} {\bibfnamefont {U.}~\bibnamefont
  {Seljak}}, \bibinfo {author} {\bibfnamefont {A.}~\bibnamefont {Slosar}}, \
  and\ \bibinfo {author} {\bibfnamefont {P.}~\bibnamefont {McDonald}},\ }\href
  {\doibase 10.1088/1475-7516/2006/10/014} {\bibfield  {journal} {\bibinfo
  {journal} {JCAP}\ }\textbf {\bibinfo {volume} {0610}},\ \bibinfo {pages}
  {014} (\bibinfo {year} {2006})},\ \Eprint
  {http://arxiv.org/abs/astro-ph/0604335} {arXiv:astro-ph/0604335 [astro-ph]}
  \BibitemShut {NoStop}%
\bibitem [{\citenamefont {Hou}\ \emph {et~al.}(2014)\citenamefont {Hou},
  \citenamefont {Reichardt}, \citenamefont {Story}, \citenamefont {Follin},
  \citenamefont {Keisler} \emph {et~al.}}]{Hou:2012xq}%
  \BibitemOpen
  \bibfield  {author} {\bibinfo {author} {\bibfnamefont {Z.}~\bibnamefont
  {Hou}}, \bibinfo {author} {\bibfnamefont {C.}~\bibnamefont {Reichardt}},
  \bibinfo {author} {\bibfnamefont {K.}~\bibnamefont {Story}}, \bibinfo
  {author} {\bibfnamefont {B.}~\bibnamefont {Follin}}, \bibinfo {author}
  {\bibfnamefont {R.}~\bibnamefont {Keisler}},  \emph {et~al.},\ }\href
  {\doibase 10.1088/0004-637X/782/2/74} {\bibfield  {journal} {\bibinfo
  {journal} {Astrophys.J.}\ }\textbf {\bibinfo {volume} {782}},\ \bibinfo
  {pages} {74} (\bibinfo {year} {2014})},\ \Eprint
  {http://arxiv.org/abs/1212.6267} {arXiv:1212.6267 [astro-ph.CO]} \BibitemShut
  {NoStop}%
\bibitem [{\citenamefont {Dvorkin}\ \emph {et~al.}(2014)\citenamefont
  {Dvorkin}, \citenamefont {Wyman}, \citenamefont {Rudd},\ and\ \citenamefont
  {Hu}}]{Dvorkin:2014lea}%
  \BibitemOpen
  \bibfield  {author} {\bibinfo {author} {\bibfnamefont {C.}~\bibnamefont
  {Dvorkin}}, \bibinfo {author} {\bibfnamefont {M.}~\bibnamefont {Wyman}},
  \bibinfo {author} {\bibfnamefont {D.~H.}\ \bibnamefont {Rudd}}, \ and\
  \bibinfo {author} {\bibfnamefont {W.}~\bibnamefont {Hu}},\ }\href {\doibase
  10.1103/PhysRevD.90.083503} {\bibfield  {journal} {\bibinfo  {journal}
  {Phys.Rev.}\ }\textbf {\bibinfo {volume} {D90}},\ \bibinfo {pages} {083503}
  (\bibinfo {year} {2014})},\ \Eprint {http://arxiv.org/abs/1403.8049}
  {arXiv:1403.8049 [astro-ph.CO]} \BibitemShut {NoStop}%
\bibitem [{\citenamefont {Bocquet}\ \emph {et~al.}(2015)\citenamefont {Bocquet}
  \emph {et~al.}}]{Bocquet:2014lmj}%
  \BibitemOpen
  \bibfield  {author} {\bibinfo {author} {\bibfnamefont {S.}~\bibnamefont
  {Bocquet}} \emph {et~al.} (\bibinfo {collaboration} {SPT Collaboration}),\
  }\href {\doibase 10.1088/0004-637X/799/2/214} {\bibfield  {journal} {\bibinfo
   {journal} {Astrophys.J.}\ }\textbf {\bibinfo {volume} {799}},\ \bibinfo
  {pages} {214} (\bibinfo {year} {2015})},\ \Eprint
  {http://arxiv.org/abs/1407.2942} {arXiv:1407.2942 [astro-ph.CO]} \BibitemShut
  {NoStop}%
\bibitem [{\citenamefont {Costanzi}\ \emph {et~al.}(2014)\citenamefont
  {Costanzi}, \citenamefont {Sartoris}, \citenamefont {Viel},\ and\
  \citenamefont {Borgani}}]{Costanzi:2014tna}%
  \BibitemOpen
  \bibfield  {author} {\bibinfo {author} {\bibfnamefont {M.}~\bibnamefont
  {Costanzi}}, \bibinfo {author} {\bibfnamefont {B.}~\bibnamefont {Sartoris}},
  \bibinfo {author} {\bibfnamefont {M.}~\bibnamefont {Viel}}, \ and\ \bibinfo
  {author} {\bibfnamefont {S.}~\bibnamefont {Borgani}},\ }\href {\doibase
  10.1088/1475-7516/2014/10/081} {\bibfield  {journal} {\bibinfo  {journal}
  {JCAP}\ }\textbf {\bibinfo {volume} {1410}},\ \bibinfo {pages} {081}
  (\bibinfo {year} {2014})},\ \Eprint {http://arxiv.org/abs/1407.8338}
  {arXiv:1407.8338 [astro-ph.CO]} \BibitemShut {NoStop}%
\bibitem [{\citenamefont {Archidiacono}\ \emph {et~al.}(2014)\citenamefont
  {Archidiacono}, \citenamefont {Fornengo}, \citenamefont {Gariazzo},
  \citenamefont {Giunti}, \citenamefont {Hannestad} \emph
  {et~al.}}]{Archidiacono:2014apa}%
  \BibitemOpen
  \bibfield  {author} {\bibinfo {author} {\bibfnamefont {M.}~\bibnamefont
  {Archidiacono}}, \bibinfo {author} {\bibfnamefont {N.}~\bibnamefont
  {Fornengo}}, \bibinfo {author} {\bibfnamefont {S.}~\bibnamefont {Gariazzo}},
  \bibinfo {author} {\bibfnamefont {C.}~\bibnamefont {Giunti}}, \bibinfo
  {author} {\bibfnamefont {S.}~\bibnamefont {Hannestad}},  \emph {et~al.},\
  }\href {\doibase 10.1088/1475-7516/2014/06/031} {\bibfield  {journal}
  {\bibinfo  {journal} {JCAP}\ }\textbf {\bibinfo {volume} {1406}},\ \bibinfo
  {pages} {031} (\bibinfo {year} {2014})},\ \Eprint
  {http://arxiv.org/abs/1404.1794} {arXiv:1404.1794 [astro-ph.CO]} \BibitemShut
  {NoStop}%
\bibitem [{\citenamefont {Hinshaw}\ \emph {et~al.}(2013)\citenamefont {Hinshaw}
  \emph {et~al.}}]{Hinshaw:2012aka}%
  \BibitemOpen
  \bibfield  {author} {\bibinfo {author} {\bibfnamefont {G.}~\bibnamefont
  {Hinshaw}} \emph {et~al.} (\bibinfo {collaboration} {WMAP}),\ }\href
  {\doibase 10.1088/0067-0049/208/2/19} {\bibfield  {journal} {\bibinfo
  {journal} {Astrophys.J.Suppl.}\ }\textbf {\bibinfo {volume} {208}},\ \bibinfo
  {pages} {19} (\bibinfo {year} {2013})},\ \Eprint
  {http://arxiv.org/abs/1212.5226} {arXiv:1212.5226 [astro-ph.CO]} \BibitemShut
  {NoStop}%
\bibitem [{\citenamefont {Hu}\ and\ \citenamefont
  {Kravtsov}(2003)}]{HuandKravtsov}%
  \BibitemOpen
  \bibfield  {author} {\bibinfo {author} {\bibfnamefont {W.}~\bibnamefont
  {Hu}}\ and\ \bibinfo {author} {\bibfnamefont {A.~V.}\ \bibnamefont
  {Kravtsov}},\ }\href {\doibase 10.1086/345846} {\bibfield  {journal}
  {\bibinfo  {journal} {Astrophys.J.}\ }\textbf {\bibinfo {volume} {584}},\
  \bibinfo {pages} {702} (\bibinfo {year} {2003})},\ \Eprint
  {http://arxiv.org/abs/astro-ph/0203169} {arXiv:astro-ph/0203169 [astro-ph]}
  \BibitemShut {NoStop}%
\end{thebibliography}%

\appendix

\section{Cluster analysis details}\label{App:clusters}

Here we give more details regarding the cluster analysis, which closely
followed one given in the \citet{Rozo} MaxBCG cosmological constraints paper.

\begin{figure}
\includegraphics[width=0.45\textwidth]{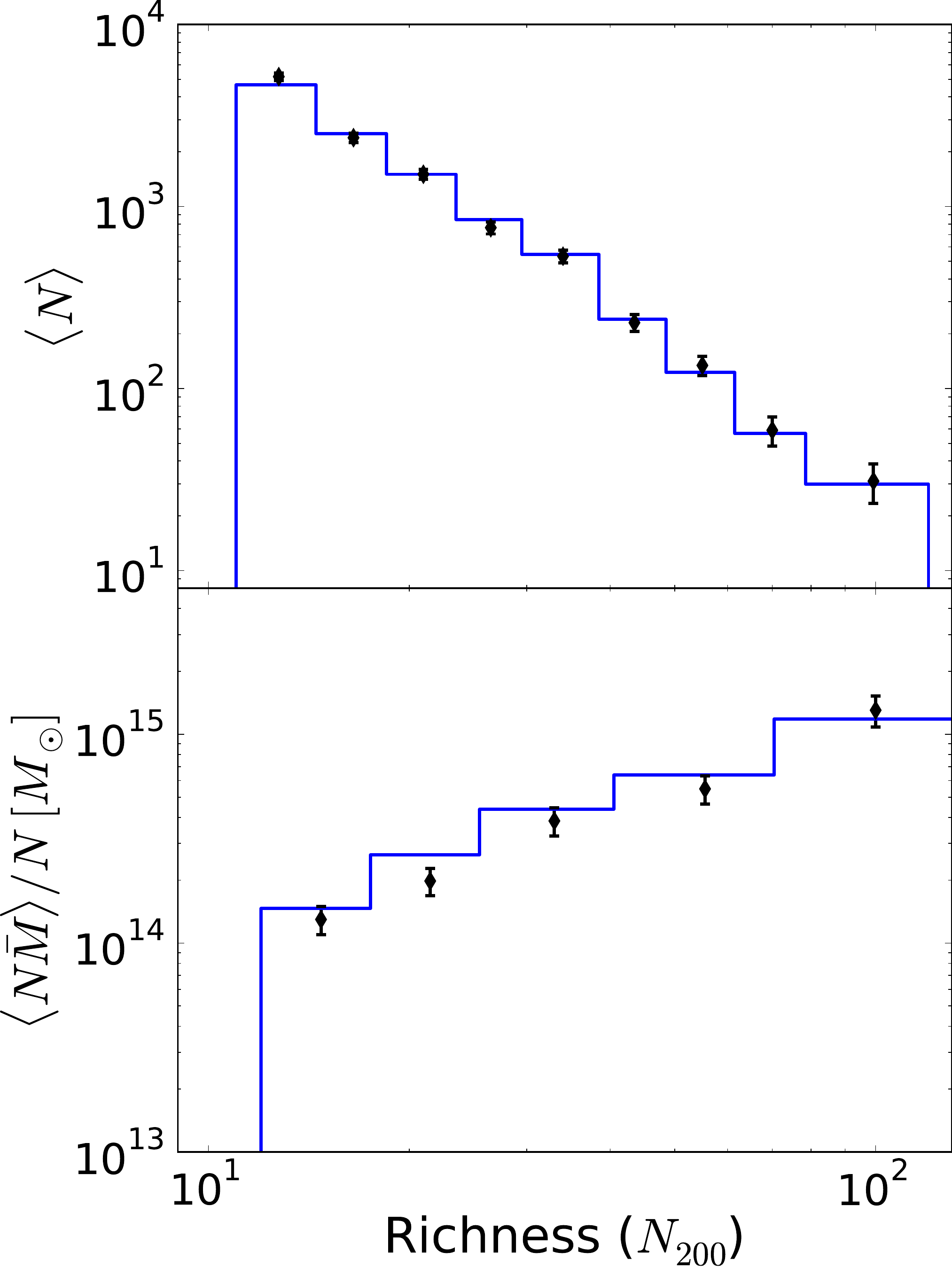}
\caption{Top: Number of galaxy clusters within a given richness bin in the MaxBCG
 data set. Errors shown are the diagonal parts of the covariance matrix. The step
  function shown uses the parameter values from the best-fit $\Lambda$CDM model 
  (Column 2 of Table \ref{tab:results}). The data are summarized in 
  Table \ref{tab:numbercount}. Bottom: Mean mass of galaxy clusters within the 
  given richness bin in the MaxBCG data set. The step function uses the same 
  parameter values as the top figure. The data are summarized in Table 
  \ref{tab:meanmass}.}
\label{fig:clusterdata}
\end{figure}

The analysis is based on assigning ``richness'' to each cluster; this is
defined as the number of galaxies in $R_{200}$, the radius at
which the average density of the cluster is 200 times that of the critical
density of the universe. Moreover, the mass is determined from richness via the
richness-mass relation which has been calibrated using weak gravitational
lensing measurements by \citet{Johnston:2007}.
The cluster numbers in each richness bin are shown in Table
\ref{tab:numbercount}, while the clusters' mean mass per bin is shown in Table
\ref{tab:meanmass} and in Fig.\ \ref{fig:clusterdata}.

In addition to the data in Table \ref{tab:numbercount}, there are also five
clusters which have $N_{200} > 120$. Due to the high richness of these
clusters, they are not analyzed with a standard $\chi^2$ approach, and are
instead included in the analysis on an individual basis.

\begin{table}[h]
\begin{center}
\begin{tabular}{c c}
\hline \hline
Richness bin & No. of Clusters \\ \hline
11-14 & 5167\\
14-18 & 2387\\
19-23 & 1504\\
24-29 & 765\\
30-38 & 533\\
39-48 & 230\\
49-61 & 134\\
62-78 & 59\\
79-120 & 31\\\hline
\end{tabular}
\caption{The number of clusters with a richness within the given bin.}
\label{tab:numbercount}
\end{center}
\end{table}

\begin{table}[h]
\begin{center}
\begin{tabular}{c c c}
\hline \hline
Richness bin & No. of Clusters & $\langle M_{200b} \rangle [10^{14} M_\odot]$ \\ \hline
12-17 & 5651 & 1.298\\
18-25 & 2269 & 1.983\\
26-40 & 1021 & 3.846\\
41-70 & 353 & 5.475\\
71+ & 55 & 13.03\\\hline
\end{tabular}
\caption{Mean mass (and their number) of clusters with a richness within the given bin.}
\label{tab:meanmass}
\end{center}
\end{table}

As already implied, the overdensity of $\Delta = 200$ is adopted to define
cluster masses. In addition, the masses measured have been assumed to be in
cosmology with $\Omega_M = 0.27$.  For other cosmologies, this leads to an
overdensity of $\Delta_v = 200 (0.27/\Omega_M)$.  To correctly account for
this, we rescale the quoted masses from Rozo et. al. for each tested cosmology
using the equations from \citet{HuandKravtsov} for mass
rescaling
\begin{equation}
\frac{M_h}{M_v} = \frac{\Delta_h}{\Delta_v}\frac{1}{c^3}\left(\frac{r_h}{r_s}\right)^3
\end{equation}
where $r$ is the radius of the halo for a given overdensity, $c$ the
concentration factor, and $\Delta$ is the overdensity. The ratio of radii can
be written as
\begin{equation}
\frac{r_s}{r_h}=x\left(\frac{\Delta_v}{\Delta_h} f\left(\frac{1}{c}\right)\right)
\end{equation}
where
\begin{equation}
f(x) = x^3 \left[\ln(1+x^{-1}) - (1+x)^{-1}\right]
\end{equation}
and its inverse can be approximated as
\begin{equation}
x(f) = \left[a_1 f^{2p} + \left(\frac{3}{4}\right)^2 \right]^{-1/2}+2f
\end{equation}
where $p=a_2+a_3 \ln f + a_4 (\ln f)^2$, and $a_i = \lbrace 0.5116,
-0.4283, -3.13 \times 10^{-3},-3.52 \times 10^{-5}\rbrace$. Finally, the
concentration can be expressed in terms of the mass as
\begin{equation}
c(M_v) = 9(1+z)^{-1} (M_v/M_*)^{-0.13},
\end{equation}
where $M_*$ is calculated at the present day.

As mentioned in Sec.~\ref{sec:Clusters}, the probability weighting functions are
\begin{align}
\langle\psi|M\rangle &= \int dN_{200} P(N_{200}|M)\psi(N_{200}), \label{app_eq:mprob}\\
\langle\phi|z\rangle &= \int dz_\text{photo} P(z_\text{photo}|z)\phi(z_\text{photo}). \label{app_eq:zprob}
\end{align}
Here $P(N_{200}|M)$ is a log-normal distribution with an unknown variance
 $\sigma^2_{NM} = \text{Var}(\ln N_{200} | M)$ and an expected value
\begin{align}
&\langle \ln N_{200} | M\rangle = \\ \nonumber
&\frac{\log_{10}\left(\frac{M}{M_1}\right)\langle \ln N_{200} | M_2\rangle -
  \log_{10}\left(\frac{M}{M_2}\right)\langle \ln N_{200} | M_1\rangle}{\log_{10}\left(\frac{M_2}{M_1}\right)}
\end{align}
where $M_1 = 1.3 \times 10^{14} M_\odot$, $M_2 = 1.3 \times 10^{15} M_\odot$,
and $\langle \ln N_{200} | M_1\rangle$, $\langle \ln N_{200} | M_2\rangle$,
and $\sigma^2_{NM}$ are nuisance parameters, which are
marginalized over during the analysis of the cluster data. Likewise, the
probability weighting function $P(z_\text{photo}|z)$ is a Gaussian distribution
with standard deviation $\sigma_z = 0.008$ and an expectation value
$\langle z_\text{photo} | z \rangle = z$. $\psi(N_{200})$ and 
$\phi(z_\text{photo})$ are once again binning functions, where the 
$z_\text{photo}$ bin is $[0.1, 0.3]$ from the range of photometric 
data from the SDSS survey.

The cluster likelihood consists of two parts \cite{Rozo}; the main part is
defined via
\begin{equation}
-2 \log \mathcal{L_\text{main}} = \Delta x^\text{T} C^{-1} \Delta x
\end{equation}
where $\Delta x = (x^\text{data} - x^\text{theory})$. 
The x vector of observables is
\begin{equation}
x = \{ N_1, ..., N_9, (N\bar{M})_1, ..., (N\bar{M})_5\}.
\end{equation}
where $N_1$ though $N_9$ are the cluster counts in the
respective richness bins, while $(N\bar{M})_1$ through $(N\bar{M})_5$ are the total 
mass of clusters in bins.

The covariance $C$ of the cluster data takes into account uncertainties due to shot noise, sample
variance, the stochasticity of the mass-richness relation, measurement error of the weak
lensing masses, and uncertainties in the purity and completeness of the sample. For more
information regarding these uncertainties, see \citet{Rozo} from which we
adopt the prescription for calculating the covariance matrix.

As previously stated, there are five clusters in the MaxBCG data set which have $N_{200} =
126$, $139$, $156$, $164$, and $188$. These clusters are added on a individual
basis to the analysis with the likelihood
\begin{equation}
\log \mathcal{L_\text{tail}} = 
\sum_{N_{200} > 120} \langle N \rangle - \sum_{N(N_{200}) = 1} \langle N \rangle + \log \langle N \rangle
\end{equation}
where the first sum is over all richnesses $> 120$, which is subtracted by
the second sum, which is for those richness bins that contain a cluster.
This additional piece is combined with the main part to obtain the full
likelihood of observing a set of cluster counts and their masses
\begin{equation}
\mathcal{L}_\text{cluster} = \mathcal{L}_\text{main} \mathcal{L}_\text{tail}.
\end{equation}

\section{RSD analysis details}\label{app:rsd-details}

\subsection{RSD correlation matrices}

For completeness, in Tables \ref{tab:rsd-corr_boss} and \ref{tab:rsd-corr_wigglez} 
we present the correlation matrices for the BOSS Low-$z$, BOSS CMASS, and
WiggleZ measurements used for the analysis. The square roots of the diagonal 
uncertainties for these measurements can be found in Table \ref{tab:rsddata}.

\begin{table}[th]
\begin{center}
\begin{tabular}{c|ccc} \hline \hline
$z=0.32$    & $H(z)$ & $D_A(z)$ & $f\sigma_8$ \\\hline
$H(z)$      & $1.00$ & $-0.32$  & $0.35$ \\
$D_A(z)$    &   ---  & $1.00$   & $0.51$ \\
$f\sigma_8$ &   ---  &   ---    & $1.00$ \\\hline
\end{tabular}
~~~
\begin{tabular}{c|ccc} \hline \hline
$z=0.57$    & $H(z)$ & $D_A(z)$ & $f\sigma_8$ \\\hline
$H(z)$      & $1.00$ & $-0.67$  & $0.05$ \\
$D_A(z)$    &   ---  & $1.00$   & $0.40$ \\
$f\sigma_8$ &   ---  &   ---    & $1.00$ \\\hline
\end{tabular}
\caption{Correlation matrices for the BOSS Low-z (left) and CMASS (right)
  samples of our RSD data set. 
}
\label{tab:rsd-corr_boss}
\end{center}
\end{table}

\begin{table}[th]
\begin{center}
\begin{tabular}{c|cccccc} \hline \hline
                & $F_a$ & $F_b$ & $F_c$ & $(f\sigma_8)_a$ & $(f\sigma_8)_b$ & $(f\sigma_8)_c$ \\\hline
$F_a$           & $1.00$ & $0.52$  & $0.00$ & $0.73$ & $0.35$ & $0.00$ \\
$F_b$           &  ---   & $1.00$  & $0.50$ & $0.38$ & $0.74$ & $0.43$ \\
$F_c$           &  ---   &  ---    & $1.00$ & $0.00$ & $0.43$ & $0.85$ \\
$(f\sigma_8)_a$ &  ---   &  ---    &  ---   & $1.00$ & $0.51$ & $0.00$ \\
$(f\sigma_8)_b$ &  ---   &  ---    &  ---   &  ---   & $1.00$ & $0.56$ \\
$(f\sigma_8)_c$ &  ---   &  ---    &  ---   &  ---   &  ---   & $1.00$ \\\hline
\end{tabular}
\caption{Correlation matrix for the WiggleZ sample of our RSD
  data set. Terms with subscript $a$ are values at $z=0.44$, subscript $b$ at
  $z=0.60$, and subscript $c$ at $z=0.73$.}
\label{tab:rsd-corr_wigglez}
\end{center}
\end{table}

\subsection{From \texorpdfstring{$(D_A, H)$}{(DA, H)} covariance to error in \texorpdfstring{$F$}{F}}

In order to make the error bars in Fig. \ref{fig:rsd-data} for the two BOSS
samples (Low-z and CMASS), we need to project the $3\times 3$ covariance
matrix in $f\sigma_8$, $H$ and $D_A$ into the $2\times 2$ space $(f\sigma_8,
F)$. Recall, $F$ is defined in Eq.~(\ref{eq:F}) and is essentially
proportional to the product of the Hubble parameter and the angular diameter
distance.

Doing this is a short exercise in statistics. First of all, note that we only
really need the variance in $F$, although computing the covariance between
$f\sigma_8$ and $F$ would be equally straightforward. 

Let us assume that we would like to calculate the variance of the product of
two Gaussian random variables $x$ and $y$. Let $X$ and $Y$ be the mean of
these two variables, and $\delta x\equiv x-X$ and $\delta y\equiv y-Y$. Then
\begin{equation}
\begin{aligned}
{\rm Var}(xy) &= {\rm Var}[(X+\delta x)(Y+\delta y)]\\[0.2cm]
&= {\rm Var}[X\delta y+Y\delta x + \delta x\delta y]
\end{aligned}
\end{equation}
where we dropped the noncontributing variance of a constant. Dropping the
three-point correlations that vanish for Gaussian variables, this evaluates to
\begin{eqnarray}
\begin{aligned}
{\rm Var}(xy) &=
X^2{\rm Var}(\delta y) + Y^2{\rm Var}(\delta x) + 2XY{\rm Cov}(\delta x, \delta y)\\[0.1cm]
& + {\rm Var}(\delta x\delta y) \\[0.13cm]
&= 
X^2{\rm Var}(\delta y) + Y^2{\rm Var}(\delta x) + 2XY{\rm Cov}(\delta x, \delta y) \\[0.1cm]
&+  {\rm Var}(\delta x){\rm Var}(\delta y) + {\rm Cov}(\delta x, \delta y)^2,
\end{aligned}
\end{eqnarray}
where in the last expression we evaluated ${\rm Var}(\delta x\delta y)$ using
Wick's theorem.  This is the expression that we need. Denoting for clarity
$D_A$ and $H$ to be the means, and $\mathcal{D}_A$ and $\mathcal{H}$ to be
fluctuations around the mean in the angular diameter distance and Hubble parameter, in
our case we have
\begin{eqnarray}
\nonumber
(1+z)^{-2} {\rm Var}(F) &=&
H^2{\rm Var}(\mathcal{D}_A) + D_A^2{\rm Var}(\mathcal{H}) \\[0.12cm]
&+& 2HD_A{\rm Cov}(\mathcal{D}_A, \mathcal{H})\\[0.12cm]
&+& {\rm Var}(\mathcal{D}_A){\rm Var}(\mathcal{H}) + 
{\rm Cov}(\mathcal{D}_A, \mathcal{H})^2.
\nonumber
\end{eqnarray}
With this equation we can evaluate the error in $F$, given the covariance
matrix in the angular diameter distance and Hubble parameter.

\section{Plots with separated contours}

\begin{figure*}[t]
\includegraphics[width=0.9\textwidth]{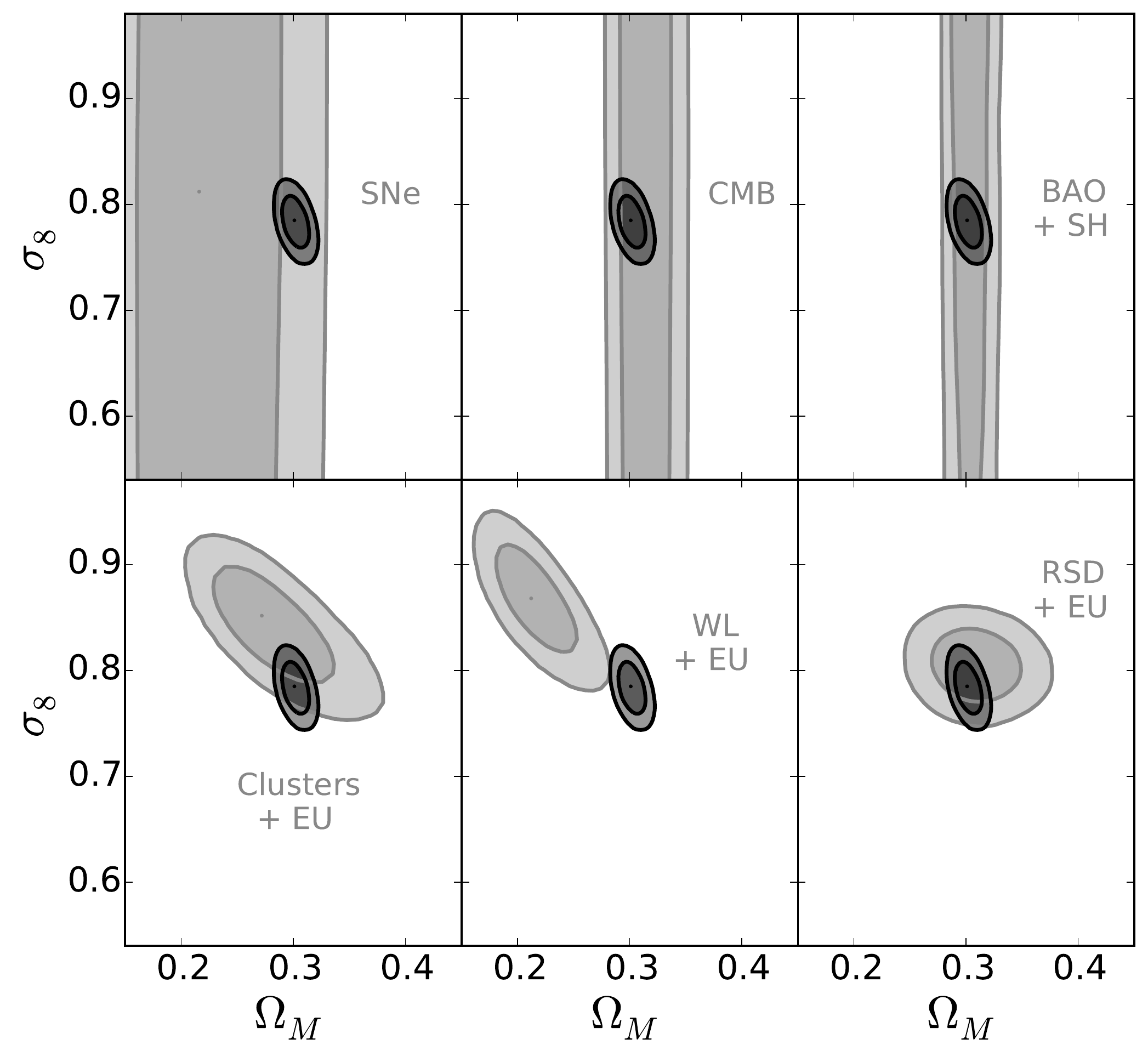}
\caption{Same as the left panel of Fig.~\ref{fig:unsplit}, but the various
  probes have been separated for easier viewing. The smaller, dark set of
  contours corresponds to all probes combined.}
\label{fig:om-s8-unsplit-separate}
\end{figure*}

\begin{figure*}[t]
\includegraphics[width=0.9\textwidth]{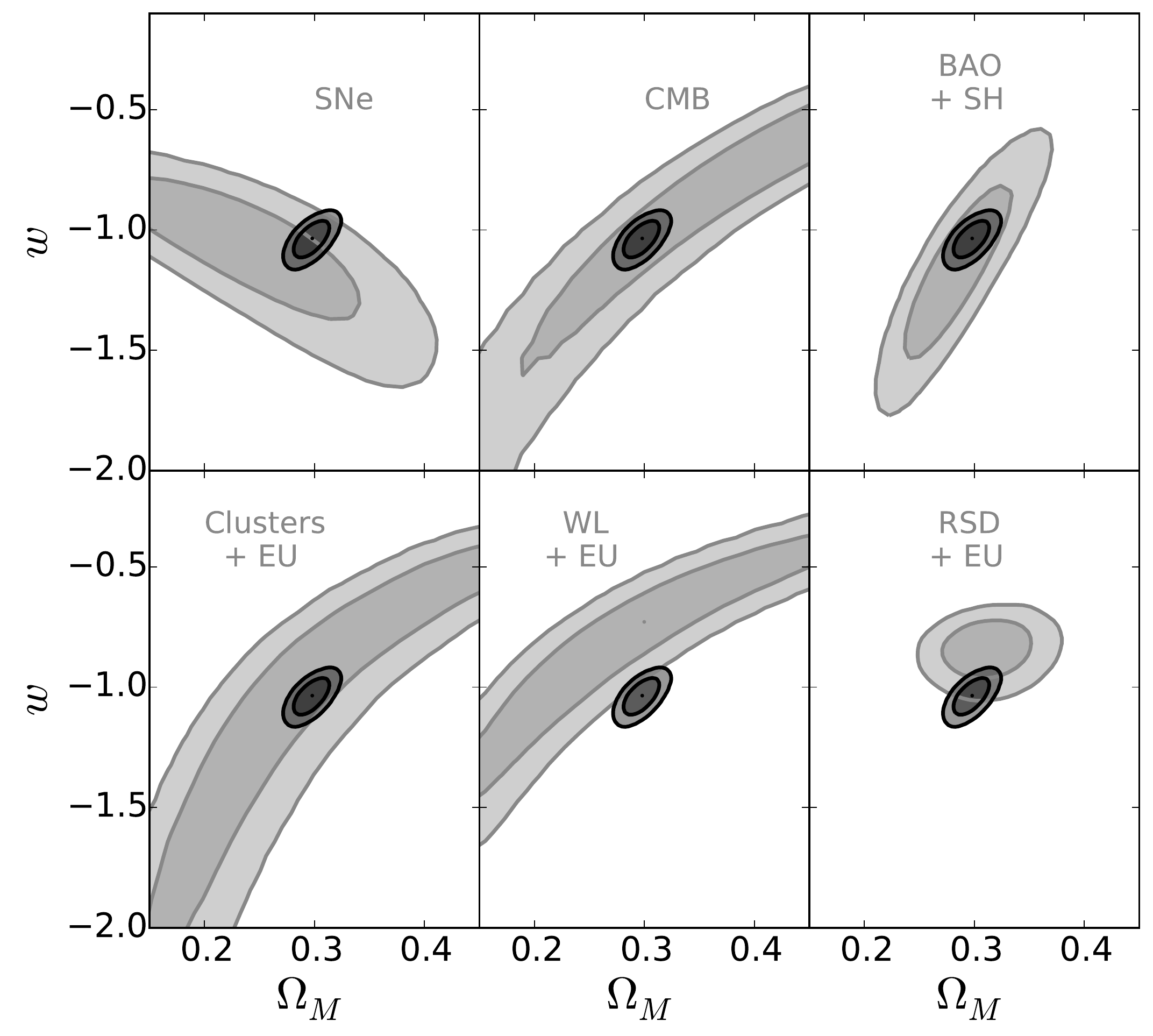}
\caption{Same as the right panel of Fig.~\ref{fig:unsplit}, but the various probes have been separated
for easier viewing. The smaller, dark set of contours
    corresponds to all probes combined.}
\label{fig:om-w-unsplit-separate}
\end{figure*}

\begin{figure*}[t]
\includegraphics[width=0.9\textwidth]{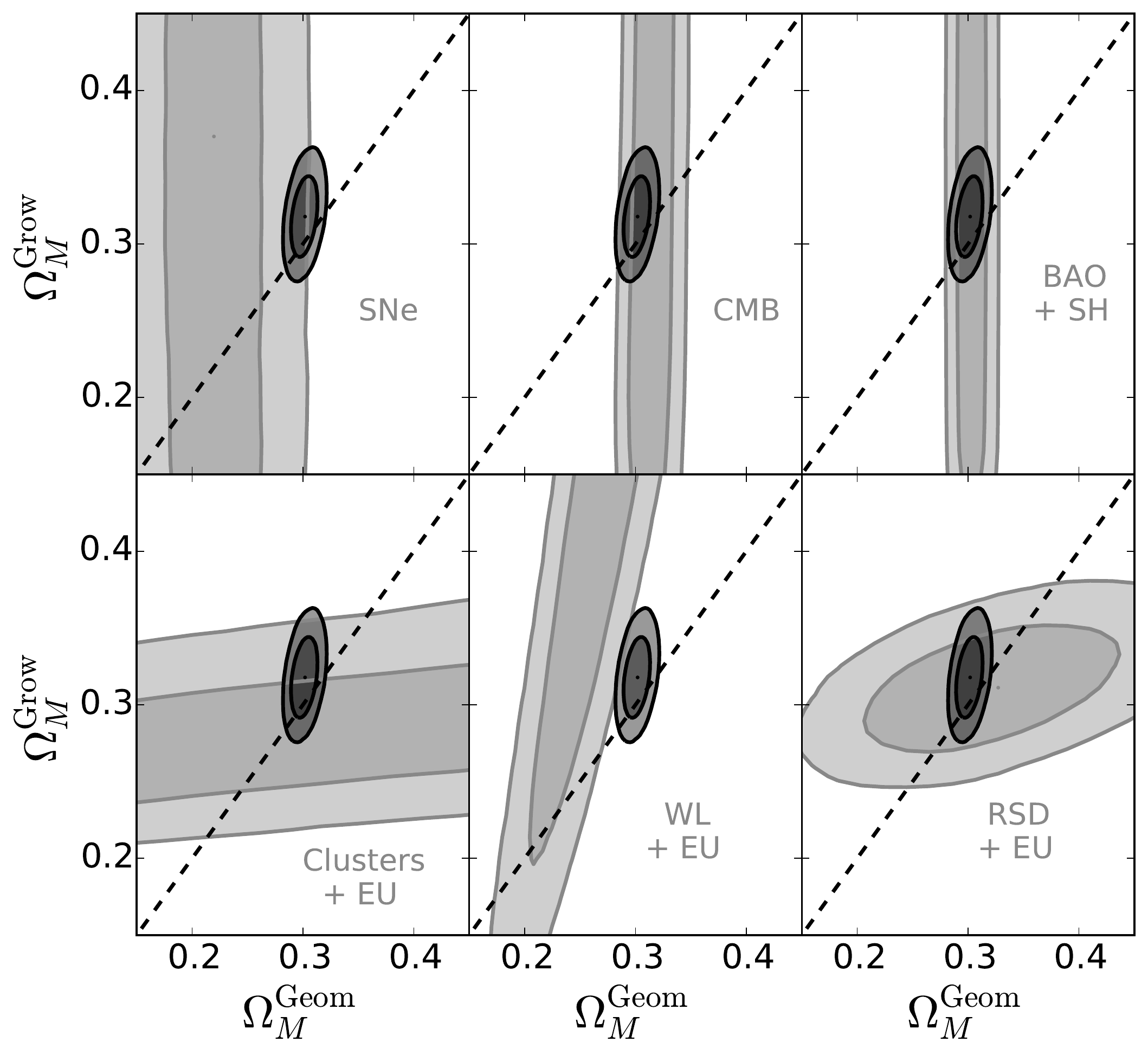}
\caption{Same as Fig.~\ref{fig:om-split}, but the various probes have been separated
for easier viewing. The smaller, dark set of contours
    corresponds to all probes combined.}
\label{fig:om-split-separate}
\end{figure*}

\begin{figure*}[t]
\includegraphics[width=0.9\textwidth]{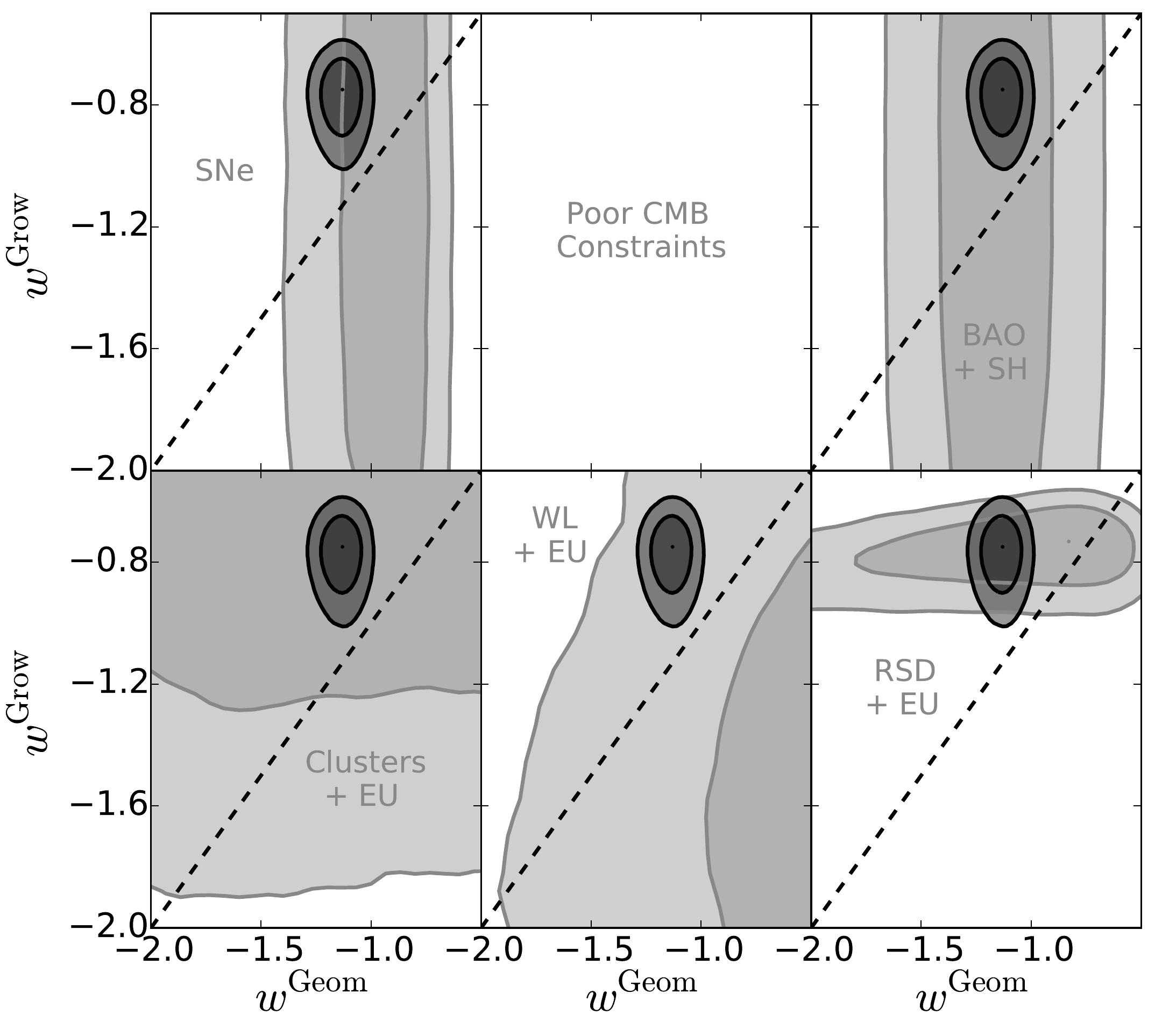}
\caption{Same as Fig.~\ref{fig:w-split}, but the various probes have been
    separated for easier viewing. The smaller, dark set of contours
    corresponds to all probes combined.}
\label{fig:wsplit-separate}
\end{figure*}

In Figs.~\ref{fig:om-s8-unsplit-separate}, \ref{fig:om-w-unsplit-separate},
and \ref{fig:om-split-separate}, and \ref{fig:wsplit-separate}, we include
alternate versions of Figs.~\ref{fig:unsplit}(a), \ref{fig:unsplit}(b),
\ref{fig:om-split}, and \ref{fig:w-split}. Here, for clarity, each probe's
constraints have been shown separately. In each case, the combined constraint
has also been shown.

\end{document}